\documentclass[twocolumn,twocolappendix,tighten]{aastex631}

\usepackage[squaren]{SIunits}	

\newcommand{\Msun}{{\rm M}_{\sun}}				
\newcommand{\Rsun}{{\rm R}_{\sun}}				

\received{April 30, 2021}
\accepted{July 9, 2021}

\submitjournal{APJ}

\shorttitle{A Catalogue of Potential Post Common Envelope Binaries}
\shortauthors{Kruckow et al. (2021)}

\graphicspath{{./}{figures/}}

\begin{document}

\title{A Catalogue of Potential Post Common Envelope Binaries}

\author[0000-0001-9331-0400]{Matthias U. Kruckow}
\affiliation{Yunnan Observatories, Chinese Academy of Sciences, Kunming 650011, China}
\affiliation{Key Laboratory for the Structure and Evolution of Celestial Objects, Chinese Academy of Sciences, Kunming 650011, China}
\correspondingauthor{Matthias U. Kruckow}
\email{mkruckow@ynao.ac.cn}

\author[0000-0001-5853-6017]{Patrick G. Neunteufel}
\affiliation{Max Planck Institut f\"{u}r Astrophysik, Karl-Schwarzschild-Stra\ss{}e 1, 85748 Garching bei M\"{u}nchen, Germany}

\author[0000-0003-0972-1376]{Rosanne Di Stefano}
\affiliation{Harvard-Smithsonian Center for Astrophysics, 60 Garden St, Cambridge, MA 02138, US}

\author[0000-0003-2467-3755]{Yan Gao}
\affiliation{Institute of Gravitational Wave Astronomy, School of Physics and Astronomy, University of Birmingham, Birmingham, B15 2TT, UK}

\author[0000-0002-4343-0487]{Chiaki Kobayashi}
\affiliation{Centre for Astrophysics Research, School of Physics, Astronomy and Mathematics, University of Hertfordshire, Hatfield, AL10 9AB, UK}




\begin{abstract}
We present a catalogue containing 839 candidate post common envelope systems. Common envelope evolution is very important in stellar astrophysics, particularly in the context of very compact and short-period binaries, including cataclysmic variables, as progenitors of e.g. supernovae type Ia or mergers of black holes and/or neutron stars. At the same time it is a barely understood process in binary evolution. Due to limitations, since partially remedied, on direct simulation, early investigations were mainly focused on providing analytic prescriptions of the outcome of common envelope evolution. In recent years, detailed hydrodynamical calculations have produced deeper insight into the previously elusive process of envelope ejection. However, a direct link between observations and theory of this relatively short-lived phase in binary evolution has not been forthcoming. Therefore, the main insight to be gained from observations has to be derived from the current state of systems likely to have gone through a common envelope. Here we present an extensive catalogue of such observations as found in the literature. The aim of this paper is to provide a reliable set of data, obtained from observations, to be used in the theoretical modelling of common envelope evolution. In this catalogue, the former common envelope donor star is commonly observed as a white dwarf star or as a hot sub-dwarf star. This catalogue includes period and mass estimates, wherever obtainable. Some binaries are border line cases to allow an investigation of the transition between a common envelope formation and other mass-transfer processes.
\end{abstract}



\section{Introduction}
\label{sec:introduction}
Observations of stellar remnants such as white dwarfs (WDs) in close binary systems pose challenging questions for the theory of binary evolution. Most notably, cataclysmic variables (CVs), in which a WD accretes mass at low rates from a hydrogen-rich companion, have orbital separations much smaller than would have been needed to accommodate the WD's progenitor star. This suggested the possibility of a common envelope (CE) phase of evolution, in which an envelope of a star that has come to fill its Roche lobe expands to encompass the binary companion \citep{pac76}. This results in a significant reduction of the binary’s orbital separation. If a binary does not merge in the course of CE evolution the shared envelope is ejected. Although many additional types of close binaries, some including neutron stars (NSs) or black holes (BHs) have since been suggested as possible CE survivors, the physics of CE evolution is still not well understood. Solving this mystery will be important, because CE evolution appears to provide the favoured viable pathway to the formation of close binaries outside of dense stellar environments that may result in supernovae (SNe) type Ia \citep[e.g.][]{hpe95,rbf09,mvd+10,tnp12,cpi2014,akx2016} or mergers of NSs and/or BHs \citep[e.g.][and many more]{bsp99,bkb02,vt03,dbf+12,mv14,es16,svm+17,ktl+18,mgs+19,kni+21}.

The CE phase has remained inaccessible to direct observation. Although there are candidates \citep[e.g.][]{ijn+2013,mmr+17}, confirmation is challenging. The CE is expected to have observable characteristics similar to that of a giant, with its photosphere hiding the binary. Because the phase probably lasts about $10^3$ to $\unit{10^5}{\mathrm{yr}}$, it is not possible to track the evolution of the envelope or the binary within. However, many close binaries could be CE end states. After the ejection of the common envelope, at least one stellar component should be hydrogen depleted and thus observable as, e.g., a WD or a subdwarf O/B star. The ejected common-envelope material is expected to cool and become dim enough to be inaccessible to direct observations. Similarly, a compact remnant will cool and become less luminous if no further nuclear burning takes place. CVs are the quintessential example, although they have evolved through tidal interactions and mass transfer, and do not therefor represent pure CE end states.

During recent years, a large number of possible systems, which are believed to have gone through a CE phase, have been identified. Observations usually focus on individual systems or specific types of stars (see references in appendix table~\ref{tab:dataObsShort}). Recently, major observational efforts went into campaigns which resulted in better characterisations of different types of stellar objects, including potential CE products \citep[e.g.][and several more]{pkh+05,hhd+09,ghp+10,cmm+11,ngs+11,zsg11,gdk+14,kgh+15,bgk+16,vkr+18}. We have compiled a ``unified'' catalogue of post CE candidates to combine systems from different observational campaigns and individual observations into a single data set which will serve as a tool of theoretical investigations on the general mechanisms of the CE phase.

A theoretical prescription based on the conservation of energy was proposed by \citet{w1984, dek90}, the so called $\alpha$ formalism. This commonly used prescription allows a prediction of the CE end state and has become more complex in recent years when taking more energy sources and sinks into account \citep[e.g.][]{ktl+16}. An alternative prescription making use of angular momentum conservation was introduced by \citet{nvyp00}, the so called $\gamma$ formalism, to explain systems which show only marginal in-spiral. For a more comprehensive summary of the CE phase we refer the reader to \citet{ijc+13,ijr20}. In recent years hydrodynamic simulations have become powerful enough, enabling simulations of the CE phase in greater detail and becoming more and more successful in ejecting the common envelope. As these simulations require significant computational resources, they only tackle a small number of cases with limited sets of initial parameters \citep[e.g.][]{rt12, pdf+12, kso+20, sos+20}.

This catalogue is intended to provide a broad overview, several times larger than previous collections of post CE binaries, of different systems. All binaries, which have not evolved significantly after a CE, should still exhibit signatures of the CE evolution. In \S~\ref{sec:catalogue} we introduce the catalogue. First, we explain our selection of systems. This is followed by a short overview and a brief statistical summary of the parameters compiled in the catalogue. At least a period and limits on the mass of the common-envelope primary, the donor of a most recent CE, and it's companion are required. We finish this section with a discussion of the limitations on the parameters derived from observations. In \S~\ref{sec:evolution} we discuss the theoretical evolution of the systems. This includes different formation paths, the evolution of the system since the end of the CE, and their future. Finally, we give a short summary in \S~\ref{sec:summary}. A shorthand version of the catalogue can be found in Appendix~\ref{sec:shorthandVersion}.


\section{The Catalogue}
\label{sec:catalogue}
This is a catalogue of binaries that are likely to have passed through a CE phase, leaving the binary in the presently observed state. The catalogue contains data obtained from a large sample of different observations. This includes detailed spectroscopic observations, eclipsing light curves, nova events, and other observations. Individual literature sources for each system are given by the references in the catalogue. We follow the guidance of the authors of each reference about which values of measured quantities are most reliable. In cases in which the authors present several solutions without a preference, the system's note indicate the chosen model. The main goal is to provide a large database with data needed for comparison to theoretical studies of CE evolution. This catalogue represents an improvement on the number of systems about an order of magnitude compared to earlier studies on CE theory \citet{nt05,dpm+11,id19}.

The catalogue can be used in different ways. First, it allows to use a maximum sized sample by combining different observations. This is the main benefit for theoretical uses of this catalogue, since comparison to theoretical models requires a large enough sample size for proper statistical analysis. Further, calibration of empirical models requires a sufficient number of limiting cases, better provided by a large sample. Second, this catalogue allows the user to use any given subsample by filtering according to specific properties which are of interest in a more detailed investigation of systems in a tight range of any parameter space of the collected quantities. Third, this catalogue also provides samples of individual observational campaigns by filtering for the corresponding references to have a sample with the same observational biases, allowing systematic comparisons between different observational campaign. Finally, this catalogue provides a comprehensive and extensive literature inspection, including several references on individual systems for focused investigations from a theoretical standpoint.

For a theoretical investigation, an observational test sample needs to provide details about all systems which went through a certain evolution independent on their observational characteristics nowadays. At the same time, it should allow to differentiate systems with similar current structure, but different evolutionary history. Additionally to this main discrepancy between theoretical needs and observational limitations, it is useful to reduce differences in observational biases, like selection biases -- naturally introduced by using different instruments and/or analysis pipelines. Consequently, each comparison between theory and observations has to find a balance between statistical and systematic uncertainties by using an appropriate subsample of this catalogue.

\subsection{Data Selection}
\label{sec:selection}
We identified an extensive set of about a thousand papers reporting on systems which are possibly related CE. The selection criteria to finally list a system in the catalogue can be summarised as: (1) the system contains a hydrogen depleted component (which lost its envelope), (2) fundamental parameters of the system have been derived from observations and given in the literature. The aforementioned parameters are the period, $P<\unit{100}{\mathrm{d}}$ (this cut off should exclude most binaries, which had no or stable mass transfer), and estimates on the masses. Here, we require at least a limit placed on both of the component masses. It should be noted, however, that the parameters provided in the catalogue are not limited to these parameters pertaining to our inclusion criteria, as will be abundantly made clear in \S~\ref{sec:overview}. Other parameters that could be useful for theoretical studies are also included.

Because a CE drains energy and angular momentum from its host binary, one common feature of all post-CE binaries is that their orbital separations are relatively small, smaller than expected from considerations of the size of the progenitor star. Speaking from an observational point of view, such binaries exhibit short orbital periods from minutes to days. A non Roche-lobe filling giant would suggest a binary orbit with a period of months or years. Additionally, there is at least one hydrogen depleted component which is assumed to be the donor star in the mass transfer phase leading up to the formation of the common envelope. This hydrogen depleted component can be a degenerate object like a WD or a Helium burning sub dwarf. The latter ones show usually a spectral class of B or O stars, thus classified as sdB/O, while being more compact and less luminous compared to main sequence stars.

We note that, for purposes of nomenclature, this study designates as the common-envelope primary component the star which was most likely the donor of the CE. This is in contrast to most observational papers, which tend to designate the more luminous star to be the primary. For some systems containing two WDs it is not always clear which component is most likely the donor of the CE. We attempt to address this via some simple comparisons, e.g. making use of given cooling ages. We split the catalogue into three main populations: double WDs (DWD), WDs with a non-WD or unknown companion (WD+), and sdB/O binaries (sdB/O+). Those are expected to differ most in terms of observational biases and are shown separately in the following. In sdB/O+WD binaries, the sdB/O star is most likely the younger star, which places them into the sdB/O+ sample instead of the WD+ sample. In the WD+ sample the companion is usually non-compact and still a main sequence star.

The identity of a post-CE candidate is usually only inferred from its current structure and compactness. This catalogue includes a note to indicate border-line cases of systems that may have formed through a different evolutionary channel, see \S~\ref{sec:NPCE}. Binaries with a neutron star or black hole as the remnant of the common envelope donor are not included in the catalogue. The reason for this lies in the consideration that the formation of a NS or BH (i.e. supernovae) are believed to expel a significant amount of mass and may impart a kick on the newly born components, changing the component masses and their orbits significantly. As a consequence, the signatures of the CE phase are not visible anymore, or are at least hidden from sight. The complex nature of supernovae prevents a solid determination of the CE end state from current observations. In most cases, an inferred post CE binary would be far from being constrained in a unique way.

We require the orbital period to be less than $100$ days and a limit for both masses to be given in the literature in order to list an observed system in this catalogue. Additionally, the system must contain a hydrogen depleted component, the donor of a potential CE. A shorthand version of the catalogue is given in appendix table~\ref{tab:dataObsShort}, while the columns in the catalogue are summarised in \S~\ref{sec:overview} and table~\ref{tab:columns} in the appendix. Some statistical properties of the catalogue are presented in \S~\ref{sec:CatalogueStatistics}. There are different sources of uncertainties and limitations on the observational side. The most common ones are summarised in \S~\ref{sec:ObservationalLimitations}. 

\subsection{Overview of the Systems and Their Recorded Quantities}
\label{sec:overview}
\begin{deluxetable}{cccc}
    \tabletypesize{\scriptsize}
    \tablecaption{
        Overview of the systems in the catalogue. The columns are the main groups of systems considered here and provide counts for different criteria. The index $1$ or $2$ refers to the common-envelope primary (the donor of the possible CE) and secondary component. The values behind a $\uparrow$ and a $\downarrow$ are the counts where lower and upper limits of the quantity given by the criteria are available, respectively. The distances are calculated form the Gaia parallax. For more details see text (\S~\ref{sec:overview}).\label{tab:dataObsOveriew}
    }
    \tablewidth{\columnwidth}
    \tablehead{
        \colhead{criteria} & \colhead{sdB/O+} & \colhead{DWD} & \colhead{WD+}
    }
    \startdata
        all & $185$ & $123$ & $531$ \\
\hline
mass, $M_1$ & $184_{\uparrow51}^{\downarrow49}$ & $122_{\uparrow96}^{\downarrow96}$ & $506_{\uparrow411}^{\downarrow389}$ \\
mass, $M_2$ & $71_{\uparrow181}^{\downarrow72}$ & $94_{\uparrow117}^{\downarrow75}$ & $505_{\uparrow450}^{\downarrow449}$ \\
masses, $M_1$ and $M_2$ & $71$ & $93$ & $492$ \\
\hline
WD companion & $126$ & $123$ & $0$ \\
sdB companion & $1$ & $0$ & $0$ \\
MS companion & $56$ & $0$ & $15$ \\
NS companion & $6$ & $0$ & $45$ \\
BH companion & $7$ & $0$ & $0$ \\
BD companion & $6$ & $0$ & $24$ \\
M type companion & $35$ & $0$ & $164$ \\
K type companion & $2$ & $0$ & $34$ \\
G type companion & $1$ & $0$ & $11$ \\
F type companion & $0$ & $0$ & $42$ \\
A type companion & $0$ & $0$ & $26$ \\
unknown companion (-) & $2$ & $0$ & $170$ \\
\hline
no flag (-) & $39$ & $64$ & $244$ \\
mass transfer (MT) & $2$ & $1$ & $22$ \\
cataclysmic variable (CV) & $0$ & $0$ & $223$ \\
statistically (S) & $0$ & $56$ & $20$ \\
assumed WD mass (SWD) & $0$ & $0$ & $82$ \\
assumed sdB mass (SsdB) & $140$ & $0$ & $0$ \\
assumed mass ratio (Sq) & $5$ & $0$ & $2$ \\
assumed companion mass (SM2) & $0$ & $1$ & $19$ \\
triple (TRI) & $1$ & $1$ & $7$ \\
\hline
mass ratio, $q$ & $40_{\uparrow31}^{\downarrow31}$ & $67_{\uparrow64}^{\downarrow69}$ & $376_{\uparrow318}^{\downarrow318}$ \\
semi-major axis, $a$ & $34_{\uparrow38}^{\downarrow33}$ & $62_{\uparrow63}^{\downarrow61}$ & $184_{\uparrow168}^{\downarrow166}$ \\
eccentricity, $e$ & $44_{\uparrow13}^{\downarrow44}$ & $0_{\uparrow0}^{\downarrow3}$ & $49_{\uparrow40}^{\downarrow44}$ \\
inclination, $i$ & $63_{\uparrow67}^{\downarrow71}$ & $27_{\uparrow23}^{\downarrow25}$ & $291_{\uparrow319}^{\downarrow314}$ \\
radius, $R_1$ & $51_{\uparrow43}^{\downarrow43}$ & $69_{\uparrow69}^{\downarrow69}$ & $209_{\uparrow155}^{\downarrow155}$ \\
radius, $R_2$ & $37_{\uparrow34}^{\downarrow35}$ & $19_{\uparrow18}^{\downarrow19}$ & $333_{\uparrow314}^{\downarrow313}$ \\
effective temperature, $T_{\mathrm{eff},1}$ & $167_{\uparrow148}^{\downarrow148}$ & $120_{\uparrow94}^{\downarrow94}$ & $296_{\uparrow266}^{\downarrow272}$ \\
effective temperature, $T_{\mathrm{eff},2}$ & $35_{\uparrow31}^{\downarrow34}$ & $35_{\uparrow28}^{\downarrow31}$ & $167_{\uparrow99}^{\downarrow103}$ \\
luminosity, $L_1$ & $18_{\uparrow18}^{\downarrow18}$ & $0_{\uparrow0}^{\downarrow0}$ & $21_{\uparrow17}^{\downarrow17}$ \\
luminosity, $L_2$ & $10_{\uparrow10}^{\downarrow10}$ & $0_{\uparrow0}^{\downarrow0}$ & $21_{\uparrow19}^{\downarrow19}$ \\
surface gravity, $\lg(g_1)$ & $161_{\uparrow143}^{\downarrow143}$ & $102_{\uparrow84}^{\downarrow84}$ & $223_{\uparrow205}^{\downarrow205}$ \\
surface gravity, $\lg(g_2)$ & $16_{\uparrow14}^{\downarrow14}$ & $16_{\uparrow12}^{\downarrow12}$ & $109_{\uparrow62}^{\downarrow62}$ \\
(cooling) age & $5_{\uparrow4}^{\downarrow6}$ & $61_{\uparrow61}^{\downarrow64}$ & $86_{\uparrow25}^{\downarrow27}$ \\
\hline
Gaia EDR3 ID & $184$ & $123$ & $529$ \\
distance from EDR3 & $183$ & $121$ & $520$ \\

    \enddata
\end{deluxetable}
The catalogue contains a growing number of systems -- 839 to date. The collected data includes the name(s) of the system in the literature, several binary and stellar parameters, a flag (assigned by us and explained in detail in \S~\ref{sec:StatisticalValues}, \S~\ref{sec:triples}, \S~\ref{sec:MT}, and \S~\ref{sec:CV}), the reference(s), and, if required, an additional note, e.g. which values are used if the corresponding reference states several independent ones. All parameters in the catalogue are given in the included references -- only unit conversions are performed. Most parameters are presented with uncertainties. The uncertainty in the orbital period is usually very small and therefore omitted in the catalogue. An overview of the numbers can be found in table~\ref{tab:dataObsOveriew}. 

The binary parameters are the orbital period, $P$, the mass ratio, $q$, the semi-major axis, $a$, the eccentricity, $e$, the inclination, $i$, and the cooling age of the system. The orbital period is recorded in days. In most cases only one of the masses is directly inferred from the observations (or is assumed) while the second mass is obtained with a measurement on the mass ratio or the mass function -- a combination of the two masses and the inclination. The mass ratio is usually given by the velocity amplitudes of the two components. Hence, this parameter is often more reliable than the individual masses. The semi-major axis is generally calculated in the literature, e.g. using Kepler’s third law, but sometimes it can be inferred from the distance in resolved binaries. Eccentricities are only calculated by modelling the orbit (generally via light curve fitting) or simply assumed to be zero in the majority of the cases. If the inclination of the binary can be constrained from observations, it is used in the determination of the individual masses from the measured mass function. Hence, systems with a given inclination usually have tighter constraints on the individual component masses than other systems. If the remnant of the common-envelope donor, the common-envelope primary WD in our catalogue\footnote{In DWDs the common-envelope primary WD is the younger one.}, is well observed, a cooling age can be estimated from models. This provides an estimated lower limit on the time since the end of the common envelope.

The stellar parameters are the mass, $M_i$, the type, the radius, $R_i$, the effective temperature, $T_{\mathrm{eff},i}$, the luminosity, $L_i$, and the surface gravity, $\lg(g_i)$, where $i\in\{1,2\}$ indicates the common-envelope primary stellar component, the remnant of the CE donor, and its companion. In theoretical modelling the masses of the stars are key parameters. Hence, only systems having at least a limit on both masses are considered in the catalogue. But those masses are difficult to measure directly. For most systems only a mass ratio can be directly obtained from the observational data. Only if the inclination is known, can the masses be directly determined. 
Hence, some authors use limits on the inclination to constrain the most likely mass of an individual system, while others fit the brighter component to a theoretical model to estimate the mass, and some resort to simply assuming a mass. The types of the stars can be constrained in different ways. First, it could be a constraint arising from the brightness and compactness of a component. On the other hand, sufficiently bright stars may be assigned to a spectral type. The radii could either be inferred from some models when constraining the mass or directly observed in eclipsing systems. The effective temperature and surface gravity are usually constrained from fitting the spectrum. The luminosity is mostly estimated from a distance measurement of the system and only rarely available at the time the source papers were written. 

For most systems at least some of the parameters are not given in the literature. Table~\ref{tab:dataObsOveriew} provides an overview of how many systems in the catalogue fulfil a certain criterion. The data is split into the main groups of systems in the catalogue: double white dwarf (DWD), a white dwarf with a non-WD or an undetermined companion (WD+), a sub dwarf B or O star in a binary with any kind of companion including WDs (sdB/O+). First, an overall count is given. In the second block, the availability of the mass values is checked –- all systems have at least an upper or lower limit for both masses, if not a value itself. Third, the different companion types are listed: a white dwarf (WD), a main sequence star (MS), a neutron star (NS), a black hole (BH), a brown dwarf (BD), or the spectral type is given. Here ``unknown'' refers to having no information about the companion in the cited reference(s). Some systems have two to three open possibilities, hence the sum of this block may exceed the total number of systems.

Fourth, depending on the circumstances, some systems ought to be excluded from certain studies conducted using our database, and are therefore marked by different flags denoting potential reasons for exclusion in each case. The first two flags are about observational features of binary interaction, for more details see \S~\ref{sec:MT} and \S~\ref{sec:CV}. All the flags starting with an ``S'' are related to assumptions about the masses, see \S~\ref{sec:StatisticalValues}. The last flag is about the possibility of a different nature of the system, e.g. in a triple system, see \S~\ref{sec:triples}. Fifth, the availability of other parameters is checked. Finally, the success with a cross match to Gaia EDR3 data\footnote{\url{https://www.cosmos.esa.int/web/gaia-users/archive}} is shown. Most of our systems are found in the Gaia EDR3, except for those only detected by other bands than optical, hence having no Gaia ID. The majority of the Gaia objects are detected at different epochs with Gaia and therefore have measured parallax values \citep{gaia2020}. The distances estimates in the catalogue are calculated from this parallax, see \S~\ref{sec:Distances}.

\subsection{Catalogue Statistics}
\label{sec:CatalogueStatistics}
\begin{figure*}
    \centering
    \includegraphics{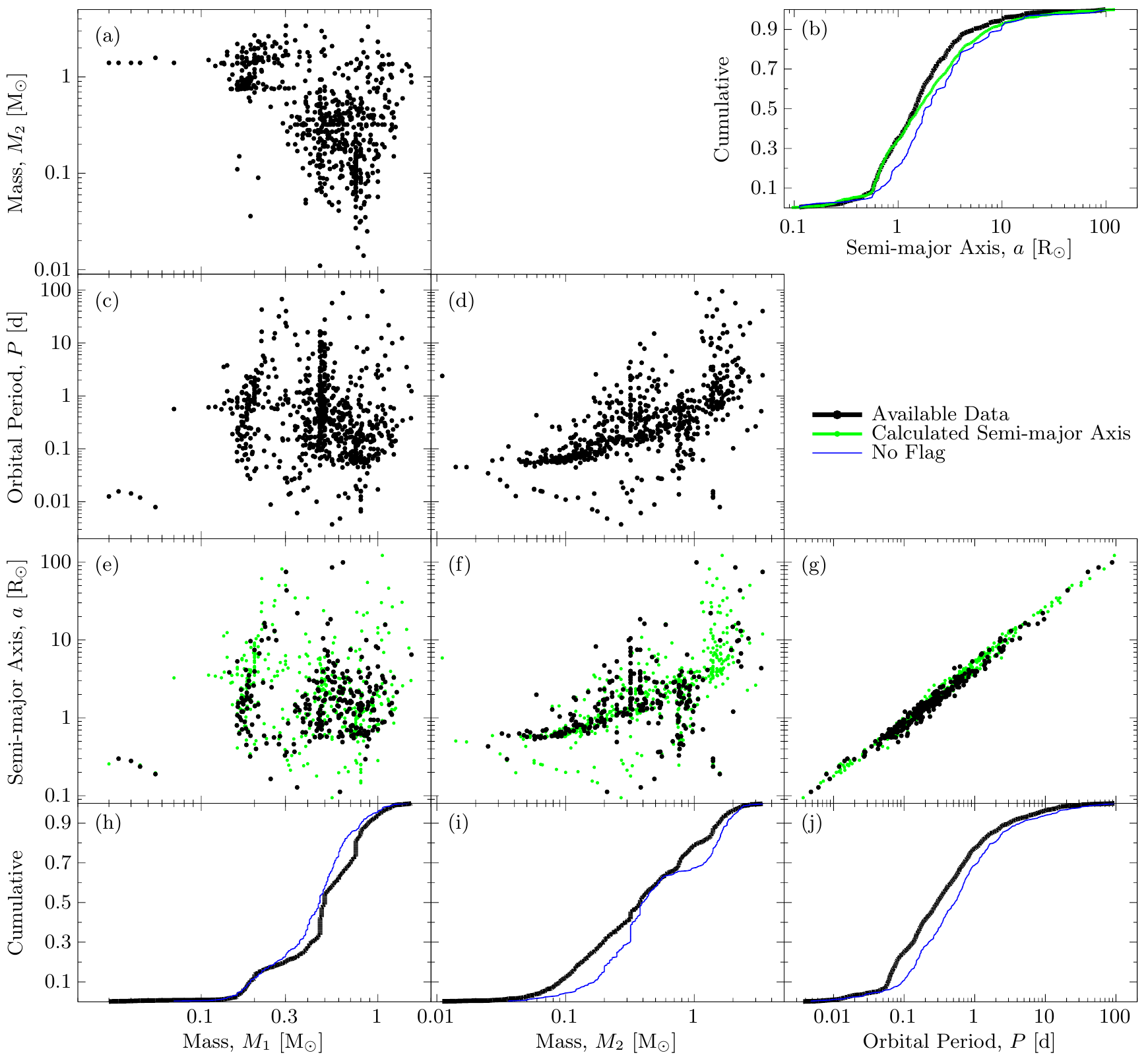}
    \caption{Masses, periods and semi-major axis of the systems in the catalogue. We show the cumulative distributions at the bottom of each column, i.e. subplots (h-j). The black/thick lines represent the full sample, while the blue/thin line excludes all systems with a flag. We show the cumulative distribution of the semi-major axis -- in green/medium values calculated with Kepler's third law -- in the top right corner (b). We note that the discrepancy between the numbers of systems is caused by missing values in the catalogue. For clarity we omit error bars in this figure. Fig.~\ref{fig:massmass} shows subplot (a) including the observational uncertainties. See \S~\ref{sec:CatalogueStatistics} and Appendix~\ref{sec:addFigures} for more details.}
    \label{fig:dataObs}
\end{figure*}
\begin{figure*}
    \centering
    \includegraphics[scale=0.95]{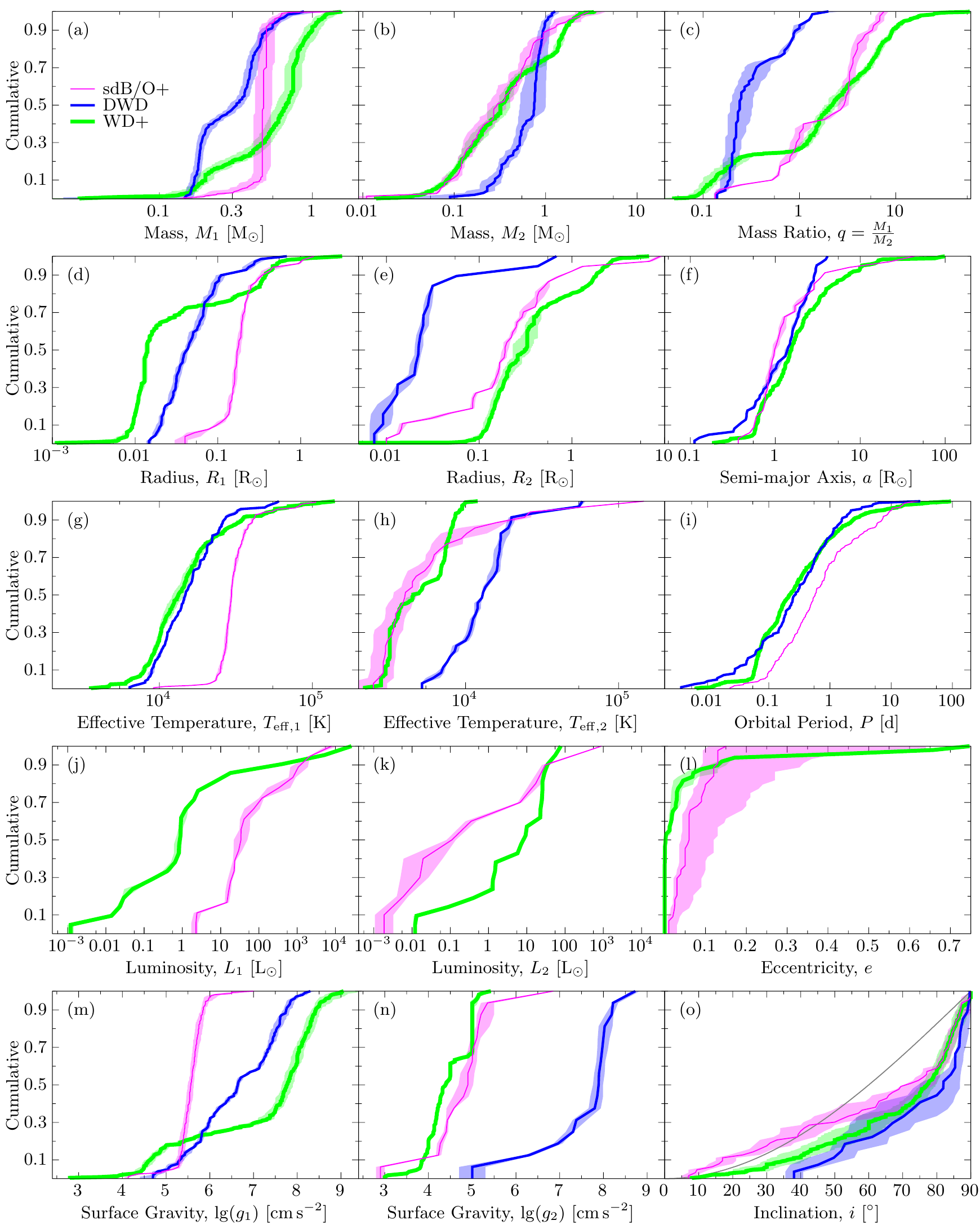}
    \caption{Cumulative distributions of the different catalogue parameters. Colour and line thickness indicate the binary type according to the key in sub figure (a). The corresponding counts of systems can be found in table~\ref{tab:dataObsOveriew}. The grey line in sub figure (o) indicates the distribution of random inclinations. To show the uncertainty contours, missing uncertainty values are assumed to be the geometric mean of the recorded ones.}
    \label{fig:cumulative}
\end{figure*}
Figure~\ref{fig:dataObs} provides an overview of the systems in the catalogue. See appendix~\ref{sec:addFigures} (Figs.~\ref{fig:dataObssdBp}, \ref{fig:dataObsDWD}, and \ref{fig:dataObsWDp}) for the three sub populations sdB/O+, DWD, and WD+ binaries individually. The cumulative distributions of the masses (Figs.~\ref{fig:dataObs}h and i) show a few jumps which are caused by the assumed values where they cannot be measured directly, see \S~\ref{sec:StatisticalValues}. The secondary masses cover a large range and do not show any preferences for systems where the masses are measured. Here, small accumulations at specific masses are caused by fits to stellar model grids, where the closest model is assigned to the observed star. Beside assumed masses, the common-envelope primary masses show an overabundance at masses $\lesssim\unit{0.2}{\Msun}$, which is caused by a large sample of binaries from an ELM survey \citep{bgk+16}, cf. Fig.~\ref{fig:cumulative}a. All sdB stars are generally accepted to possess masses higher than \unit{0.3}{\Msun} as lower mass objects are unable to meet the conditions required for helium ignition \citep{kww12}. These lower mass objects will therefore immediately thermally contract to become He WDs. Shortly after their progenitor cores are exposed, such proto-WDs still look similar to sdB stars. There are very few stars below \unit{0.3}{\Msun} classified as sdBs in the literature. Usually, WDs, especially low mass ones, are very faint and therefore difficult to observe.

The relation of the black dots visible in Fig.~\ref{fig:dataObs}g is mainly following Kepler's third law, cf. green dots. Hence, it shows a slope of $2/3$ on a logarithmic scale. The spread of the relation is caused by the different total masses of the binaries. As a consequence, the period and semi-major axis plots (Figs.~\ref{fig:dataObs}c and d vs. Figs.~\ref{fig:dataObs}e and f) show similar distributions of the systems, with the main difference being that certain systems, for which semi-major axis values are not available, are absent from the corresponding plots. It should be noted that the systems having a brown dwarf as a companion look like a natural extension of the binaries with hydrogen burning stars with a mass above $\unit{0.08}{\Msun}$.

Both, the period and the semi-major axis distribution (Figs.~\ref{fig:dataObs}j and b), show a change in the slope at about $\unit{0.06}{\mathrm{d}}\approx \unit{1.4}{\hour}\approx \unit{90}{\minute}\approx \unit{5\,000}{\second}$. This can be interpreted as the range where systems decay via gravitational wave radiation at a similar rate as such systems are created. Nearly all systems below the aforementioned limit of about $\unit{0.06}{\mathrm{d}}$ are double WD or sdB/O+WD binaries, cf. Fig.~\ref{fig:cumulative}i. Hence, this gives a second interpretation of that limit to be the limiting case for hydrogen rich companions to fill their Roche lobe and initiate mass transfer in the post-CE system and become a CV, see \S~\ref{sec:CV}. Additionally, it should be noted here, that double WD and sdB/O+WD binaries may have experienced more than one CE phase during their prior evolution.

The tightest binaries are double WD systems as one would expect. The binaries with the shortest periods in the WD+ sample have either undetermined or NS companions. More massive WDs have smaller radii. Hence, it would be expected that the tightest orbits short of a merger event of two WDs are only reachable for massive WDs. In the catalogue the tightest double WD binaries contain intermediate mass WDs, perhaps indicating some biases in the collected sample. The distributions for the common-envelope primary star -- the donor star of the potential common envelope -- do not show such clear trends. For some reason the double WD systems show a significantly lower average mass of the common-envelope primary than the systems having any other companion to the common-envelope primary WD. It is probably related to the fact that the less massive WD usually forms later than the more massive one, hence being the donor of a most recent mass transfer.

From Fig.~\ref{fig:cumulative}, it is evident that the three main groups we differentiate here cover different ranges in most of the parameters collected in the catalogue. The sdB/O stars cover a smaller mass range than the WDs. WDs can have lower masses because those stars are too low in mass to become an sdB/O and directly cool to become WDs after they have lost their envelope. In principle, sdO stars may have masses similar or even larger than WDs, but here the observational sample is biased by the shorter lifetime of massive sdO stars which makes them less common. In the sdB/O+ and WD+ sample most companions are low mass main sequence stars, hence these distributions in Fig.~\ref{fig:cumulative}b look very similar. The DWD population differs here because the companion is the first formed and usually more massive WD, but it is classified in the catalogue as the common-envelope secondary. This is confirmed by the mass ratio distribution.

The radius distributions resemble the known relations, cf. Figs.~\ref{fig:cumulative}d and e. First, the most massive WDs are most compact. Second, WDs are more compact than sdB/O stars. Third, Brown dwarfs have a similar size as sdB/O stars. In this sample (as would be expected), hydrogen burning stars have physically the largest radii. There are barely any giant like stars in the sample as they require a longer orbital period. Similarly, sdB/O stars are hotter than WDs, while low mass main sequence stars and brown dwarfs are the coolest stars in the sample, cf. Figs.~\ref{fig:cumulative}g and h. Again, such trends are visible in the surface gravity, cf. Figs.~\ref{fig:cumulative}m and n. Massive WDs are most compact, hence having the highest surface gravity. Next in line are the low mass WDs the sdB/O stars, which have lower surface gravity, followed by main sequence stars and the brown dwarfs, which have the lowest surface gravity of them all.

The differences in the distribution of the inclination (Fig.~\ref{fig:cumulative}o) is probably caused by observational biases. Eclipsing systems allow better constraints on the system’s parameters from the observations. Additionally, the projected velocities are larger and therefore easier to observe. It is evident that this observational selection effect impacts the three main sub populations differently. Most systems have low eccentricities, see Fig.~\ref{fig:cumulative}l, or are assumed to be circular due to their short periods. Additionally, some eccentricity effects have only small imprints in the observations, hence eccentricity values are omitted in most observational papers.

Figure~\ref{fig:age}a shows the cumulative distribution of the cooling ages of the common-envelope primary star, which corresponds to a lower limit of the time since the end of the common envelope. We do not show the distribution for the sdB/O+ category since it contains only five systems. These five systems show very short cooling ages. The WD+ systems tend to be younger than DWD systems, which is probably due to the smaller secondary mass $M_2$. Although half of the DWD systems are older than $\unit{1}{\giga\mathrm{yr}}$, the age distribution peaks at a much younger age, than predicted in a model of single stars; the grey line shows the predicted WD age distribution in the galactic chemical evolution model for the solar neighbourhood \citep{kkl20}, where the metallicity dependent stellar lifetimes and upper limits of carbon-oxygen WDs are assumed but no binary effects are included. In binaries, the companion stars could cause the young ages of WDs. We should also note that the initial-mass to final mass relation, and hence the stellar lifetimes of both stars, can be different. It is also possible that some DWD systems have merged and exploded as SNe type Ia, see \S~\ref{sec:SN}. Hence, it is expected that the observed distributions are different from the model line.

\subsection{Observational Limitations}
\label{sec:ObservationalLimitations}
We can observe the stars only as we can see them with our telescopes from earth. Hence, there are limitations on the information we get. Often we cannot resolve binaries, especially the tight ones in this catalogue. The observed light is therefore a combination of the emission of the two components. This causes the problem that the information of the dimmer star is hidden in the light of the brighter one.

To get information on the tightly orbiting masses a system needs to be either eclipsing or show clear spectral lines -- preferentially for both stars. The catalogue systems show an overabundance of eclipsing systems -- $i\approx\unit{90}{\degree}$, cf. Fig.~\ref{fig:cumulative}o. To get a spectrum of a faint star or binary a certain integration time is required. This becomes problematic when this integration time covers a longer part of the orbit and therefore leads to line broadening by averaging over different orbital phases. Additionally, spectroscopic surveys do not revisit binaries often enough to get a good orbital coverage to constrain the orbital period well.

\subsubsection{Statistically Inferred and Assumed Mass Values}
\label{sec:StatisticalValues}
Often, the required theoretical quantities cannot be inferred from the observation directly. Some values are assumed in order to make full use of information about observational knowledge of combinations of quantities. One of the most prominent cases is the observable mass function which relates the masses with the usually unknown inclination. The inclination enters as a geometrical effect and therefore as $\sin(i)$ and powers of it. The limitation of the trigonometrical functions in a Euclidean space allows one to determine limits on the masses. The limit is usually a lower limit on the mass via $\sin(i)\leq 1$.

If the orientation is isotropic the most likely inclination is $\unit{60}{\degree}$. Some observational papers use this to determine the detailed mass values of an individual system to be the most likely. In some cases the observations allow the exclusion of some inclinations, hence the most likely value is then determined with Monte Carlo techniques \citep[e.g.][]{bgk+16}. Those cases are marked in the catalogue with the flag for statistical determination of the values (S).

The mass ratio is the most fundamental quantity which is usually determined by spectroscopic observations. It is directly related to the relative velocity of the two components. For eclipsing binaries the mass ratio can be determined as well. Hence, most observations will provide a good measurement of the mass ratio. It is a bit more tricky to get individual masses. Here are different approaches used, e.g. the herein before mentioned mass function when the inclination could be determined. On the other hand the masses can be determined via the total mass of the binary, $M_1+M_2$. One way to determine the total mass uses the distance of a resolved system, which provides an independent measurement of the semi-major axis and provides a total mass via Kepler's third law. To resolve a system, the orbital separation usually needs to be wider, hence sometimes for those systems the mass ratio is uncertain, and consequently in rare cases even set to a certain value -- in the catalogue those are marked with a flag (Sq).

Another common way to determine individual masses is a fit of the brighter component in the binary to theoretical models. Here the values in the catalogue are model dependent and usually come with large uncertainties. The crudest way to get the individual masses is to assume the mass of one component to have a typical value for this kind of object. Those cases are marked with a flag (SWD, SsdB, and SM2) of assumed values which are simply set by the authors. The most commonly assumed mass for WDs is $\unit{0.75}{\Msun}$ \citep[e.g.][]{pkh+05}\footnote{In recent years, literary consensus has caused the canonical mass of a a WD to decrease to $\unit{0.6}{\Msun}$, down from the $\unit{0.75}{\Msun}$ assumed in older papers. It should be noted that the mass of the other component will need to be changed accordingly if a different canonical mass is assumed.} and for sdB stars it is in the range from $\unit{0.47}{\Msun}$ \citep[e.g.][]{kgh+15} through $\unit{0.48}{\Msun}$ to $\unit{0.5}{\Msun}$ \citep[e.g.][]{mmm+03}. This results in an overdensity at those assumed masses in the full sample, cf. Figs.~\ref{fig:dataObs} and \ref{fig:cumulative}.

\subsubsection{Uncertainties and Inconsistencies}
\label{sec:UncertaintiesInconsistencies}
\begin{figure*}
    \centering
    \includegraphics{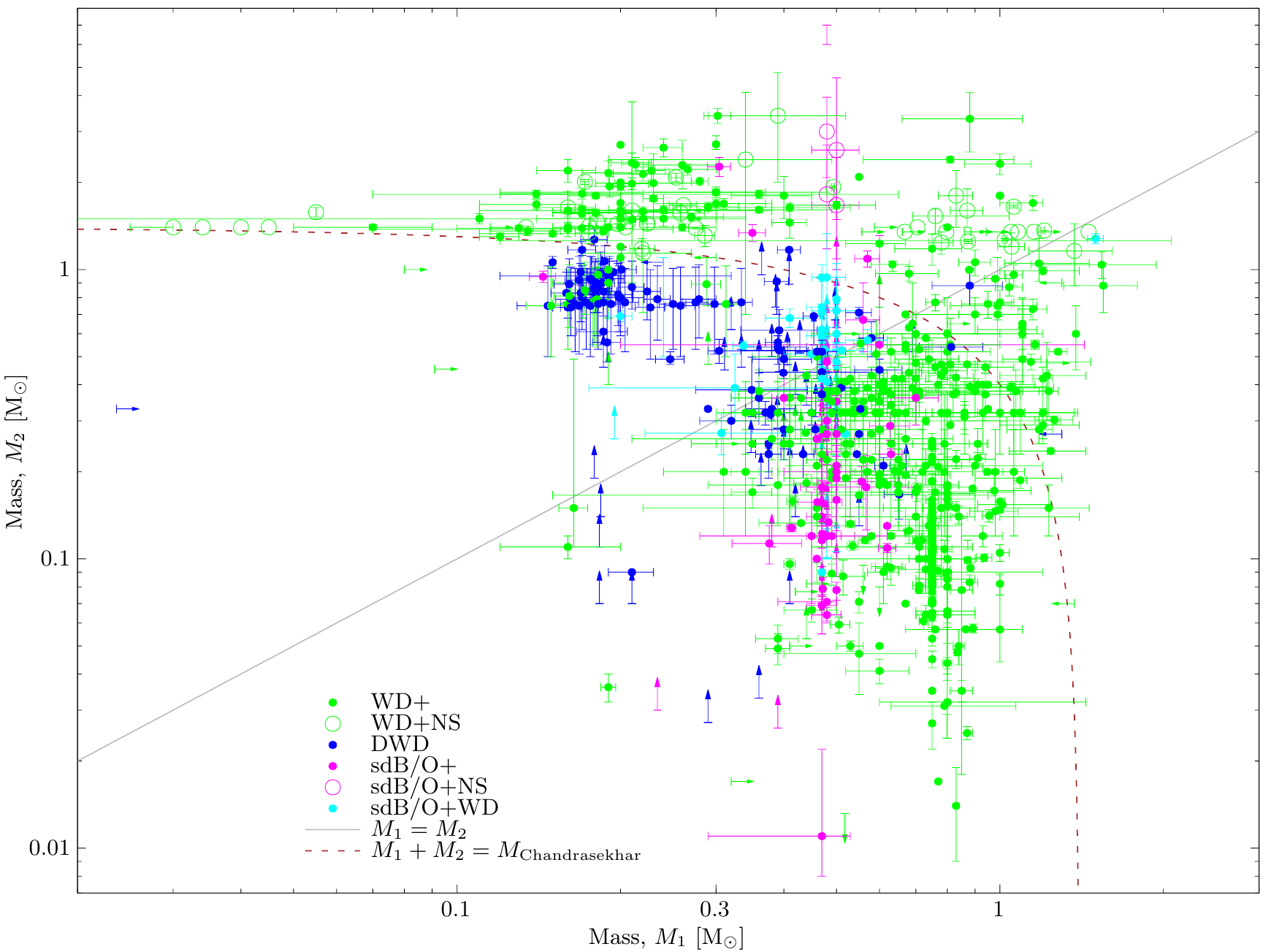}
    \caption{Mass-mass plot like in Fig.~\ref{fig:dataObs}a. Additionally, it contains error bars. If there is no value the error bars indicate the possible range, while an arrow is a limit -- usually a lower limit. The colours show the type of the system, where neutron star companions are highlighted by the open circles. The grey line separates the regions where the common-envelope primary or secondary is more massive. The brown dashed line indicates a binary mass of $\unit{1.4}{\Msun}$.}
    \label{fig:massmass}
\end{figure*}
Where available, the estimated uncertainties are included in the catalogue, which replace the upper and lower limit column in such cases. Figure~\ref{fig:massmass} shows the uncertainties on the masses as error bars. Compared to Fig.~\ref{fig:dataObs}a it contains additionally the systems for which only a limit or a range for a mass is available. The arrows indicate an upper or lower limit, while the ranges represent systems where both limits are given. It should be noted that binaries with an assumed mass are usually displayed as sole data points without an uncertainty. At the same time the figure gives an impression on how different the constraints from different observations are.

A few systems are included twice in the catalogue (not twice counted in table~\ref{tab:dataObsOveriew} and only shown once in all plots), but the second occurrence is marked as a comment with a \# at the beginning of the line and in the flag column. In those cases it is unclear which reference provides the more reliable values while they are inconsistent between different authors. If a newer determination explains an inconsistency to older ones usually the newer values are assumed to be reliable and are used to replace older ones.

\subsubsection{Distances and Possible Detection Limits}
\label{sec:Distances}
\begin{figure*}
    \centering
    \includegraphics{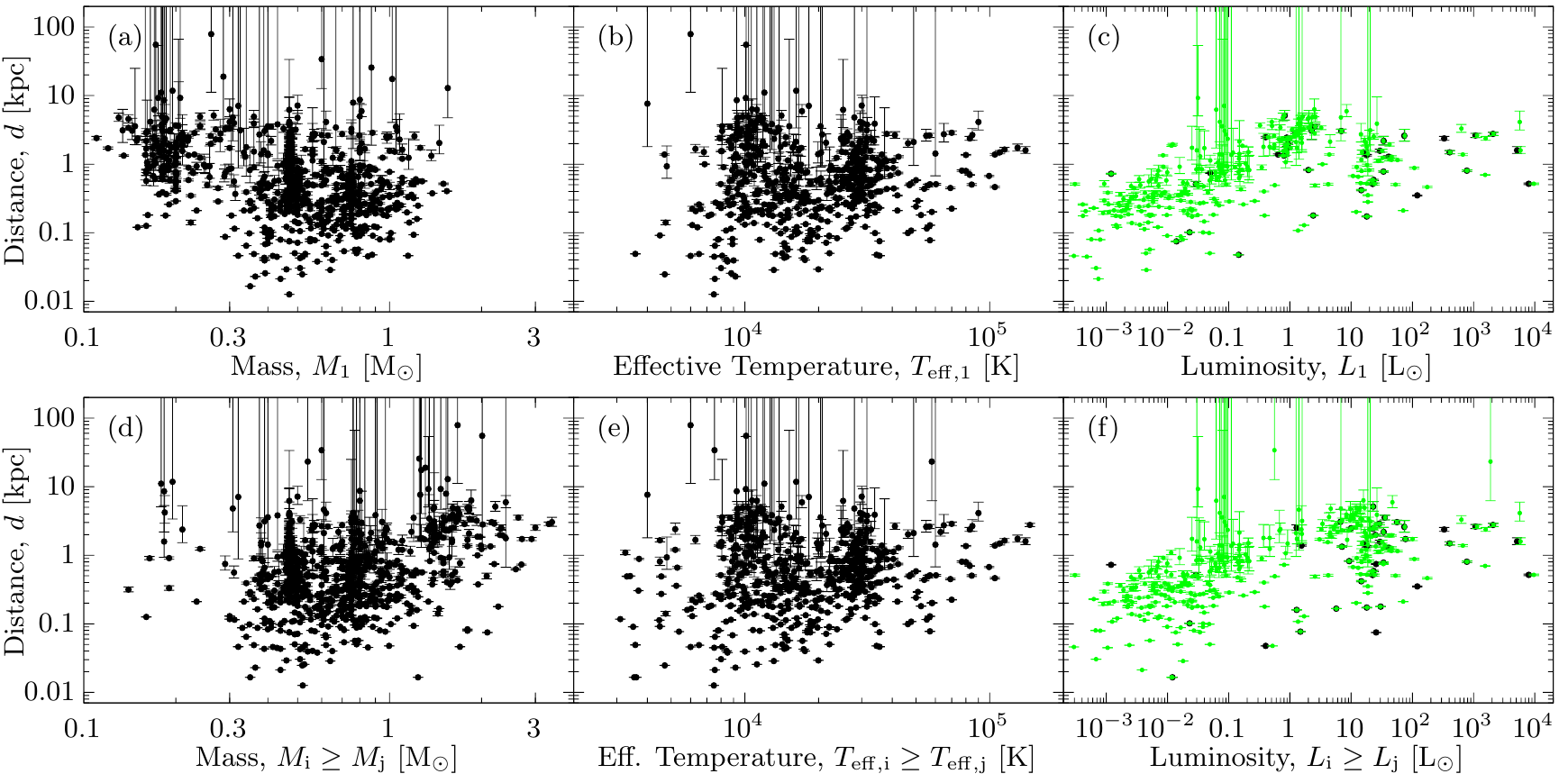}
    \caption{Distance depending on the mass, effective temperature and the luminosity. Upper row (a-c): the values of the common-envelope primary component. Lower row (d-f): the values of the more massive, hotter, and more luminous star, respectively. The large/black dots are the catalogue values. The medium/green ones are the luminosity calculated via the Stefan-Boltzmann law from the radius and effective temperature. The partially large uncertainties on the distances, especially for distant systems, are shown as error bars, while the error bars in the other quantities are not shown for clarity.
    }
    \label{fig:distance}
\end{figure*}
From the cross matching with the Gaia EDR3, we have obtained the distances, $d$, of most of the systems (see Fig.~\ref{fig:location} in the appendix for the spatial distribution). Distances are calculated from the parallaxes given by Gaia by using the general zero-point offset of $\unit{-0.017}{\mathrm{mas}}$ \citep{lbb+21}. It should be noted that for some Gaia objects the real offset could be up to $\unit{-0.15}{\mathrm{mas}}$ \citep{brf+21}. If the renormalised unit weight error (RUWE) of the systems is larger or equal to $1.4$ \citep{lbb+21} the system is explicitly identified as having an uncertain parallax value in the catalogue. Figure~\ref{fig:distance} shows the relations of the distances to some stellar parameters. There are 8 systems with $d>\unit{20}{\kilo\mathrm{pc}}$, which is probably due to large uncertainties in the Gaia parallaxes. Therefore, all of our systems are within $\unit{20}{\kilo\mathrm{pc}}$ with the majority located within $\unit{1}{\kilo\mathrm{pc}}$. In terms of mass, we do not find an obvious selection limit for the common-envelope primary mass (panel a) or the more massive component mass (panel d); WDs as small as $\unit{0.15}{\Msun}$, and massive WDs close to the Chandrasekhar-mass limit ($\unit{1.4}{\Msun}$) are included in our catalogue. The sample of the small WDs are mainly from one reference \citet{bgk+16}. The massive companions are mostly neutron stars (see also Fig.~\ref{fig:massmass}). However, we find a detection limit in terms of luminosity with no systems in the top-left regions of the panels c and f. We emphasise, however, that only the large black dots are provided in the catalogue. The rest of the dots are estimated from the radii and effective temperatures by making use of the Stefan-Boltzmann law.

\subsection{Comments on Outstanding Systems}
\label{sec:ParticularSystems}
\emph{Hen 2-428} \citep[taken from][]{src+15} stands out among DWD systems. Its mass ratio is measured to be very close to 1, hence containing two WDs with a similar mass. Additionally, their effective temperature and radii are stated to be similar, too. Thus, both components may have formed in quick succession. This may indicate either a formation channel via stable mass transfer \citep{src+15} or the CE did not have a single donor -- putting this system into a very distinct category of CEs. When both stars would overfill their Roche-lobe at the same time both would contribute to a CE and therefore eject both envelopes together. But this is not the only outstanding feature of Hen 2-428. \citet{src+15} derived radii of $\unit{0.68}{\Rsun}$ for both WDs. With this radius, which is too large for a usual WD, the WDs would even fill their Roche-lobes and should be in contact.

\emph{PSR J1807-2500B} \citep[taken from][]{lfr+12} has by far the largest eccentricity in the catalogue. It is a WD+NS binary located in the globular cluster NGC 6544. This very large eccentricity indicates that the binary orbit was probably changed significantly after the end of the CE. This could be related to the fact that the system is within a globular cluster and has had dynamical interactions with other stars within the cluster. Alternatively, the WD is not the last formed compact object in this binary. If the NS is the remnant of the donor of the CE it might have received a large kick upon its formation leading to the observed high eccentricity.

\emph{0935+4411} is a system for which the component masses are under debate. While \citet{hhd+09} find the WD to be the more massive star \citet{bgk+16} present a WD mass of less than half of the companion’s. Interestingly, the inferred effective temperature and surface gravity are similar in both references. This system has the largest inconsistency between two different references used for this catalogue and has therefore two entries in the catalogue -- the second one is commented out. Additionally, this system has a very short orbital period while not currently observed to be undergoing mass transfer. It has a period more than an order of magnitude shorter than usual CVs with such a massive hydrogen-rich star, cf. Fig.~\ref{fig:dataObsWDp}d. This might indicate that the companion is not a main-sequence star -- \citet{hhd+09} identified the companion as an M-type star. A number of systems are found a bit below the band of CVs in the WD+ sample, see Fig.~\ref{fig:dataObsWDp}, which are candidates for having a hydrogen depleted and unseen companion, e.g. another WD (in that case they would be in the wrong subsample), or a NS (if massive enough).

Ultra compact X-ray Binaries (UCXBs) are thought to consist of a NS accreting material from a WD, which has already lost most of its mass to the NS companion. Consequently, these systems are therefore likely to have evolved significantly since a potential CE stage, see \S~\ref{sec:NPCE}.


\section{Discussion of the Evolution of the Binaries in the Catalogue}
\label{sec:evolution}
This catalogue is intended to facilitate the study of common envelope phase by providing a sample of candidate post CE systems. Additionally, it allows to study the future evolution of tight binaries as an initial population, see \S~\ref{sec:SN}.

\subsection{The Origin of the Binaries}
\label{sec:origin}
Consider for example, the first category of binary in our catalogue, WD+. These are binaries that consist of a WD orbiting a non-compact or unknown companion, generally a MS star. The orbital separations between the WD and its companion are too small to accommodate the giant progenitor of the WD, and this is the characteristic that makes these systems candidates for post common envelope systems. The WD was once the core of its progenitor, usually a giant. The progenitor filled its Roche lobe. Possibly because it was more massive than its companion, dynamically stable mass transfer was not possible, and there was a CE instead. The orbital periods prior to the CE would have been on the order of months or years, but are presently on the order of days or hours.

The second category we use to organise the catalogue consists of double-white-dwarf binaries, DWDs. Again, those WDs are believed to be the remnant core of their progenitors. Most likely both progenitors would have been at least massive enough to produce one of the present-day WDs. Because of binary interactions the last formed WD could be more massive than its main sequence progenitor could have evolved into in single star evolution when it accreted enough material during the mass transfer leading to the formation of the first WD. It should be noted that the first formed WD is labelled as the secondary star in our catalogue, while the common-envelope primary is the last formed WD. If a DWD went through two CE phases the intermediate system after the first WD formation would show up in the herein before mentioned WD+ sample. Because we indicate the star according to the remnant of the last CE donor, the stars swap their roles between a first and a second CE. The tightest DWD systems are best candidates for binaries, which experienced two CEs.

The third category consists of an sdB/O star and its companion. An sdB/O star is usually a Helium core burning star that lost its hydrogen rich envelope. Hence, those stars are more compact than normal main sequence stars but have a similar or even larger luminosity. When the companion is an unevolved star it is clear that the sdB/O star is the remnant of a mass transfer episode where its envelope got lost. In the other case of a WD companion the sdB/O star is most likely the younger compact star because of the relatively short time of the core Helium burning phase compared to the cooling time of the WD.

\subsubsection{Non Common Envelope-end States}
\label{sec:NPCE}
The selection from observations is mainly determined by having a short period. But there is no guarantee that all those systems went through a CE phase. There might be binaries which evolve into tight orbits even without the need of a CE evolution. This mainly applies to wider systems in the catalogue and/or for systems with massive companions.

If, for example, the donor is a subgiant filling its Roche lobe, its size may be small enough that the final orbital separation is on the order of $\sim\unit{10}{\Rsun}$. Depending on the uncertainties in the measurements of the masses and orbital period, such mass-transfer end states could be mistaken for CE end states. In such cases the donor star would be Roche-lobe filling until the envelope collapses when its mass becomes negligibly small with respect to the core mass. The value of the donor's radius at the end of mass transfer is not, however, uniquely determined. If it were, then it would be straightforward to know whether or not the condition of an end state of stable mass transfer is satisfied. Instead, we can determine whether it is satisfied for reasonable values of the radius that are consistent with what we know about stellar evolution and mass transfer.

The way we have explored the possibility that a given binary experienced stable mass transfer is to start with a set of stellar evolutionary models, conducted with the BEC code \citep{ywl10}, for single stars with zero-age main sequence (ZAMS) masses ranging from $0.5$ to $\unit{100}{\Msun}$ at Milky Way metallicity, $Z=0.0088$ \citep{bdc+11,ktl+18}. For each ZAMS mass\footnote{We use only the models from $0.5$ to $\unit{9}{\Msun}$ here.}, we have interpolated the values of the core mass, envelope mass and radius on intervals of $\unit{0.001}{\Msun}$ in core mass. In this simple model the envelope mass is not necessarily close to the value of collapse, which introduces our primary source of uncertainty here. For example, the end state should correspond to the donor having lost most of its envelope; we therefore expect there to be some difference between the actual radius of the donor at the end of mass transfer and the radius we have employed.

For each post CE candidate, we take the common-envelope primary's mass and identify the state of each BEC star with this mass as its core mass. We therefore know the radius. We compare it with the radius of the Roche lobe associated with the core of the donor star in a circular orbit around its companion. If the stellar radius is smaller than the Roche lobe, but comparable to it in size, then it is likely that stable mass transfer occurred. We add an additional condition. We consider only donor stars with mass $> \unit{0.7}{\Msun}$, so that they will have had a chance to evolve during the lifetime of the Galaxy. For most reasonable mass combinations of the two stars, this results in a period limit of a few to several days. Less than $10\%$ of the systems in the catalogue may be above this limit. It should be recalled that not passing this condition increases the likelihood that the system formed without involvement of a CE but does not exclude this possibility. Those systems, which had stable mass transfer instead of a CE formation, would tend to contribute to the high portion of the distribution of orbital separations/periods.

\smallskip

In binaries with more then one compact star, the assignment of the common envelope primary is not necessarily unique. Especially if the companion is a neutron star and the systems shows a non-zero eccentricity, i.e. $e\gtrsim 0.2$, the last formed compact object might not be the WD. In those cases, it is unclear how much the system differs from the end state of a potential CE. The possibility of showing no signature of the end state can not be excluded, thus such systems have to be taken with additional caution.

For extremely low mass WDs with a massive but compact companion, e.g. a neutron star, there might be other formation pathways beside a CE. If magnetic breaking and tides are strong enough those systems may sustain very tight during stable mass transfer \citep{cth+21}.

\smallskip

Whenever we spot that any of the here mentioned conditions (systems that may avoided a CE evolution) is met we indicate this by a note in the data set.

\subsubsection{Inner Binaries of Triples}
\label{sec:triples}
A few of the binaries in the sample have a detected outer tertiary orbiting around the inner binary. Therefore, we indicate systems with a confirmed tertiary with a flag to indicate its triple nature (TRI). The fact that these systems exist in our catalogue should come as no surprise. It is well known that about $10\%$ of all stellar systems are triples in which the presence of the third star cannot be ignored \citep{rmh+10,mp14,md17}, and $20\%$ of all binaries, such as those hereby presented, are expected to have a tertiary at birth. Thus, the apparent paucity of such systems in our catalogue might speak more of the observational biases inherent to the data that led to this catalogue, than of an actual scarcity of triples in post-CE systems.

A mechanism known as Lidov-Kozai Resonance \citep{koz62,nao16} can drive the eccentricity of the inner binary into an oscillation with the inclination between the inner and outer orbit. The resulting decrease of the periastron coupled with other mechanisms, such as gravitational radiation \citep[e.g.][]{tho11} or tidal effects \citep[e.g.][]{ft07}, may cause a permanent decrease of the inner orbital separation to significant amount. Even without these other effects, an inner binary can be driven close enough to interact at periastron in ways that it would otherwise not, given the same initial semi-major axis. The timescale and therefore importance of this phenomenon depends on the distance and mass of the tertiary.

To date, there are many unknowns in the understanding of triple evolution, some further discussion can be found in Appendix~\ref{sec:CloseTriples}. Even effects of a CE phase in the inner binary on the outer orbit cannot be excluded. While a tertiary could have a long term influence during the evolution similar effects can act in a cluster environment by other stars which spend some time in the vicinity.

\subsection{The Evolution Between the end of a Common Envelope and the Observation}
\label{sec:usage}
\begin{figure*}
    \centering
    \includegraphics{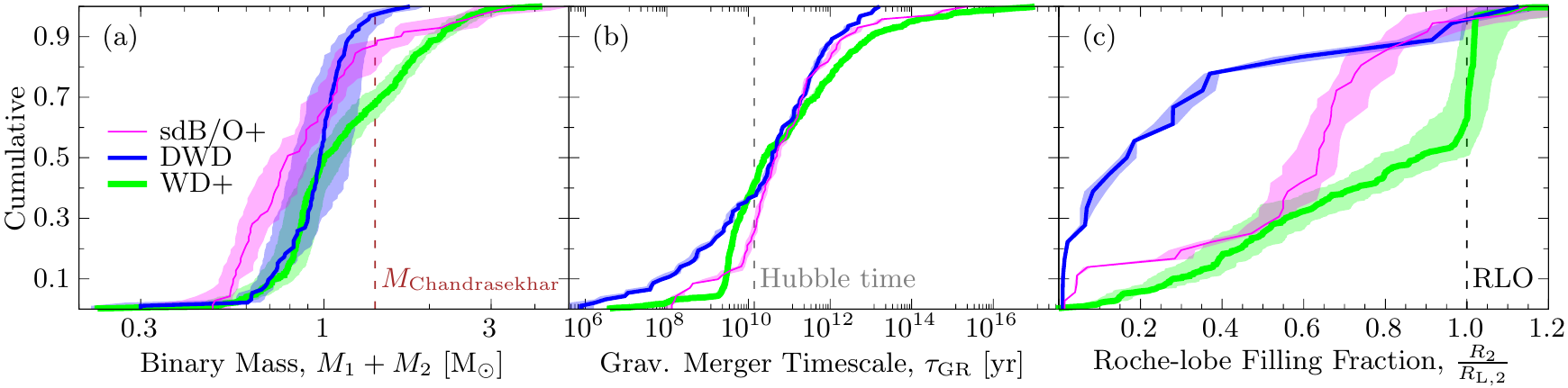}
    \caption{Cumulative distributions of calculated parameters. The colour and line thickness indicates the binary type according to the key in sub figure (a). Sub figures: (a) the binary mass -- a mass of $M_\mathrm{Chandrasekhar}\approx\unit{1.4}{\Msun}$ is marked by the dashed line, (b) the timescale to merge the system by pure GWR -- cf. Eq.~(\ref{eq:tGR}) and $\unit{13.81}{\giga\mathrm{yr}}$ is marked by the dashed line, (c) the fraction of the secondaries radius and it's Roche lobe -- the equality of the two radii is marked, hence whether the companion undergoes Roche-lobe overflow (RLO). The required semi-major axis is always calculated via Kepler's third law and the Roche lobe with the fitting formula of \citet{egg83}. To show the uncertainty contours, missing uncertainty values are assumed to be the geometric mean of the recorded ones. Additionally we had to assume that the errors of the underlying quantities, e.g. the masses, are uncorrelated because of missing information on the correlation factors.}
    \label{fig:cumulative2}
\end{figure*}
It is also possible that a binary which has emerged from a CE phase has had time to evolve before the present day. There is no definitive way to establish that the state we observe is different from the original post CE state. Instead we rely on general notions of probability.

Consider for example two WDs emerging from a CE. It is possible for the final orbital separation to be very small, small enough for the WDs to undergo gravitational-radiation induced merger in a few thousand or a few million years. In such cases, however, the probability is low that we will detect the binary pre-merger, at least when compared with the probability we will discover a systems that lasts tens to thousands of times as long.

When the post CE system consists of a WD and an extended star, similar considerations apply. In such systems the future evolution will involve an episode in which the extended star comes to fill its Roche lobe, see \S~\ref{sec:MT} and \ref{sec:CV}. For the reasons given above, it is unlikely that we will discover the binary shortly before the extended star fills its Roche lobe. If therefore it is close to Roche-lobe filling, we judge that it, the binary, is not likely to represent the unevolved end state of CE evolution.

\subsubsection{Gravitational Wave Radiation}
\label{sec:GW}
\begin{figure}
    \centering
    \includegraphics{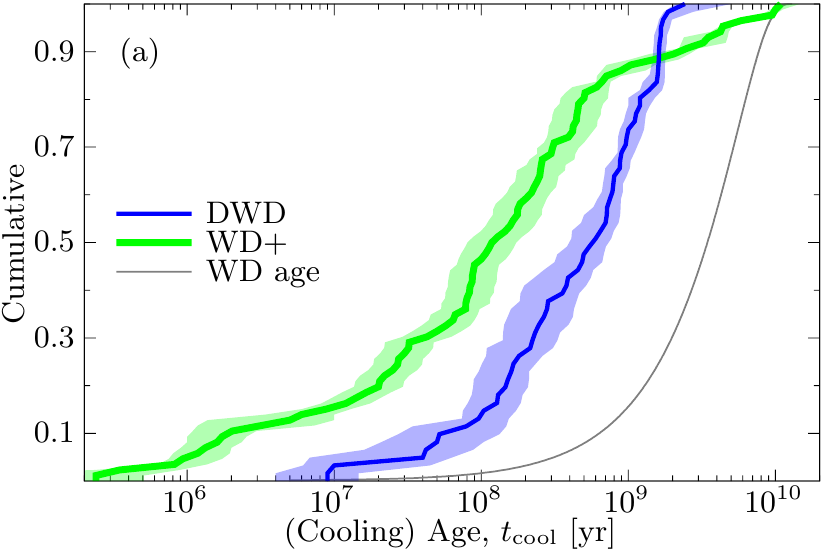}
    \includegraphics{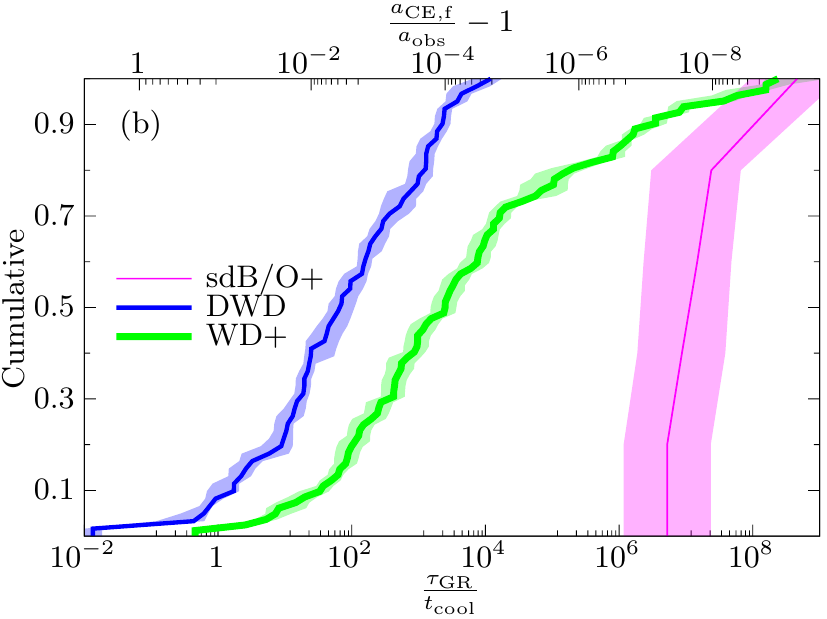}
    \caption{Top panel (a): Cumulative distribution of the (cooling) age of the catalogue systems. The colour and line thickness indicates the binary type. For comparison, the grey line shows the WD age distribution in the solar neighbourhood \citep{kkl20}. Bottom panel (b): Cumulative distribution of the gravitational merger timescale, $\tau_\mathrm{GR}$, in units of the cooling age, $t_\mathrm{cool}$, of the catalogue systems. The top axis shows the fractional change of the semi-major axis between the end of the CE, $a_\mathrm{CE,f}$, and the observation, $a_\mathrm{obs}$, if the orbital change is purely caused by GWR. The colour and line thickness indicates the binary type. To show the uncertainty contours, missing uncertainty values are assumed to be the geometric mean of the recorded ones. Additionally we had to assume that the errors of the underlying quantities, e.g. the masses, are uncorrelated because of missing information on the correlation factors.}
    \label{fig:age}
\end{figure}
All binary or higher multiple star systems will lose orbital angular momentum through emission of gravitational radiation (GWR). In circular non-interacting binaries, such post-CE binaries, this angular momentum loss will proceed on a timescale of \citep[e.g.][]{pet64,ll1975,e2006}
\begin{equation} \label{eq:tGR}
    \tau_\text{GR} =\frac{5}{32}\, \frac{c^5\, a^4}{G^3\, (M_1+M_2)\, M_1\, M_2}.
\end{equation}
This expression is called gravitational merger timescale, with $c$ the speed of light in a vacuum, $G$ the gravitational constant, $M_1$ and $M_2$ the two involved masses and $a$ the semi-major axis of the system.
We note, without going into detail, since most post-CE systems are expected to be circular or near circular, that the gravitational wave merger timescale depends on the eccentricity of the system.
Accordingly, the semi-major axis of the system obeys 
\begin{equation}
	\frac{\dot{a}}{a}=-\frac{2}{\tau_\text{GR}}.
\end{equation}
As has been widely noted \citep[e.g. recently][]{bkm+2016,pll+2019}, the dependence on the fourth power of the semi-major axis of the system prevents merging of wide binaries within the Hubble time unless some mechanism, such as mass transfer, or common envelope evolution is invoked to produce a closer binary. However, in post-CE systems, especially those involving more compact stars, such as He-sdBs and WDs \citep[][]{wjh2013,nyl2016,n2020}, as opposed to hydrogen-rich main sequence stars, can be in such tight orbits, that angular momentum loss due to GWR potentially results in the system being observed in a significantly closer configuration than at the conclusion of the most recent CE phase. Depending on the configuration of the system, the inferred lifetime of the most recently produced component may give some insight into the elapsed time since the most recent CE phase. In Fig.~\ref{fig:cumulative2}b we show the cumulative distribution of gravitational merger timescales of the systems contained in this catalogue in relation to the Hubble time. We note that about $(36.6^{+2.1}_{-0.0})\%$ of the DWD systems can be expected to merge within the Hubble time, making these systems, if their total mass is (roughly) higher than the Chandrasekhar mass, putative SN progenitors \citep[see e.g.][]{pkr2010,r2020} or progenitors of R Coronae Borealis stars (for total binary masses significantly below the Chandrasekhar mass) in the double degenerate merger channel \citep[see e.g.][]{w1984,zj2012}. Figure~\ref{fig:cumulative2}a shows the total binary mass distribution of the systems in this catalogue. We note that out of all sdB/O+ systems, about $(25.4^{+4.2}_{-4.2})\%$ of which will merge within the Hubble time. If $\tau_\mathrm{GR}>\unit{10^9}{\mathrm{yr}}$, which is the case for most of them, they likely also evolve into DWD systems. Where obtainable, we show the gravitational merger timescale, $\tau_\mathrm{GR}$, in relation to the cooling age, $t_\mathrm{cool}$, of the degenerate component in Fig.~\ref{fig:age}b. This is intended to give an indication of the time elapsed since the most recent CE phase. Additionally, this ratio can be transferred into a semi-major axis change if purely GWR influenced the orbit post CE.

\subsubsection{Mass Transfer}
\label{sec:MT}
Post-CE orbits will be subject to subsequent modification if the system undergoes stable mass transfer when one of the component stars fills its Roche-lobe. The material overfilling the star's Roche-lobe will then be transferred to the accreting component, carrying with it a corresponding amount of orbital angular momentum. Generally, the semi-major axis of the system will react to mass transfer according to \citep[see e.g.][chapter 3]{e2006}
\begin{equation} \label{eq:adot-mt}
    \frac{\dot{a}}{a}=2\, \frac{\dot{M}_1}{M_1+M_2}\, \frac{q^2-1}{q},
\end{equation}
assuming conservative mass transfer, which is expected for donor/accretor combinations (excepting WDs) of the stellar types contained in this catalogue. If mass transfer is not conservative, then angular momentum is lost from the system, causing a further decrease in orbital separation. Here, $\dot{M}_1$ is the mass loss rate from the donor star and $q =\frac{M_1}{M_2}$ is the system's mass ratio. Accordingly, for cases of the less massive component transferring mass to the more massive one, the system's semi-major axis will increase, while cases of the more massive component transferring mass to the less massive, the system's semi-major axis will decrease. Mass transfer due to Roche-lobe overflow (trivially) requires the mass donating component to fill its Roche-lobe. Stars overfilling their Roche-lobe correspondingly are expected to exhibit higher mass loss rates \citep[][]{r1988,kr1990}. In Fig.~\ref{fig:cumulative2}c we show the cumulative Roche-lobe filling fraction of the systems contained in this catalogue. We emphasise the large jump for WD+ systems at unity, which is due to CVs being assigned to this class of binaries. Roche-lobe filling fractions exceeding unity by a significant margin may be caused by observational uncertainties. We expect this uncertainty to mainly arise from the radius determination and the fact that mass transferring donors are non-spherical. This uncertainty becomes smaller for smaller Roche-lobe filling fractions.

Mass transfer has the potential to stabilise a binary system against decreasing orbital separation due to GWR depending on whether a critical mass transfer rate $\dot{M}_\text{crit}$ is exceeded. The condition for this effect can be written \citep{ty1979, n2020}
\begin{equation} \label{eq:adot-cond}
    \dot{M}_\mathrm{1} < \dot{M}_\text{crit} = - \frac{32}{5}\, \frac{G^3}{c^5\, a^4} \frac{(M_1+M_2)\, M_1^2\, M_2^2}{M_1-M_2}.
\end{equation}
Due to the dependence on $a^{-4}$, which places restrictions on the radius of the components via their Roche-lobe radius, this condition is usually fulfilled by default in most H-rich and evolved stars with extended envelopes, but not necessarily in interacting systems containing compact objects such as WDs or He sdBs and has to be checked for independently. Most notably, interacting systems with a He sdB transferring material to a more massive WD may still exhibit decreasing orbital separations due to mass transfer being unable to overcome GWR emission. 

In this catalogue, systems known to currently undergo mass transfer are given the mass transfer flag (MT-flag). Systems containing a WD currently accreting H-rich material will be observationally distinguished as cataclysmic variables and, hence, are designated by the CV-flag in this catalogue.
However, we note that cataclysmic variables, see \S~\ref{sec:CV} may, other than RLOF-induced mass transfer, be fuelled by wind accretion.

Since none of the WDs in this catalogue happen to be accreting He-rich material and none of the non-degenerate stars happen to be accreting H-rich material, the MT-flag is equivalent to a system with He-rich mass transfer onto a non-degenerate companion, while the CV flag is equivalent to a system transferring H-rich material onto a WD. Of the sample contained in the catalogue, 25 systems are currently known to be undergoing mass transfer as defined in this section.
We note that a number of sources in this catalogue are observable as X-ray (NS accretors) and/or supersoft X-ray (WD accretors) sources. 

\subsubsection{Cataclysmic Variables}
\label{sec:CV}
A WD in a compact binary with a H-rich MS star accreting H-rich material from a non-degenerate companion will undergo successive episodes of unstable hydrogen ignition at low mass transfer rates, leading to the accreted material and burning products being ejected from the system. As the ejected material carries a certain amount of orbital angular momentum, this will affect the orbital evolution of the system. The frequency and magnitude of these eruptions depends on the accretion rate, with rates in the range $10^{-11}-\unit{10^{-8}}{\Msun\,\mathrm{yr}^{-1}}$ commonly considered in literature. Here, higher accretion rates are connected to a higher outburst frequency and lower ejecta masses per individual outburst. \citep[CVs are extensively discussed in the literature. See, e.g.][and others.]{r1976,sr1983,saa+2002,rk03,ssv+05,hsp+2020}

Of our sample, 223 systems were observationally confirmed as CVs. Most of these can be found in catalogues of CVs \citep[e.g.][]{rk03} as well. Due to uncertainties imposed by the orbital evolution in the post-CE phase, CVs should only be used as limiting cases.

\subsubsection{Tidal Locking and Magnetic Braking}
\label{sec:Tides}
An extended star in a binary will be deformed by the gravitational field of its companion. This deformation and associated rotational shears will impact the rotation of the star as well as, due to the quadrupole moment exerted by the tidal bulges, the orbital angular momentum of the system. Consequently, most post-CE systems will have only small eccentricities, cf. Fig.~\ref{fig:cumulative}l, and an extended companion will get tidally locked. \citep[see e.g.][]{zb1989}

Magnetic braking acts through interaction of a star's magnetic field with its own wind, causing a star to spin down as angular momentum is exchanged between a rotating star and its wind, linearly moving outward. For binary stars, this loss of spin angular momentum affects the system's total orbital angular momentum via tidal interaction. Magnetic braking will cause a star to rotate slower than demanded by pure tidal effects, leading to a loss of total orbital angular momentum as tides cause the star to rotate faster. The efficiency of magnetic braking is currently a matter of debate \citep[see e.g.][]{m2000,e2006,fb+2019}.


\section{Summary}
\label{sec:summary}
We have compiled a unified catalogue of 839 post-CE candidates taken from the literature. The data set is probably incomplete in several different ways. First, the various observations of individual binaries and surveys come all with their own uncertainties and limitations. Second, there are systems missing in the catalogue because they do not match the criterion of having derived masses or some might not have been found in the literature search. Nevertheless, this catalogue represents a unique resource toward unravelling the physics of the CE phase. Even prior to using its systems as inputs for calculations, we have been able to explore the relationships between the component masses and their orbital separation. Interesting patterns emerge in the figures shown in \S~\ref{sec:catalogue}. These patterns surely provide clues to the physical processes producing these end states.

Individual post-CE candidates and small groups of them can provide reliable input for interesting studies. We anticipate that one of the most significant uses of this catalogue will be to improve the parameterized methods currently used (such as the $\alpha$ and $\gamma$ formalisms) that map initial CE states to the CE end states. Another important use of the catalogue is as a testing ground for hydrodynamic simulations which are currently taking on the challenge of computing CE evolution.

Observations of close binaries are ongoing, and the results are being published at an increasing rate. Thus, while this catalogue is currently the best resource for the study of post-CEs, it will be useful to supplement its contents with new systems, as they are discovered. There is, however, a crucial caveat: we have surveyed the vast literature on potential post-CEs and developed a set of uniform criteria for the selection of the candidates included here. While some post-CEs may have been missed, and some of the candidates we list may represent ``false positives'', the value of the careful selection of systems is that it provides a level of uniformity, important for subsequent uses of the catalogue. Therefore, additions to the catalogue should be made by employing similar selection criteria, see \S~\ref{sec:selection}. This catalogue and its future extensions will provide insight into the evolution of close binaries through the CE phase.


\section*{Data availability}
The catalogue data set will be available in the online version. Additionally, the newest version of the data can be requested via Email to \href{mailto:mkruckow@ynao.ac.cn}{mkruckow@ynao.ac.cn}.


\begin{acknowledgments}
\section*{Acknowledgements}
We thank Jiao Li for his support on the Gaia data.
We are grateful for the improvement suggestions by the anonymous referee.
MK is partly supported by Grant No 11521303, 11733008, 12090040, and 12090043 of the Natural Science Foundation of China.
RD acknowledges support from the National Science Foundation, through NSF AST-2009520.
YG is funded by the Royal Society and the K.C. Wong Education Foundation.
CK acknowledges funding from the UK Science and Technology Facility Council (STFC) through grant ST/ R000905/1 and ST/V000632/1, and a travel support from Z. Han.
This team collaboration started at a conference ``Progenitors of Type Ia Supernovae'' in Lijiang, 2019.

This research has made use of the VizieR catalogue access tool, CDS, Strasbourg, France (DOI: \url{https://doi.org/10.26093/cds/vizier}). The original description of the VizieR service was published in \citet{obm00}. This research has made use of the SIMBAD database, operated at CDS, Strasbourg, France \citep{woe+00}.
\end{acknowledgments}

\bibliography{PCEC_refs.bib}{}

\begin{thebibliography}{}
\expandafter\ifx\csname natexlab\endcsname\relax\def\natexlab#1{#1}\fi
\providecommand{\url}[1]{\href{#1}{#1}}
\providecommand{\dodoi}[1]{doi:~\href{http://doi.org/#1}{\nolinkurl{#1}}}
\providecommand{\doeprint}[1]{\href{http://ascl.net/#1}{\nolinkurl{http://ascl.net/#1}}}
\providecommand{\doarXiv}[1]{\href{https://arxiv.org/abs/#1}{\nolinkurl{https://arxiv.org/abs/#1}}}

\bibitem[{{Ablimit} {et~al.}(2016){Ablimit}, {Maeda}, \& {Li}}]{akx2016}
{Ablimit}, I., {Maeda}, K., \& {Li}, X.-D. 2016, \apj, 826, 53,
  \dodoi{10.3847/0004-637X/826/1/53}

\bibitem[{{Af{\textcommabelow s}ar} \& {Ibano{\v{g}}lu}(2008)}]{ai08}
{Af{\textcommabelow s}ar}, M., \& {Ibano{\v{g}}lu}, C. 2008, \mnras, 391, 802,
  \dodoi{10.1111/j.1365-2966.2008.13927.x}

\bibitem[{{Almeida} {et~al.}(2017){Almeida}, {Damineli}, {Rodrigues},
  {Pereira}, \& {Jablonski}}]{adr+17}
{Almeida}, L.~A., {Damineli}, A., {Rodrigues}, C.~V., {Pereira}, M.~G., \&
  {Jablonski}, F. 2017, \mnras, 472, 3093, \dodoi{10.1093/mnras/stx2150}

\bibitem[{{Almenara} {et~al.}(2012){Almenara}, {Alonso}, {Rabus}, {L{\'a}zaro},
  {Ar{\'e}valo}, {Belmonte}, {Deeg}, {Brown}, \& {V{\'a}zquez
  Rami{\'o}}}]{aar+12}
{Almenara}, J.~M., {Alonso}, R., {Rabus}, M., {et~al.} 2012, \mnras, 420, 3017,
  \dodoi{10.1111/j.1365-2966.2011.20175.x}

\bibitem[{{Antoniadis} {et~al.}(2016){Antoniadis}, {Kaplan}, {Stovall},
  {Freire}, {Deneva}, {Koester}, {Jenet}, \& {Martinez}}]{aks+16}
{Antoniadis}, J., {Kaplan}, D.~L., {Stovall}, K., {et~al.} 2016, \apj, 830, 36,
  \dodoi{10.3847/0004-637X/830/1/36}

\bibitem[{{Antoniadis} {et~al.}(2012){Antoniadis}, {van Kerkwijk}, {Koester},
  {Freire}, {Wex}, {Tauris}, {Kramer}, \& {Bassa}}]{avk+12}
{Antoniadis}, J., {van Kerkwijk}, M.~H., {Koester}, D., {et~al.} 2012, \mnras,
  423, 3316, \dodoi{10.1111/j.1365-2966.2012.21124.x}

\bibitem[{{Antoniadis} {et~al.}(2013){Antoniadis}, {Freire}, {Wex}, {Tauris},
  {Lynch}, {van Kerkwijk}, {Kramer}, {Bassa}, {Dhillon}, {Driebe}, {Hessels},
  {Kaspi}, {Kondratiev}, {Langer}, {Marsh}, {McLaughlin}, {Pennucci}, {Ransom},
  {Stairs}, {van Leeuwen}, {Verbiest}, \& {Whelan}}]{afw+13}
{Antoniadis}, J., {Freire}, P. C.~C., {Wex}, N., {et~al.} 2013, Science, 340,
  448, \dodoi{10.1126/science.1233232}

\bibitem[{{Araujo-Betancor} {et~al.}(2003){Araujo-Betancor}, {Knigge}, {Long},
  {Hoard}, {Szkody}, {Rodgers}, {Krisciunas}, {Dhillon}, {Hynes}, {Patterson},
  \& {Kemp}}]{akl+03}
{Araujo-Betancor}, S., {Knigge}, C., {Long}, K.~S., {et~al.} 2003, \apj, 583,
  437, \dodoi{10.1086/345098}

\bibitem[{{Arenas} {et~al.}(2000){Arenas}, {Catal{\'a}n}, {Augusteijn}, \&
  {Retter}}]{aca+00}
{Arenas}, J., {Catal{\'a}n}, M.~S., {Augusteijn}, T., \& {Retter}, A. 2000,
  \mnras, 311, 135, \dodoi{10.1046/j.1365-8711.2000.03061.x}

\bibitem[{{Armstrong} {et~al.}(2012){Armstrong}, {Patterson}, \&
  {Kemp}}]{apk12}
{Armstrong}, E., {Patterson}, J., \& {Kemp}, J. 2012, \mnras, 421, 2310,
  \dodoi{10.1111/j.1365-2966.2012.20463.x}

\bibitem[{{Arnold} {et~al.}(1976){Arnold}, {Berg}, \& {Duthie}}]{abd76}
{Arnold}, S., {Berg}, R.~A., \& {Duthie}, J.~G. 1976, \apj, 206, 790,
  \dodoi{10.1086/154440}

\bibitem[{{Aungwerojwit} {et~al.}(2007){Aungwerojwit}, {G{\"a}nsicke},
  {Rodr{\'\i}guez-Gil}, {Hagen}, {Giannakis}, {Papadimitriou}, {Allende
  Prieto}, \& {Engels}}]{agr+07}
{Aungwerojwit}, A., {G{\"a}nsicke}, B.~T., {Rodr{\'\i}guez-Gil}, P., {et~al.}
  2007, \aap, 469, 297, \dodoi{10.1051/0004-6361:20077276}

\bibitem[{{Badenes} {et~al.}(2013){Badenes}, {van Kerkwijk}, {Kilic},
  {Bickerton}, {Mazeh}, {Mullally}, {Tal-Or}, \& {Thompson}}]{bvk+13}
{Badenes}, C., {van Kerkwijk}, M.~H., {Kilic}, M., {et~al.} 2013, \mnras, 429,
  3596, \dodoi{10.1093/mnras/sts646}

\bibitem[{{Bai} {et~al.}(2016){Bai}, {Justham}, {Liu}, {Guo}, {Gao}, \&
  {Gong}}]{bjl+16}
{Bai}, Y., {Justham}, S., {Liu}, J., {et~al.} 2016, \apj, 828, 39,
  \dodoi{10.3847/0004-637X/828/1/39}

\bibitem[{{Bailer-Jones} {et~al.}(2021){Bailer-Jones}, {Rybizki}, {Fouesneau},
  {Demleitner}, \& {Andrae}}]{brf+21}
{Bailer-Jones}, C.~A.~L., {Rybizki}, J., {Fouesneau}, M., {Demleitner}, M., \&
  {Andrae}, R. 2021, \aj, 161, 147, \dodoi{10.3847/1538-3881/abd806}

\bibitem[{{Baptista} {et~al.}(2003){Baptista}, {Borges}, {Bond}, {Jablonski},
  {Steiner}, \& {Grauer}}]{bbb+03}
{Baptista}, R., {Borges}, B.~W., {Bond}, H.~E., {et~al.} 2003, \mnras, 345,
  889, \dodoi{10.1046/j.1365-8711.2003.07014.x}

\bibitem[{{Baptista} {et~al.}(1998){Baptista}, {Catalan}, {Horne}, \&
  {Zilli}}]{bch+98}
{Baptista}, R., {Catalan}, M.~S., {Horne}, K., \& {Zilli}, D. 1998, \mnras,
  300, 233, \dodoi{10.1046/j.1365-8711.1998.01887.x}

\bibitem[{{Baptista} {et~al.}(1994){Baptista}, {Steiner}, \&
  {Cieslinski}}]{bsc94}
{Baptista}, R., {Steiner}, J.~E., \& {Cieslinski}, D. 1994, \apj, 433, 332,
  \dodoi{10.1086/174648}

\bibitem[{{Baran} {et~al.}(2019){Baran}, {Telting}, {Jeffery}, {{\O}stensen},
  {Vos}, {Reed}, {V{\r{A}}­ckovi{\'c}}, \& {}}]{btj+19}
{Baran}, A.~S., {Telting}, J.~H., {Jeffery}, C.~S., {et~al.} 2019, \mnras, 489,
  1556, \dodoi{10.1093/mnras/stz2209}

\bibitem[{{Barlow} {et~al.}(2011){Barlow}, {Dunlap}, \& {Clemens}}]{bdc11}
{Barlow}, B.~N., {Dunlap}, B.~H., \& {Clemens}, J.~C. 2011, \apjl, 737, L2,
  \dodoi{10.1088/2041-8205/737/1/L2}

\bibitem[{{Barlow} {et~al.}(2013){Barlow}, {Kilkenny}, {Drechsel}, {Dunlap},
  {O'Donoghue}, {Geier}, {O'Steen}, {Clemens}, {LaCluyze}, {Reichart},
  {Haislip}, {Nysewander}, \& {Ivarsen}}]{bkd+13}
{Barlow}, B.~N., {Kilkenny}, D., {Drechsel}, H., {et~al.} 2013, \mnras, 430,
  22, \dodoi{10.1093/mnras/sts271}

\bibitem[{{Bassa} {et~al.}(2006{\natexlab{a}}){Bassa}, {van Kerkwijk},
  {Koester}, \& {Verbunt}}]{bvk+06}
{Bassa}, C.~G., {van Kerkwijk}, M.~H., {Koester}, D., \& {Verbunt}, F.
  2006{\natexlab{a}}, \aap, 456, 295, \dodoi{10.1051/0004-6361:20065181}

\bibitem[{{Bassa} {et~al.}(2003){Bassa}, {van Kerkwijk}, \& {Kulkarni}}]{bvk03}
{Bassa}, C.~G., {van Kerkwijk}, M.~H., \& {Kulkarni}, S.~R. 2003, \aap, 403,
  1067, \dodoi{10.1051/0004-6361:20030384}

\bibitem[{{Bassa} {et~al.}(2006{\natexlab{b}}){Bassa}, {van Kerkwijk}, \&
  {Kulkarni}}]{bvk06}
---. 2006{\natexlab{b}}, \aap, 450, 295, \dodoi{10.1051/0004-6361:20054316}

\bibitem[{{Belczynski} {et~al.}(2002){Belczynski}, {Kalogera}, \&
  {Bulik}}]{bkb02}
{Belczynski}, K., {Kalogera}, V., \& {Bulik}, T. 2002, \apj, 572, 407,
  \dodoi{10.1086/340304}

\bibitem[{{Bell} {et~al.}(1994){Bell}, {Pollacco}, \& {Hilditch}}]{bph94}
{Bell}, S.~A., {Pollacco}, D.~L., \& {Hilditch}, R.~W. 1994, \mnras, 270, 449,
  \dodoi{10.1093/mnras/270.2.449}

\bibitem[{{Bergeron} {et~al.}(1994){Bergeron}, {Wesemael}, {Beauchamp}, {Wood},
  {Lamontagne}, {Fontaine}, \& {Liebert}}]{bwb+94}
{Bergeron}, P., {Wesemael}, F., {Beauchamp}, A., {et~al.} 1994, \apj, 432, 305,
  \dodoi{10.1086/174571}

\bibitem[{{Bergeron} {et~al.}(1989){Bergeron}, {Wesemael}, {Liebert}, \&
  {Fontaine}}]{bwl+89}
{Bergeron}, P., {Wesemael}, F., {Liebert}, J., \& {Fontaine}, G. 1989, \apjl,
  345, L91, \dodoi{10.1086/185560}

\bibitem[{{Beuermann} {et~al.}(2004){Beuermann}, {Harrison}, {McArthur},
  {Benedict}, \& {G{\"a}nsicke}}]{bhm+04}
{Beuermann}, K., {Harrison}, T.~E., {McArthur}, B.~E., {Benedict}, G.~F., \&
  {G{\"a}nsicke}, B.~T. 2004, \aap, 419, 291,
  \dodoi{10.1051/0004-6361:20034424}

\bibitem[{{Beuermann} \& {Reinsch}(2008)}]{br08}
{Beuermann}, K., \& {Reinsch}, K. 2008, \aap, 480, 199,
  \dodoi{10.1051/0004-6361:20079010}

\bibitem[{{Beuermann} {et~al.}(1992){Beuermann}, {Stasiewski}, \&
  {Schwope}}]{bss92}
{Beuermann}, K., {Stasiewski}, U., \& {Schwope}, A.~D. 1992, \aap, 256, 433

\bibitem[{{Beuermann} \& {Thomas}(1990)}]{bt90}
{Beuermann}, K., \& {Thomas}, H.~C. 1990, \aap, 230, 326

\bibitem[{{Beuermann} {et~al.}(2013){Beuermann}, {Dreizler}, {Hessman},
  {Backhaus}, {Boesch}, {Husser}, {Nortmann}, {Schmelev}, \&
  {Springer}}]{bdh+13}
{Beuermann}, K., {Dreizler}, S., {Hessman}, F.~V., {et~al.} 2013, \aap, 558,
  A96, \dodoi{10.1051/0004-6361/201322241}

\bibitem[{{Bhat} {et~al.}(2008){Bhat}, {Bailes}, \& {Verbiest}}]{bbv08}
{Bhat}, N.~D.~R., {Bailes}, M., \& {Verbiest}, J. P.~W. 2008, \prd, 77, 124017,
  \dodoi{10.1103/PhysRevD.77.124017}

\bibitem[{{Bianchini}(1980)}]{bia80}
{Bianchini}, A. 1980, \mnras, 192, 127, \dodoi{10.1093/mnras/192.2.127}

\bibitem[{{Bitner} {et~al.}(2007){Bitner}, {Robinson}, \& {Behr}}]{brb07}
{Bitner}, M.~A., {Robinson}, E.~L., \& {Behr}, B.~B. 2007, \apj, 662, 564,
  \dodoi{10.1086/517496}

\bibitem[{{Bleach} {et~al.}(2000){Bleach}, {Wood}, {Catal{\'a}n}, {Welsh},
  {Robinson}, \& {Skidmore}}]{bwc+00}
{Bleach}, J.~N., {Wood}, J.~H., {Catal{\'a}n}, M.~S., {et~al.} 2000, \mnras,
  312, 70, \dodoi{10.1046/j.1365-8711.2000.03218.x}

\bibitem[{{Bloemen} {et~al.}(2011){Bloemen}, {Marsh}, {{\O}stensen},
  {Charpinet}, {Fontaine}, {Degroote}, {Heber}, {Kawaler}, {Aerts}, {Green},
  {Telting}, {Brassard}, {G{\"a}nsicke}, {Handler}, {Kurtz}, {Silvotti}, {Van
  Grootel}, {Lindberg}, {Pursimo}, {Wilson}, {Gilliland}, {Kjeldsen},
  {Christensen-Dalsgaard}, {Borucki}, {Koch}, {Jenkins}, \& {Klaus}}]{bmo+11}
{Bloemen}, S., {Marsh}, T.~R., {{\O}stensen}, R.~H., {et~al.} 2011, \mnras,
  410, 1787, \dodoi{10.1111/j.1365-2966.2010.17559.x}

\bibitem[{{Bloemen} {et~al.}(2012){Bloemen}, {Marsh}, {Degroote},
  {{\O}stensen}, {P{\'a}pics}, {Aerts}, {Koester}, {G{\"a}nsicke}, {Breedt},
  {Lombaert}, {Pyrzas}, {Copperwheat}, {Exter}, {Raskin}, {Van Winckel},
  {Prins}, {Pessemier}, {Fr{\'e}mat}, {Hensberge}, {Jorissen}, \& {Van
  Eck}}]{bmd+12}
{Bloemen}, S., {Marsh}, T.~R., {Degroote}, P., {et~al.} 2012, \mnras, 422,
  2600, \dodoi{10.1111/j.1365-2966.2012.20818.x}

\bibitem[{{Bloom} {et~al.}(1999){Bloom}, {Sigurdsson}, \& {Pols}}]{bsp99}
{Bloom}, J.~S., {Sigurdsson}, S., \& {Pols}, O.~R. 1999, \mnras, 305, 763,
  \dodoi{10.1046/j.1365-8711.1999.02437.x}

\bibitem[{{Borges} \& {Baptista}(2005)}]{bb05}
{Borges}, B.~W., \& {Baptista}, R. 2005, \aap, 437, 235,
  \dodoi{10.1051/0004-6361:20042045}

\bibitem[{{Bours} {et~al.}(2014){Bours}, {Marsh}, {Parsons}, {Copperwheat},
  {Dhillon}, {Littlefair}, {G{\"a}nsicke}, {Gianninas}, \& {Tremblay}}]{bmp+14}
{Bours}, M.~C.~P., {Marsh}, T.~R., {Parsons}, S.~G., {et~al.} 2014, \mnras,
  438, 3399, \dodoi{10.1093/mnras/stt2453}

\bibitem[{{Bours} {et~al.}(2015){Bours}, {Marsh}, {G{\"a}nsicke}, {Tauris},
  {Istrate}, {Badenes}, {Dhillon}, {Gal-Yam}, {Hermes}, {Kengkriangkrai},
  {Kilic}, {Koester}, {Mullally}, {Prasert}, {Steeghs}, {Thompson}, \&
  {Thorstensen}}]{bmg+15}
{Bours}, M.~C.~P., {Marsh}, T.~R., {G{\"a}nsicke}, B.~T., {et~al.} 2015,
  \mnras, 450, 3966, \dodoi{10.1093/mnras/stv889}

\bibitem[{{Breedt} {et~al.}(2012{\natexlab{a}}){Breedt}, {G{\"a}nsicke},
  {Girven}, {Drake}, {Copperwheat}, {Parsons}, \& {Marsh}}]{bgg+12}
{Breedt}, E., {G{\"a}nsicke}, B.~T., {Girven}, J., {et~al.} 2012{\natexlab{a}},
  \mnras, 423, 1437, \dodoi{10.1111/j.1365-2966.2012.20965.x}

\bibitem[{{Breedt} {et~al.}(2012{\natexlab{b}}){Breedt}, {G{\"a}nsicke},
  {Marsh}, {Steeghs}, {Drake}, \& {Copperwheat}}]{bgm+12}
{Breedt}, E., {G{\"a}nsicke}, B.~T., {Marsh}, T.~R., {et~al.}
  2012{\natexlab{b}}, \mnras, 425, 2548,
  \dodoi{10.1111/j.1365-2966.2012.21724.x}

\bibitem[{{Breedt} {et~al.}(2017){Breedt}, {Steeghs}, {Marsh}, {Gentile
  Fusillo}, {Tremblay}, {Green}, {De Pasquale}, {Hermes}, {G{\"a}nsicke},
  {Parsons}, {Bours}, {Longa-Pe{\~n}a}, \& {Rebassa-Mansergas}}]{bsm+17}
{Breedt}, E., {Steeghs}, D., {Marsh}, T.~R., {et~al.} 2017, \mnras, 468, 2910,
  \dodoi{10.1093/mnras/stx430}

\bibitem[{{Breton} {et~al.}(2012){Breton}, {Rappaport}, {van Kerkwijk}, \&
  {Carter}}]{brv+12}
{Breton}, R.~P., {Rappaport}, S.~A., {van Kerkwijk}, M.~H., \& {Carter}, J.~A.
  2012, \apj, 748, 115, \dodoi{10.1088/0004-637X/748/2/115}

\bibitem[{{Brott} {et~al.}(2011){Brott}, {de Mink}, {Cantiello}, {Langer}, {de
  Koter}, {Evans}, {Hunter}, {Trundle}, \& {Vink}}]{bdc+11}
{Brott}, I., {de Mink}, S.~E., {Cantiello}, M., {et~al.} 2011, \aap, 530, A115,
  \dodoi{10.1051/0004-6361/201016113}

\bibitem[{{Brown} {et~al.}(2011{\natexlab{a}}){Brown}, {Kilic}, {Brown}, \&
  {Kenyon}}]{bkb+11}
{Brown}, J.~M., {Kilic}, M., {Brown}, W.~R., \& {Kenyon}, S.~J.
  2011{\natexlab{a}}, \apj, 730, 67, \dodoi{10.1088/0004-637X/730/2/67}

\bibitem[{{Brown} {et~al.}(2016{\natexlab{a}}){Brown}, {Gianninas}, {Kilic},
  {Kenyon}, \& {Allende Prieto}}]{bgk+16}
{Brown}, W.~R., {Gianninas}, A., {Kilic}, M., {Kenyon}, S.~J., \& {Allende
  Prieto}, C. 2016{\natexlab{a}}, \apj, 818, 155,
  \dodoi{10.3847/0004-637X/818/2/155}

\bibitem[{{Brown} {et~al.}(2013){Brown}, {Kilic}, {Allende Prieto},
  {Gianninas}, \& {Kenyon}}]{bka+13}
{Brown}, W.~R., {Kilic}, M., {Allende Prieto}, C., {Gianninas}, A., \&
  {Kenyon}, S.~J. 2013, \apj, 769, 66, \dodoi{10.1088/0004-637X/769/1/66}

\bibitem[{{Brown} {et~al.}(2010){Brown}, {Kilic}, {Allende Prieto}, \&
  {Kenyon}}]{bka+10}
{Brown}, W.~R., {Kilic}, M., {Allende Prieto}, C., \& {Kenyon}, S.~J. 2010,
  \apj, 723, 1072, \dodoi{10.1088/0004-637X/723/2/1072}

\bibitem[{{Brown} {et~al.}(2012){Brown}, {Kilic}, {Allende Prieto}, \&
  {Kenyon}}]{bka+12}
---. 2012, \apj, 744, 142, \dodoi{10.1088/0004-637X/744/2/142}

\bibitem[{{Brown} {et~al.}(2011{\natexlab{b}}){Brown}, {Kilic}, {Hermes},
  {Allende Prieto}, {Kenyon}, \& {Winget}}]{bkh+11}
{Brown}, W.~R., {Kilic}, M., {Hermes}, J.~J., {et~al.} 2011{\natexlab{b}},
  \apjl, 737, L23, \dodoi{10.1088/2041-8205/737/1/L23}

\bibitem[{{Brown} {et~al.}(2016{\natexlab{b}}){Brown}, {Kilic}, {Kenyon}, \&
  {Gianninas}}]{bkm+2016}
{Brown}, W.~R., {Kilic}, M., {Kenyon}, S.~J., \& {Gianninas}, A.
  2016{\natexlab{b}}, \apj, 824, 46, \dodoi{10.3847/0004-637X/824/1/46}

\bibitem[{{Brown} {et~al.}(2017){Brown}, {Kilic}, {Kosakowski}, \&
  {Gianninas}}]{bkk+17}
{Brown}, W.~R., {Kilic}, M., {Kosakowski}, A., \& {Gianninas}, A. 2017, \apj,
  847, 10, \dodoi{10.3847/1538-4357/aa8724}

\bibitem[{{Bruch}(2003)}]{bru03}
{Bruch}, A. 2003, \aap, 409, 647, \dodoi{10.1051/0004-6361:20031041}

\bibitem[{{Bruch} {et~al.}(2001){Bruch}, {Vaz}, \& {Diaz}}]{bvd01}
{Bruch}, A., {Vaz}, L.~P.~R., \& {Diaz}, M.~P. 2001, \aap, 377, 898,
  \dodoi{10.1051/0004-6361:20011092}

\bibitem[{{Burdge} {et~al.}(2019{\natexlab{a}}){Burdge}, {Coughlin}, {Fuller},
  {Kupfer}, {Bellm}, {Bildsten}, {Graham}, {Kaplan}, {Roestel}, {Dekany},
  {Duev}, {Feeney}, {Giomi}, {Helou}, {Kaye}, {Laher}, {Mahabal}, {Masci},
  {Riddle}, {Shupe}, {Soumagnac}, {Smith}, {Szkody}, {Walters}, {Kulkarni}, \&
  {Prince}}]{bcf+19}
{Burdge}, K.~B., {Coughlin}, M.~W., {Fuller}, J., {et~al.} 2019{\natexlab{a}},
  \nat, 571, 528, \dodoi{10.1038/s41586-019-1403-0}

\bibitem[{{Burdge} {et~al.}(2019{\natexlab{b}}){Burdge}, {Fuller}, {Phinney},
  {van Roestel}, {Claret}, {Cukanovaite}, {Gentile Fusillo}, {Coughlin},
  {Kaplan}, {Kupfer}, {Tremblay}, {Dekany}, {Duev}, {Feeney}, {Riddle},
  {Kulkarni}, \& {Prince}}]{bfp+19}
{Burdge}, K.~B., {Fuller}, J., {Phinney}, E.~S., {et~al.} 2019{\natexlab{b}},
  \apjl, 886, L12, \dodoi{10.3847/2041-8213/ab53e5}

\bibitem[{{Burdge} {et~al.}(2020{\natexlab{a}}){Burdge}, {Coughlin}, {Fuller},
  {Kaplan}, {Kulkarni}, {Marsh}, {Bellm}, {Dekany}, {Duev}, {Graham},
  {Mahabal}, {Masci}, {Laher}, {Riddle}, {Soumagnac}, \& {Prince}}]{bcf+20}
{Burdge}, K.~B., {Coughlin}, M.~W., {Fuller}, J., {et~al.} 2020{\natexlab{a}},
  \apjl, 905, L7, \dodoi{10.3847/2041-8213/abca91}

\bibitem[{{Burdge} {et~al.}(2020{\natexlab{b}}){Burdge}, {Prince}, {Fuller},
  {Kaplan}, {Marsh}, {Tremblay}, {Zhuang}, {Bellm}, {Caiazzo}, {Coughlin},
  {Dhillon}, {Gaensicke}, {Rodr{\'\i}guez-Gil}, {Graham}, {Hermes}, {Kupfer},
  {Littlefair}, {Mr{\'o}z}, {Phinney}, {van Roestel}, {Yao}, {Dekany}, {Drake},
  {Duev}, {Hale}, {Feeney}, {Helou}, {Kaye}, {Mahabal}, {Masci}, {Riddle},
  {Smith}, {Soumagnac}, \& {Kulkarni}}]{bpf+20}
{Burdge}, K.~B., {Prince}, T.~A., {Fuller}, J., {et~al.} 2020{\natexlab{b}},
  \apj, 905, 32, \dodoi{10.3847/1538-4357/abc261}

\bibitem[{{Burwitz} {et~al.}(1999){Burwitz}, {Reinsch}, {Beuermann}, \&
  {Thomas}}]{brb+99}
{Burwitz}, V., {Reinsch}, K., {Beuermann}, K., \& {Thomas}, H.-C. 1999, in
  Astronomical Society of the Pacific Conference Series, Vol. 157, Annapolis
  Workshop on Magnetic Cataclysmic Variables, ed. C.~{Hellier} \& K.~{Mukai},
  127.
\newblock \doarXiv{astro-ph/9810437}

\bibitem[{{Burwitz} {et~al.}(1998){Burwitz}, {Reinsch}, {Schwope}, {Hakala},
  {Beuermann}, {Rousseau}, {Thomas}, {G{\"a}nsicke}, {Piirola}, \&
  {Vilhu}}]{brs+98}
{Burwitz}, V., {Reinsch}, K., {Schwope}, A.~D., {et~al.} 1998, \aap, 331, 262

\bibitem[{{Callanan} {et~al.}(1998){Callanan}, {Garnavich}, \&
  {Koester}}]{cgk98}
{Callanan}, P.~J., {Garnavich}, P.~M., \& {Koester}, D. 1998, \mnras, 298, 207,
  \dodoi{10.1046/j.1365-8711.1998.01634.x}

\bibitem[{{Cameron} {et~al.}(2020){Cameron}, {Champion}, {Bailes},
  {Balakrishnan}, {Barr}, {Bassa}, {Bates}, {Bhandari}, {Bhat}, {Burgay},
  {Burke-Spolaor}, {Flynn}, {Jameson}, {Johnston}, {Keith}, {Kramer}, {Levin},
  {Lyne}, {Ng}, {Petroff}, {Possenti}, {Smith}, {Stappers}, {van Straten},
  {Tiburzi}, \& {Wu}}]{ccb+20}
{Cameron}, A.~D., {Champion}, D.~J., {Bailes}, M., {et~al.} 2020, \mnras, 493,
  1063, \dodoi{10.1093/mnras/staa039}

\bibitem[{{Camilo} {et~al.}(2001){Camilo}, {Lyne}, {Manchester}, {Bell},
  {Stairs}, {D'Amico}, {Kaspi}, {Possenti}, {Crawford}, \& {McKay}}]{clm+01}
{Camilo}, F., {Lyne}, A.~G., {Manchester}, R.~N., {et~al.} 2001, \apjl, 548,
  L187, \dodoi{10.1086/319120}

\bibitem[{{Carter} {et~al.}(2011){Carter}, {Rappaport}, \& {Fabrycky}}]{crf11}
{Carter}, J.~A., {Rappaport}, S., \& {Fabrycky}, D. 2011, \apj, 728, 139,
  \dodoi{10.1088/0004-637X/728/2/139}

\bibitem[{{Casewell} {et~al.}(2012){Casewell}, {Burleigh}, {Wynn}, {Alexander},
  {Napiwotzki}, {Lawrie}, {Dobbie}, {Jameson}, \& {Hodgkin}}]{cbw+12}
{Casewell}, S.~L., {Burleigh}, M.~R., {Wynn}, G.~A., {et~al.} 2012, \apjl, 759,
  L34, \dodoi{10.1088/2041-8205/759/2/L34}

\bibitem[{{Casewell} {et~al.}(2020){Casewell}, {Belardi}, {Parsons},
  {Littlefair}, {Braker}, {Hermes}, {Debes}, {Vanderbosch}, {Burleigh},
  {G{\"a}nsicke}, {Dhillon}, {Marsh}, {Winget}, \& {Winget}}]{cbpl+20}
{Casewell}, S.~L., {Belardi}, C., {Parsons}, S.~G., {et~al.} 2020, \mnras, 497,
  3571, \dodoi{10.1093/mnras/staa1608}

\bibitem[{{Chen} {et~al.}(1995){Chen}, {O'Donoghue}, {Stobie}, {Kilkenny},
  {Roberts}, \& {van Wyk}}]{cos+95}
{Chen}, A., {O'Donoghue}, D., {Stobie}, R.~S., {et~al.} 1995, \mnras, 275, 100,
  \dodoi{10.1093/mnras/275.1.100}

\bibitem[{{Chen} {et~al.}(2021){Chen}, {Tauris}, {Han}, \& {Chen}}]{cth+21}
{Chen}, H.-L., {Tauris}, T.~M., {Han}, Z., \& {Chen}, X. 2021, \mnras, 503,
  3540, \dodoi{10.1093/mnras/stab670}

\bibitem[{{Claeys} {et~al.}(2014){Claeys}, {Pols}, {Izzard}, {Vink}, \&
  {Verbunt}}]{cpi2014}
{Claeys}, J.~S.~W., {Pols}, O.~R., {Izzard}, R.~G., {Vink}, J., \& {Verbunt},
  F.~W.~M. 2014, \aap, 563, A83, \dodoi{10.1051/0004-6361/201322714}

\bibitem[{{Comerford} \& {Izzard}(2020)}]{ci20}
{Comerford}, T.~A.~F., \& {Izzard}, R.~G. 2020, \mnras, 498, 2957,
  \dodoi{10.1093/mnras/staa2539}

\bibitem[{{Copperwheat} {et~al.}(2010){Copperwheat}, {Marsh}, {Dhillon},
  {Littlefair}, {Hickman}, {G{\"a}nsicke}, \& {Southworth}}]{cmd+10}
{Copperwheat}, C.~M., {Marsh}, T.~R., {Dhillon}, V.~S., {et~al.} 2010, \mnras,
  402, 1824, \dodoi{10.1111/j.1365-2966.2009.16010.x}

\bibitem[{{Copperwheat} {et~al.}(2011{\natexlab{a}}){Copperwheat},
  {Morales-Rueda}, {Marsh}, {Maxted}, \& {Heber}}]{cmm+11}
{Copperwheat}, C.~M., {Morales-Rueda}, L., {Marsh}, T.~R., {Maxted}, P.~F.~L.,
  \& {Heber}, U. 2011{\natexlab{a}}, \mnras, 415, 1381,
  \dodoi{10.1111/j.1365-2966.2011.18786.x}

\bibitem[{{Copperwheat} {et~al.}(2011{\natexlab{b}}){Copperwheat}, {Marsh},
  {Littlefair}, {Dhillon}, {Ramsay}, {Drake}, {G{\"a}nsicke}, {Groot},
  {Hakala}, {Koester}, {Nelemans}, {Roelofs}, {Southworth}, {Steeghs}, \&
  {Tulloch}}]{cml+11}
{Copperwheat}, C.~M., {Marsh}, T.~R., {Littlefair}, S.~P., {et~al.}
  2011{\natexlab{b}}, \mnras, 410, 1113,
  \dodoi{10.1111/j.1365-2966.2010.17508.x}

\bibitem[{{Corcoran} {et~al.}(2021{\natexlab{a}}){Corcoran}, {Barlow},
  {Schaffenroth}, {Heber}, {Walser}, \& {Irgang}}]{cbs+21}
{Corcoran}, K.~A., {Barlow}, B.~N., {Schaffenroth}, V., {et~al.}
  2021{\natexlab{a}}, arXiv e-prints, arXiv:2106.08328.
\newblock \doarXiv{2106.08328}

\bibitem[{{Corcoran} {et~al.}(2021{\natexlab{b}}){Corcoran}, {Lewis},
  {Anguiano}, {Majewski}, {Kounkel}, {McDonald}, {Stassun}, {Cunha}, {Smith},
  {Allende Prieto}, {Badenes}, {De Lee}, {Mazzola}, {Longa-Pe{\~n}a}, \&
  {Roman-Lopes}}]{cla+21}
{Corcoran}, K.~A., {Lewis}, H.~M., {Anguiano}, B., {et~al.} 2021{\natexlab{b}},
  \aj, 161, 143, \dodoi{10.3847/1538-3881/abd62e}

\bibitem[{{Coughlin} {et~al.}(2020){Coughlin}, {Burdge}, {Phinney}, {van
  Roestel}, {Bellm}, {Dekany}, {Delacroix}, {Duev}, {Feeney}, {Graham},
  {Kulkarni}, {Kupfer}, {Laher}, {Masci}, {Prince}, {Riddle}, {Rosnet},
  {Smith}, {Serabyn}, \& {Walters}}]{cbp+20}
{Coughlin}, M.~W., {Burdge}, K., {Phinney}, E.~S., {et~al.} 2020, \mnras, 494,
  L91, \dodoi{10.1093/mnrasl/slaa044}

\bibitem[{{Crampton} \& {Cowley}(1977)}]{cc77}
{Crampton}, D., \& {Cowley}, A.~P. 1977, \pasp, 89, 374, \dodoi{10.1086/130135}

\bibitem[{{{\c{S}}ener} \& {Jeffery}(2014)}]{sj14}
{{\c{S}}ener}, H.~T., \& {Jeffery}, C.~S. 2014, \mnras, 440, 2676,
  \dodoi{10.1093/mnras/stu397}

\bibitem[{{Cui} {et~al.}(2020){Cui}, {Guo}, {Gao}, {Ren}, {Zhang}, {Zhou}, \&
  {Liu}}]{cgg+20}
{Cui}, K., {Guo}, Z., {Gao}, Q., {et~al.} 2020, \apj, 898, 136,
  \dodoi{10.3847/1538-4357/ab9b85}

\bibitem[{{Dai} {et~al.}(2017){Dai}, {Smith}, {Wang}, {Okamoto}, {Xu}, {Yue},
  \& {Liu}}]{dsw+17}
{Dai}, S., {Smith}, M.~C., {Wang}, S., {et~al.} 2017, \apj, 842, 105,
  \dodoi{10.3847/1538-4357/aa7209}

\bibitem[{{Dai} \& {Qian}(2010)}]{dq10}
{Dai}, Z., \& {Qian}, S. 2010, \na, 15, 380,
  \dodoi{10.1016/j.newast.2009.11.003}

\bibitem[{{Danziger} {et~al.}(1993){Danziger}, {Baade}, \& {della
  Valle}}]{dbd93}
{Danziger}, I.~J., {Baade}, D., \& {della Valle}, M. 1993, \aap, 276, 382

\bibitem[{{de Kool}(1990)}]{dek90}
{de Kool}, M. 1990, \apj, 358, 189, \dodoi{10.1086/168974}

\bibitem[{{De Marco} {et~al.}(2008){De Marco}, {Hillwig}, \& {Smith}}]{dhs08}
{De Marco}, O., {Hillwig}, T.~C., \& {Smith}, A.~J. 2008, \aj, 136, 323,
  \dodoi{10.1088/0004-6256/136/1/323}

\bibitem[{{De Marco} {et~al.}(2011){De Marco}, {Passy}, {Moe}, {Herwig}, {Mac
  Low}, \& {Paxton}}]{dpm+11}
{De Marco}, O., {Passy}, J.-C., {Moe}, M., {et~al.} 2011, \mnras, 411, 2277,
  \dodoi{10.1111/j.1365-2966.2010.17891.x}

\bibitem[{{de Martino} {et~al.}(2006){de Martino}, {Bonnet-Bidaud}, {Mouchet},
  {G{\"a}nsicke}, {Haberl}, \& {Motch}}]{dbm+06}
{de Martino}, D., {Bonnet-Bidaud}, J.~M., {Mouchet}, M., {et~al.} 2006, \aap,
  449, 1151, \dodoi{10.1051/0004-6361:20053877}

\bibitem[{{Debes} {et~al.}(2015){Debes}, {Kilic}, {Tremblay},
  {L{\'o}pez-Morales}, {Anglada-Escude}, {Napiwotzki}, {Osip}, \&
  {Weinberger}}]{dkt+15}
{Debes}, J.~H., {Kilic}, M., {Tremblay}, P.-E., {et~al.} 2015, \aj, 149, 176,
  \dodoi{10.1088/0004-6256/149/5/176}

\bibitem[{{Delfosse} {et~al.}(1999){Delfosse}, {Forveille}, {Beuzit}, {Udry},
  {Mayor}, \& {Perrier}}]{dfb+99}
{Delfosse}, X., {Forveille}, T., {Beuzit}, J.~L., {et~al.} 1999, \aap, 344,
  897.
\newblock \doarXiv{astro-ph/9812008}

\bibitem[{{Deller} {et~al.}(2016){Deller}, {Vigeland}, {Kaplan}, {Goss},
  {Brisken}, {Chatterjee}, {Cordes}, {Janssen}, {Lazio}, {Petrov}, {Stappers},
  \& {Lyne}}]{dvk+16}
{Deller}, A.~T., {Vigeland}, S.~J., {Kaplan}, D.~L., {et~al.} 2016, \apj, 828,
  8, \dodoi{10.3847/0004-637X/828/1/8}

\bibitem[{{Derekas} {et~al.}(2015){Derekas}, {N{\'e}meth}, {Southworth},
  {Borkovits}, {S{\'a}rneczky}, {P{\'a}l}, {Cs{\'a}k}, {Garcia-Alvarez},
  {Maxted}, {Kiss}, {Vida}, {Szab{\'o}}, \& {Kriskovics}}]{dns+15}
{Derekas}, A., {N{\'e}meth}, P., {Southworth}, J., {et~al.} 2015, \apj, 808,
  179, \dodoi{10.1088/0004-637X/808/2/179}

\bibitem[{{Desvignes} {et~al.}(2016){Desvignes}, {Caballero}, {Lentati},
  {Verbiest}, {Champion}, {Stappers}, {Janssen}, {Lazarus}, {Os{\l}owski},
  {Babak}, {Bassa}, {Brem}, {Burgay}, {Cognard}, {Gair}, {Graikou},
  {Guillemot}, {Hessels}, {Jessner}, {Jordan}, {Karuppusamy}, {Kramer},
  {Lassus}, {Lazaridis}, {Lee}, {Liu}, {Lyne}, {McKee}, {Mingarelli},
  {Perrodin}, {Petiteau}, {Possenti}, {Purver}, {Rosado}, {Sanidas}, {Sesana},
  {Shaifullah}, {Smits}, {Taylor}, {Theureau}, {Tiburzi}, {van Haasteren}, \&
  {Vecchio}}]{dcl+16}
{Desvignes}, G., {Caballero}, R.~N., {Lentati}, L., {et~al.} 2016, \mnras, 458,
  3341, \dodoi{10.1093/mnras/stw483}

\bibitem[{{Dhillon} {et~al.}(1991){Dhillon}, {Marsh}, \& {Jones}}]{dmj91}
{Dhillon}, V.~S., {Marsh}, T.~R., \& {Jones}, D.~H.~P. 1991, \mnras, 252, 342,
  \dodoi{10.1093/mnras/252.3.342}

\bibitem[{{Di Stefano}(2019)}]{dis19}
{Di Stefano}, R. 2019, in American Astronomical Society Meeting Abstracts, Vol.
  233, American Astronomical Society Meeting Abstracts \#233, 414.05

\bibitem[{{Di Stefano}(2020)}]{dis20}
{Di Stefano}, R. 2020, \mnras, 491, 495, \dodoi{10.1093/mnras/stz2572}

\bibitem[{{Diaz} \& {Ribeiro}(2003)}]{dr03}
{Diaz}, M.~P., \& {Ribeiro}, F.~M.~A. 2003, \aj, 125, 3359,
  \dodoi{10.1086/375328}

\bibitem[{{Dominik} {et~al.}(2012){Dominik}, {Belczynski}, {Fryer}, {Holz},
  {Berti}, {Bulik}, {Mandel}, \& {O'Shaughnessy}}]{dbf+12}
{Dominik}, M., {Belczynski}, K., {Fryer}, C., {et~al.} 2012, \apj, 759, 52,
  \dodoi{10.1088/0004-637X/759/1/52}

\bibitem[{{Drake} {et~al.}(2009){Drake}, {Djorgovski}, {Mahabal}, {Beshore},
  {Larson}, {Graham}, {Williams}, {Christensen}, {Catelan}, {Boattini},
  {Gibbs}, {Hill}, \& {Kowalski}}]{ddm+09}
{Drake}, A.~J., {Djorgovski}, S.~G., {Mahabal}, A., {et~al.} 2009, \apj, 696,
  870, \dodoi{10.1088/0004-637X/696/1/870}

\bibitem[{{Drake} {et~al.}(2014){Drake}, {Djorgovski},
  {Garc{\'\i}a-{\'A}lvarez}, {Graham}, {Catelan}, {Mahabal}, {Donalek},
  {Prieto}, {Torrealba}, {Abraham}, {Williams}, {Larson}, \&
  {Christensen}}]{ddg+14}
{Drake}, A.~J., {Djorgovski}, S.~G., {Garc{\'\i}a-{\'A}lvarez}, D., {et~al.}
  2014, \apj, 790, 157, \dodoi{10.1088/0004-637X/790/2/157}

\bibitem[{{Drechsel} {et~al.}(2001){Drechsel}, {Heber}, {Napiwotzki},
  {{\O}stensen}, {Solheim}, {Johannessen}, {Schuh}, {Deetjen}, \&
  {Zola}}]{dhn+01}
{Drechsel}, H., {Heber}, U., {Napiwotzki}, R., {et~al.} 2001, \aap, 379, 893,
  \dodoi{10.1051/0004-6361:20011376}

\bibitem[{{Dunford} {et~al.}(2012){Dunford}, {Watson}, \& {Smith}}]{dws12}
{Dunford}, A., {Watson}, C.~A., \& {Smith}, R.~C. 2012, \mnras, 422, 3444,
  \dodoi{10.1111/j.1365-2966.2012.20854.x}

\bibitem[{{Echevarr{\'\i}a} {et~al.}(2007{\natexlab{a}}){Echevarr{\'\i}a}, {de
  la Fuente}, \& {Costero}}]{edc07}
{Echevarr{\'\i}a}, J., {de la Fuente}, E., \& {Costero}, R. 2007{\natexlab{a}},
  \aj, 134, 262, \dodoi{10.1086/518562}

\bibitem[{{Echevarr{\'\i}a} {et~al.}(2007{\natexlab{b}}){Echevarr{\'\i}a},
  {Michel}, {Costero}, \& {Zharikov}}]{emc+07}
{Echevarr{\'\i}a}, J., {Michel}, R., {Costero}, R., \& {Zharikov}, S.
  2007{\natexlab{b}}, \aap, 462, 1069, \dodoi{10.1051/0004-6361:20052906}

\bibitem[{{Echevarr{\'\i}a} {et~al.}(2016){Echevarr{\'\i}a},
  {Ram{\'\i}rez-Torres}, {Michel}, \& {Hern{\'a}ndez Santisteban}}]{erm+16}
{Echevarr{\'\i}a}, J., {Ram{\'\i}rez-Torres}, A., {Michel}, R., \&
  {Hern{\'a}ndez Santisteban}, J.~V. 2016, \mnras, 461, 1576,
  \dodoi{10.1093/mnras/stw1425}

\bibitem[{{Echevarr{\'\i}a} {et~al.}(2008){Echevarr{\'\i}a}, {Smith},
  {Costero}, {Zharikov}, \& {Michel}}]{esc+08}
{Echevarr{\'\i}a}, J., {Smith}, R.~C., {Costero}, R., {Zharikov}, S., \&
  {Michel}, R. 2008, \mnras, 387, 1563,
  \dodoi{10.1111/j.1365-2966.2008.13248.x}

\bibitem[{{Echevarr{\'\i}a} {et~al.}(2019){Echevarr{\'\i}a}, {de Miguel},
  {Hern{\'a}ndez Santisteban}, {Costero}, {Michel}, {S{\'a}nchez}, {Olivares},
  {Ruelas-Mayorga}, {Gonz{\'a}lez-Buitrago}, {Jones}, {Oskanen}, {Goff},
  {Ulowetz}, {Bolt}, {Sabo}, {Hambsch}, {Slauson}, \& {Stein}}]{edh+19}
{Echevarr{\'\i}a}, J., {de Miguel}, E., {Hern{\'a}ndez Santisteban}, J.~V.,
  {et~al.} 2019, \rmxaa, 55, 21.
\newblock \doarXiv{1810.09864}

\bibitem[{{Edelmann}(2008)}]{ede08}
{Edelmann}, H. 2008, in Astronomical Society of the Pacific Conference Series,
  Vol. 392, Hot Subdwarf Stars and Related Objects, ed. U.~{Heber}, C.~S.
  {Jeffery}, \& R.~{Napiwotzki}, 187.
\newblock \doarXiv{0809.3260}

\bibitem[{{Edelmann} {et~al.}(2006){Edelmann}, {Altmann}, \& {Heber}}]{eah06}
{Edelmann}, H., {Altmann}, M., \& {Heber}, U. 2006, Baltic Astronomy, 15, 191

\bibitem[{{Edelmann} {et~al.}(2005){Edelmann}, {Heber}, {Altmann}, {Karl}, \&
  {Lisker}}]{eha+05}
{Edelmann}, H., {Heber}, U., {Altmann}, M., {Karl}, C., \& {Lisker}, T. 2005,
  \aap, 442, 1023, \dodoi{10.1051/0004-6361:20053267}

\bibitem[{{Edelmann} {et~al.}(2004){Edelmann}, {Heber}, {Lisker}, \&
  {Green}}]{ehl+04}
{Edelmann}, H., {Heber}, U., {Lisker}, T., \& {Green}, E.~M. 2004, \apss, 291,
  315, \dodoi{10.1023/B:ASTR.0000044338.27638.9f}

\bibitem[{{Edwards} \& {Bailes}(2001)}]{eb01}
{Edwards}, R.~T., \& {Bailes}, M. 2001, \apjl, 547, L37, \dodoi{10.1086/318893}

\bibitem[{{Eggleton} {et~al.}(2006){Eggleton}, {King}, {Lin}, {Maran},
  {Pringle}, \& {Ward}}]{e2006}
{Eggleton}, P., {King}, A., {Lin}, D., {et~al.} 2006, {Evolutionary Processes
  in Binary and Multiple Stars} (Cambridge: Cambridge Univ. Press)

\bibitem[{{Eggleton}(1983)}]{egg83}
{Eggleton}, P.~P. 1983, \apj, 268, 368, \dodoi{10.1086/160960}

\bibitem[{{El-Badry} {et~al.}(2021){El-Badry}, {Quataert}, {Rix}, {Weisz},
  {Kupfer}, {Shen}, {Xiang}, {Yang}, \& {Liu}}]{eqr+21}
{El-Badry}, K., {Quataert}, E., {Rix}, H.-W., {et~al.} 2021, arXiv e-prints,
  arXiv:2104.07033.
\newblock \doarXiv{2104.07033}

\bibitem[{{Eldridge} \& {Stanway}(2016)}]{es16}
{Eldridge}, J.~J., \& {Stanway}, E.~R. 2016, \mnras, 462, 3302,
  \dodoi{10.1093/mnras/stw1772}

\bibitem[{{Esmer} {et~al.}(2021){Esmer}, {Ba{\c{s}}t{\"u}rk}, {Hinse}, {Selam},
  \& {Correia}}]{ebo+21}
{Esmer}, E.~M., {Ba{\c{s}}t{\"u}rk}, {\"O}., {Hinse}, T.~C., {Selam}, S.~O., \&
  {Correia}, A. C.~M. 2021, \aap, 648, A85, \dodoi{10.1051/0004-6361/202038640}

\bibitem[{{Espaillat} {et~al.}(2005){Espaillat}, {Patterson}, {Warner}, \&
  {Woudt}}]{epw+05}
{Espaillat}, C., {Patterson}, J., {Warner}, B., \& {Woudt}, P. 2005, \pasp,
  117, 189, \dodoi{10.1086/427959}

\bibitem[{{Fabrycky} \& {Tremaine}(2007)}]{ft07}
{Fabrycky}, D., \& {Tremaine}, S. 2007, \apj, 669, 1298, \dodoi{10.1086/521702}

\bibitem[{{Faigler} {et~al.}(2015){Faigler}, {Kull}, {Mazeh}, {Kiefer},
  {Latham}, \& {Bloemen}}]{fkm+15}
{Faigler}, S., {Kull}, I., {Mazeh}, T., {et~al.} 2015, \apj, 815, 26,
  \dodoi{10.1088/0004-637X/815/1/26}

\bibitem[{{Farihi} {et~al.}(2008){Farihi}, {Burleigh}, \& {Hoard}}]{fbh08}
{Farihi}, J., {Burleigh}, M.~R., \& {Hoard}, D.~W. 2008, \apj, 674, 421,
  \dodoi{10.1086/524933}

\bibitem[{{Feline} {et~al.}(2004{\natexlab{a}}){Feline}, {Dhillon}, {Marsh}, \&
  {Brinkworth}}]{fdmb04}
{Feline}, W.~J., {Dhillon}, V.~S., {Marsh}, T.~R., \& {Brinkworth}, C.~S.
  2004{\natexlab{a}}, \mnras, 355, 1, \dodoi{10.1111/j.1365-2966.2004.08302.x}

\bibitem[{{Feline} {et~al.}(2004{\natexlab{b}}){Feline}, {Dhillon}, {Marsh},
  {Stevenson}, {Watson}, \& {Brinkworth}}]{fdm+04}
{Feline}, W.~J., {Dhillon}, V.~S., {Marsh}, T.~R., {et~al.} 2004{\natexlab{b}},
  \mnras, 347, 1173, \dodoi{10.1111/j.1365-2966.2004.07281.x}

\bibitem[{{Ferdman} {et~al.}(2010){Ferdman}, {Stairs}, {Kramer}, {McLaughlin},
  {Lorimer}, {Nice}, {Manchester}, {Hobbs}, {Lyne}, {Camilo}, {Possenti},
  {Demorest}, {Cognard}, {Desvignes}, {Theureau}, {Faulkner}, \&
  {Backer}}]{fsk+10}
{Ferdman}, R.~D., {Stairs}, I.~H., {Kramer}, M., {et~al.} 2010, \apj, 711, 764,
  \dodoi{10.1088/0004-637X/711/2/764}

\bibitem[{{Ferguson} {et~al.}(1999){Ferguson}, {Liebert}, {Haas}, {Napiwotzki},
  \& {James}}]{flh+99}
{Ferguson}, D.~H., {Liebert}, J., {Haas}, S., {Napiwotzki}, R., \& {James},
  T.~A. 1999, \apj, 518, 866, \dodoi{10.1086/307289}

\bibitem[{{Fink} {et~al.}(2007){Fink}, {Hillebrandt}, \& {R{\"o}pke}}]{FHR2007}
{Fink}, M., {Hillebrandt}, W., \& {R{\"o}pke}, F.~K. 2007, \aap, 476, 1133,
  \dodoi{10.1051/0004-6361:20078438}

\bibitem[{{Fink} {et~al.}(2010){Fink}, {R{\"o}pke}, {Hillebrandt},
  {Seitenzahl}, {Sim}, \& {Kromer}}]{FRH2010}
{Fink}, M., {R{\"o}pke}, F.~K., {Hillebrandt}, W., {et~al.} 2010, \aap, 514,
  A53, \dodoi{10.1051/0004-6361/200913892}

\bibitem[{{Fleming} {et~al.}(2019){Fleming}, {Barnes}, {Davenport}, \&
  {Luger}}]{fb+2019}
{Fleming}, D.~P., {Barnes}, R., {Davenport}, J. R.~A., \& {Luger}, R. 2019,
  \apj, 881, 88, \dodoi{10.3847/1538-4357/ab2ed2}

\bibitem[{{Fonseca} {et~al.}(2016){Fonseca}, {Pennucci}, {Ellis}, {Stairs},
  {Nice}, {Ransom}, {Demorest}, {Arzoumanian}, {Crowter}, {Dolch}, {Ferdman},
  {Gonzalez}, {Jones}, {Jones}, {Lam}, {Levin}, {McLaughlin}, {Stovall},
  {Swiggum}, \& {Zhu}}]{fpe+16}
{Fonseca}, E., {Pennucci}, T.~T., {Ellis}, J.~A., {et~al.} 2016, \apj, 832,
  167, \dodoi{10.3847/0004-637X/832/2/167}

\bibitem[{{Fonseca} {et~al.}(2021){Fonseca}, {Cromartie}, {Pennucci}, {Ray},
  {Kirichenko}, {Ransom}, {Demorest}, {Stairs}, {Arzoumanian}, {Guillemot},
  {Parthasarathy}, {Kerr}, {Cognard}, {Baker}, {Blumer}, {Brook}, {DeCesar},
  {Dolch}, {Dong}, {Ferrara}, {Fiore}, {Garver-Daniels}, {Good}, {Jennings},
  {Jones}, {Kaspi}, {Lam}, {Lorimer}, {Luo}, {McEwen}, {McKee}, {McLaughlin},
  {McMann}, {Meyers}, {Naidu}, {Ng}, {Nice}, {Pol}, {Radovan},
  {Shapiro-Albert}, {Tan}, {Tendulkar}, {Swiggum}, {Wahl}, \& {Zhu}}]{fcp+21}
{Fonseca}, E., {Cromartie}, H.~T., {Pennucci}, T.~T., {et~al.} 2021, arXiv
  e-prints, arXiv:2104.00880.
\newblock \doarXiv{2104.00880}

\bibitem[{{Fontaine} {et~al.}(2011){Fontaine}, {Brassard}, {Green},
  {Charpinet}, {Dufour}, {Hubeny}, {Steeghs}, {Aerts}, {Randall}, {Bergeron},
  {Guvenen}, {O'Malley}, {Van Grootel}, {{\O}stensen}, {Bloemen}, {Silvotti},
  {Howell}, {Baran}, {Kepler}, {Marsh}, {Montgomery}, {Oreiro}, {Provencal},
  {Telting}, {Winget}, {Zima}, {Christensen-Dalsgaard}, \& {Kjeldsen}}]{fbg+11}
{Fontaine}, G., {Brassard}, P., {Green}, E.~M., {et~al.} 2011, \apj, 726, 92,
  \dodoi{10.1088/0004-637X/726/2/92}

\bibitem[{{For} {et~al.}(2008){For}, {Edelmann}, {Green}, {Drechsel},
  {Nesslinger}, \& {Fontaine}}]{feg+08}
{For}, B.~Q., {Edelmann}, H., {Green}, E.~M., {et~al.} 2008, in Astronomical
  Society of the Pacific Conference Series, Vol. 392, Hot Subdwarf Stars and
  Related Objects, ed. U.~{Heber}, C.~S. {Jeffery}, \& R.~{Napiwotzki}, 203.
\newblock \doarXiv{0809.4517}

\bibitem[{{For} {et~al.}(2010){For}, {Green}, {Fontaine}, \& {Shaw}}]{fgf+10}
{For}, B.-Q., {Green}, E.~M., {Fontaine}, G., \& {Shaw}, S. 2010, \apss, 329,
  87, \dodoi{10.1007/s10509-010-0280-7}

\bibitem[{{Friend} {et~al.}(1990{\natexlab{a}}){Friend}, {Martin}, {Smith}, \&
  {Jones}}]{fms+90b}
{Friend}, M.~T., {Martin}, J.~S., {Smith}, R.~C., \& {Jones}, D.~H.~P.
  1990{\natexlab{a}}, \mnras, 246, 637

\bibitem[{{Friend} {et~al.}(1990{\natexlab{b}}){Friend}, {Martin}, {Smith}, \&
  {Jones}}]{fms+90c}
---. 1990{\natexlab{b}}, \mnras, 246, 654

\bibitem[{{Fuller} {et~al.}(2013){Fuller}, {Derekas}, {Borkovits}, {Huber},
  {Bedding}, \& {Kiss}}]{fdb+13}
{Fuller}, J., {Derekas}, A., {Borkovits}, T., {et~al.} 2013, \mnras, 429, 2425,
  \dodoi{10.1093/mnras/sts511}

\bibitem[{{Gaia Collaboration} {et~al.}(2020){Gaia Collaboration}, {Brown},
  {Vallenari}, {Prusti}, {de Bruijne}, {Babusiaux}, \& {Biermann}}]{gaia2020}
{Gaia Collaboration}, {Brown}, A.~G.~A., {Vallenari}, A., {et~al.} 2020, arXiv
  e-prints, arXiv:2012.01533, \dodoi{10.1051/0004-6361/202039657}

\bibitem[{{G{\"a}nsicke} {et~al.}(2004){G{\"a}nsicke}, {Araujo-Betancor},
  {Hagen}, {Harlaftis}, {Kitsionas}, {Dreizler}, \& {Engels}}]{gah+04}
{G{\"a}nsicke}, B.~T., {Araujo-Betancor}, S., {Hagen}, H.~J., {et~al.} 2004,
  \aap, 418, 265, \dodoi{10.1051/0004-6361:20035591}

\bibitem[{{G{\"a}nsicke} {et~al.}(2006){G{\"a}nsicke}, {Long}, {Barstow}, \&
  {Hubeny}}]{glb+06}
{G{\"a}nsicke}, B.~T., {Long}, K.~S., {Barstow}, M.~A., \& {Hubeny}, I. 2006,
  \apj, 639, 1039, \dodoi{10.1086/499358}

\bibitem[{{Gao} {et~al.}(2018){Gao}, {Correia}, {Eggleton}, \& {Han}}]{gce+18}
{Gao}, Y., {Correia}, A. C.~M., {Eggleton}, P.~P., \& {Han}, Z. 2018, \mnras,
  479, 3604, \dodoi{10.1093/mnras/sty1558}

\bibitem[{{Gao} {et~al.}(2020){Gao}, {Toonen}, {Grishin}, {Comerford}, \&
  {Kruckow}}]{gtg+20}
{Gao}, Y., {Toonen}, S., {Grishin}, E., {Comerford}, T., \& {Kruckow}, M.~U.
  2020, \mnras, 491, 264, \dodoi{10.1093/mnras/stz3035}

\bibitem[{{Geier} {et~al.}(2009){Geier}, {Edelmann}, {Heber}, \&
  {Morales-Rueda}}]{geh+09}
{Geier}, S., {Edelmann}, H., {Heber}, U., \& {Morales-Rueda}, L. 2009, \apjl,
  702, L96, \dodoi{10.1088/0004-637X/702/1/L96}

\bibitem[{{Geier} {et~al.}(2013{\natexlab{a}}){Geier}, {Heber}, {Edelmann},
  {Morales-Rueda}, {Kilkenny}, {O'Donoghue}, {Marsh}, \&
  {Copperwheat}}]{ghe+13}
{Geier}, S., {Heber}, U., {Edelmann}, H., {et~al.} 2013{\natexlab{a}}, \aap,
  557, A122, \dodoi{10.1051/0004-6361/201322057}

\bibitem[{{Geier} {et~al.}(2010{\natexlab{a}}){Geier}, {Heber}, {Kupfer}, \&
  {Napiwotzki}}]{ghk+10}
{Geier}, S., {Heber}, U., {Kupfer}, T., \& {Napiwotzki}, R. 2010{\natexlab{a}},
  \aap, 515, A37, \dodoi{10.1051/0004-6361/200912545}

\bibitem[{{Geier} {et~al.}(2010{\natexlab{b}}){Geier}, {Heber},
  {Podsiadlowski}, {Edelmann}, {Napiwotzki}, {Kupfer}, \&
  {M{\"u}ller}}]{ghp+10}
{Geier}, S., {Heber}, U., {Podsiadlowski}, P., {et~al.} 2010{\natexlab{b}},
  \aap, 519, A25, \dodoi{10.1051/0004-6361/201014465}

\bibitem[{{Geier} {et~al.}(2008){Geier}, {Karl}, {Edelmann}, {Heber}, \&
  {Napiwotzki}}]{gke+08}
{Geier}, S., {Karl}, C., {Edelmann}, H., {Heber}, U., \& {Napiwotzki}, R. 2008,
  \memsai, 79, 608.
\newblock \doarXiv{0804.1282}

\bibitem[{{Geier} {et~al.}(2011{\natexlab{a}}){Geier}, {Napiwotzki}, {Heber},
  \& {Nelemans}}]{gnh+11}
{Geier}, S., {Napiwotzki}, R., {Heber}, U., \& {Nelemans}, G.
  2011{\natexlab{a}}, \aap, 528, L16, \dodoi{10.1051/0004-6361/201116641}

\bibitem[{{Geier} {et~al.}(2007){Geier}, {Nesslinger}, {Heber}, {Przybilla},
  {Napiwotzki}, \& {Kudritzki}}]{gnh+07}
{Geier}, S., {Nesslinger}, S., {Heber}, U., {et~al.} 2007, \aap, 464, 299,
  \dodoi{10.1051/0004-6361:20066098}

\bibitem[{{Geier} {et~al.}(2011{\natexlab{b}}){Geier}, {Maxted}, {Napiwotzki},
  {{\O}stensen}, {Heber}, {Hirsch}, {Kupfer}, {M{\"u}ller}, {Tillich},
  {Barlow}, {Oreiro}, {Ottosen}, {Copperwheat}, {G{\"a}nsicke}, \&
  {Marsh}}]{gmn+11}
{Geier}, S., {Maxted}, P.~F.~L., {Napiwotzki}, R., {et~al.} 2011{\natexlab{b}},
  \aap, 526, A39, \dodoi{10.1051/0004-6361/201015794}

\bibitem[{{Geier} {et~al.}(2012){Geier}, {Classen}, {Br{\"u}nner}, {Nagel},
  {Schaffenroth}, {Heuser}, {Heber}, {Drechsel}, {Edelmann}, {Koen}, {O'Toole},
  \& {Morales-Rueda}}]{gcb+12}
{Geier}, S., {Classen}, L., {Br{\"u}nner}, P., {et~al.} 2012, in Astronomical
  Society of the Pacific Conference Series, Vol. 452, Fifth Meeting on Hot
  Subdwarf Stars and Related Objects, ed. D.~{Kilkenny}, C.~S. {Jeffery}, \&
  C.~{Koen}, 153.
\newblock \doarXiv{1112.2929}

\bibitem[{{Geier} {et~al.}(2013{\natexlab{b}}){Geier}, {Marsh}, {Wang},
  {Dunlap}, {Barlow}, {Schaffenroth}, {Chen}, {Irrgang}, {Maxted}, {Ziegerer},
  {Kupfer}, {Miszalski}, {Heber}, {Han}, {Shporer}, {Telting}, {G{\"a}nsicke},
  {{\O}stensen}, {O'Toole}, \& {Napiwotzki}}]{gmw+13}
{Geier}, S., {Marsh}, T.~R., {Wang}, B., {et~al.} 2013{\natexlab{b}}, \aap,
  554, A54, \dodoi{10.1051/0004-6361/201321395}

\bibitem[{{Geier} {et~al.}(2014){Geier}, {{\O}stensen}, {Heber}, {Kupfer},
  {Maxted}, {Barlow}, {Vu{\v{c}}kovi{\'c}}, {Tillich}, {M{\"u}ller},
  {Edelmann}, {Classen}, \& {McLeod}}]{goh+14}
{Geier}, S., {{\O}stensen}, R.~H., {Heber}, U., {et~al.} 2014, \aap, 562, A95,
  \dodoi{10.1051/0004-6361/201323115}

\bibitem[{{Gianninas} {et~al.}(2014){Gianninas}, {Dufour}, {Kilic}, {Brown},
  {Bergeron}, \& {Hermes}}]{gdk+14}
{Gianninas}, A., {Dufour}, P., {Kilic}, M., {et~al.} 2014, \apj, 794, 35,
  \dodoi{10.1088/0004-637X/794/1/35}

\bibitem[{{Gies} {et~al.}(2008){Gies}, {Dieterich}, {Richardson}, {Riedel},
  {B.~L. Team}, {McAlister}, {Bagnuolo}, {Grundstrom}, {{\v{S}}tefl},
  {Rivinius}, \& {Baade}}]{gdr+08}
{Gies}, D.~R., {Dieterich}, S., {Richardson}, N.~D., {et~al.} 2008, \apjl, 682,
  L117, \dodoi{10.1086/591148}

\bibitem[{{Gizis}(1998)}]{giz98}
{Gizis}, J.~E. 1998, \aj, 115, 2053, \dodoi{10.1086/300325}

\bibitem[{{Glanz} \& {Perets}(2021)}]{gp21}
{Glanz}, H., \& {Perets}, H.~B. 2021, \mnras, 500, 1921,
  \dodoi{10.1093/mnras/staa3242}

\bibitem[{{Godon} \& {Sion}(2021)}]{gs21}
{Godon}, P., \& {Sion}, E.~M. 2021, \apj, 908, 173,
  \dodoi{10.3847/1538-4357/abda47}

\bibitem[{{Green} {et~al.}(2005){Green}, {For}, \& {Hyde}}]{gfh05}
{Green}, E.~M., {For}, B.~Q., \& {Hyde}, E.~A. 2005, in Astronomical Society of
  the Pacific Conference Series, Vol. 334, 14th European Workshop on White
  Dwarfs, ed. D.~{Koester} \& S.~{Moehler}, 363

\bibitem[{{Green} {et~al.}(2004){Green}, {For}, {Hyde}, {Seitenzahl},
  {Callerame}, {White}, {Young}, {Huff}, {Mills}, \& {Steinfadt}}]{gfh+04}
{Green}, E.~M., {For}, B., {Hyde}, E.~A., {et~al.} 2004, \apss, 291, 267,
  \dodoi{10.1023/B:ASTR.0000044332.46251.5f}

\bibitem[{{Green} {et~al.}(2018{\natexlab{a}}){Green}, {Hermes}, {Marsh},
  {Steeghs}, {Bell}, {Littlefair}, {Parsons}, {Dennihy}, {Fuchs}, {Reding},
  {Kaiser}, {Ashley}, {Breedt}, {Dhillon}, {Gentile Fusillo}, {Kerry}, \&
  {Sahman}}]{ghm+18}
{Green}, M.~J., {Hermes}, J.~J., {Marsh}, T.~R., {et~al.} 2018{\natexlab{a}},
  \mnras, 477, 5646, \dodoi{10.1093/mnras/sty1032}

\bibitem[{{Green} {et~al.}(2018{\natexlab{b}}){Green}, {Marsh}, {Steeghs},
  {Kupfer}, {Ashley}, {Bloemen}, {Breedt}, {Campbell}, {Chakpor},
  {Copperwheat}, {Dhillon}, {Hallinan}, {Hardy}, {Hermes}, {Kerry},
  {Littlefair}, {Milburn}, {Parsons}, {Prasert}, {van Roestel}, {Sahman}, \&
  {Singh}}]{gms+18}
{Green}, M.~J., {Marsh}, T.~R., {Steeghs}, D.~T.~H., {et~al.}
  2018{\natexlab{b}}, \mnras, 476, 1663, \dodoi{10.1093/mnras/sty299}

\bibitem[{{Guo} {et~al.}(2015){Guo}, {Zhao}, {Tziamtzis}, {Liu}, {Li}, {Zhang},
  {Hou}, \& {Wang}}]{gzt+15}
{Guo}, J., {Zhao}, J., {Tziamtzis}, A., {et~al.} 2015, \mnras, 454, 2787,
  \dodoi{10.1093/mnras/stv2104}

\bibitem[{{Guo} {et~al.}(2017){Guo}, {Gies}, {Matson}, {Garc{\'\i}a
  Hern{\'a}ndez}, {Han}, \& {Chen}}]{ggm+17}
{Guo}, Z., {Gies}, D.~R., {Matson}, R.~A., {et~al.} 2017, \apj, 837, 114,
  \dodoi{10.3847/1538-4357/aa61a4}

\bibitem[{{G{\"u}ver} {et~al.}(2010){G{\"u}ver}, {Wroblewski}, {Camarota}, \&
  {{\"O}zel}}]{gwc+10}
{G{\"u}ver}, T., {Wroblewski}, P., {Camarota}, L., \& {{\"O}zel}, F. 2010,
  \apj, 719, 1807, \dodoi{10.1088/0004-637X/719/2/1807}

\bibitem[{{Hallakoun} {et~al.}(2016){Hallakoun}, {Maoz}, {Kilic}, {Mazeh},
  {Gianninas}, {Agol}, {Bell}, {Bloemen}, {Brown}, {Debes}, {Faigler}, {Kull},
  {Kupfer}, {Loeb}, {Morris}, \& {Mullally}}]{hmk+16}
{Hallakoun}, N., {Maoz}, D., {Kilic}, M., {et~al.} 2016, \mnras, 458, 845,
  \dodoi{10.1093/mnras/stw364}

\bibitem[{{Hamilton} \& {Sion}(2004)}]{hs04}
{Hamilton}, R.~T., \& {Sion}, E.~M. 2004, \pasp, 116, 926,
  \dodoi{10.1086/425082}

\bibitem[{{Han} {et~al.}(1995){Han}, {Podsiadlowski}, \& {Eggleton}}]{hpe95}
{Han}, Z., {Podsiadlowski}, P., \& {Eggleton}, P.~P. 1995, \mnras, 272, 800,
  \dodoi{10.1093/mnras/272.4.800}

\bibitem[{{Haswell} {et~al.}(1997){Haswell}, {Patterson}, {Thorstensen},
  {Hellier}, \& {Skillman}}]{hpt+97}
{Haswell}, C.~A., {Patterson}, J., {Thorstensen}, J.~R., {Hellier}, C., \&
  {Skillman}, D.~R. 1997, \apj, 476, 847, \dodoi{10.1086/303630}

\bibitem[{{Hawkins} {et~al.}(1990){Hawkins}, {Smith}, \& {Jones}}]{hsj90}
{Hawkins}, N.~A., {Smith}, R.~C., \& {Jones}, D.~H.~P. 1990, in
  Accretion-Powered Compact Binaries, ed. C.~W. {Mauche}, 113--116

\bibitem[{{Heber} {et~al.}(2004){Heber}, {Drechsel}, {{\O}stensen}, {Karl},
  {Napiwotzki}, {Altmann}, {Cordes}, {Solheim}, {Voss}, {Koester}, \&
  {Folkes}}]{hdo+04}
{Heber}, U., {Drechsel}, H., {{\O}stensen}, R., {et~al.} 2004, \aap, 420, 251,
  \dodoi{10.1051/0004-6361:20041043}

\bibitem[{{Heller} {et~al.}(2009){Heller}, {Homeier}, {Dreizler}, \&
  {Oestensen}}]{hhd+09}
{Heller}, R., {Homeier}, D., {Dreizler}, S., \& {Oestensen}, R. 2009, VizieR
  Online Data Catalog, J/A+A/496/191

\bibitem[{{Hellier}(1997)}]{hel97}
{Hellier}, C. 1997, \mnras, 291, 71, \dodoi{10.1093/mnras/291.1.71}

\bibitem[{{Hermes} {et~al.}(2012){Hermes}, {Kilic}, {Brown}, {Winget}, {Allende
  Prieto}, {Gianninas}, {Mukadam}, {Cabrera-Lavers}, \& {Kenyon}}]{hkb+12}
{Hermes}, J.~J., {Kilic}, M., {Brown}, W.~R., {et~al.} 2012, \apjl, 757, L21,
  \dodoi{10.1088/2041-8205/757/2/L21}

\bibitem[{{Hermes} {et~al.}(2013){Hermes}, {Montgomery}, {Winget}, {Brown},
  {Gianninas}, {Kilic}, {Kenyon}, {Bell}, \& {Harrold}}]{hmw+13}
{Hermes}, J.~J., {Montgomery}, M.~H., {Winget}, D.~E., {et~al.} 2013, \apj,
  765, 102, \dodoi{10.1088/0004-637X/765/2/102}

\bibitem[{{Hermes} {et~al.}(2014){Hermes}, {Brown}, {Kilic}, {Gianninas},
  {Chote}, {Sullivan}, {Winget}, {Bell}, {Falcon}, {Winget}, {Mason},
  {Harrold}, \& {Montgomery}}]{hbk+14}
{Hermes}, J.~J., {Brown}, W.~R., {Kilic}, M., {et~al.} 2014, \apj, 792, 39,
  \dodoi{10.1088/0004-637X/792/1/39}

\bibitem[{{Hermes} {et~al.}(2015){Hermes}, {G{\"a}nsicke}, {Bischoff-Kim},
  {Kawaler}, {Fuchs}, {Dunlap}, {Clemens}, {Montgomery}, {Chote}, {Barclay},
  {Marsh}, {Gianninas}, {Koester}, {Winget}, {Armstrong}, {Rebassa-Mansergas},
  \& {Schreiber}}]{hgb+15}
{Hermes}, J.~J., {G{\"a}nsicke}, B.~T., {Bischoff-Kim}, A., {et~al.} 2015,
  \mnras, 451, 1701, \dodoi{10.1093/mnras/stv1053}

\bibitem[{{Hernandez} {et~al.}(2017){Hernandez}, {Zharikov}, {Neustroev}, \&
  {Tovmassian}}]{hzn+17}
{Hernandez}, M.~S., {Zharikov}, S., {Neustroev}, V., \& {Tovmassian}, G. 2017,
  \mnras, 470, 1960, \dodoi{10.1093/mnras/stx1341}

\bibitem[{{Hernandez} {et~al.}(2021){Hernandez}, {Schreiber}, {Parsons},
  {G{\"a}nsicke}, {Lagos}, {Raddi}, {Toloza}, {Tovmassian}, {Zorotovic},
  {Irawati}, {Past{\'e}n}, {Rebassa-Mansergas}, {Ren}, {Rittipruk}, \&
  {Tappert}}]{hsp+20}
{Hernandez}, M.~S., {Schreiber}, M.~R., {Parsons}, S.~G., {et~al.} 2021,
  \mnras, 501, 1677, \dodoi{10.1093/mnras/staa3815}

\bibitem[{{Hessman}(1988)}]{hes88}
{Hessman}, F.~V. 1988, \aaps, 72, 515

\bibitem[{{Hilditch} {et~al.}(1996){Hilditch}, {Harries}, \& {Hill}}]{hhh96}
{Hilditch}, R.~W., {Harries}, T.~J., \& {Hill}, G. 1996, \mnras, 279, 1380,
  \dodoi{10.1093/mnras/279.4.1380}

\bibitem[{{Hillebrandt} \& {Niemeyer}(2000)}]{hn2000}
{Hillebrandt}, W., \& {Niemeyer}, J.~C. 2000, \araa, 38, 191,
  \dodoi{10.1146/annurev.astro.38.1.191}

\bibitem[{{Hillman} {et~al.}(2020){Hillman}, {Shara}, {Prialnik}, \&
  {Kovetz}}]{hsp+2020}
{Hillman}, Y., {Shara}, M.~M., {Prialnik}, D., \& {Kovetz}, A. 2020, Nature
  Astronomy, 4, 886, \dodoi{10.1038/s41550-020-1062-y}

\bibitem[{{Hillwig} {et~al.}(2010){Hillwig}, {Bond}, {Af{\textcommabelow s}ar},
  \& {De Marco}}]{hba+10}
{Hillwig}, T.~C., {Bond}, H.~E., {Af{\textcommabelow s}ar}, M., \& {De Marco},
  O. 2010, \aj, 140, 319, \dodoi{10.1088/0004-6256/140/2/319}

\bibitem[{{Hillwig} {et~al.}(2015){Hillwig}, {Frew}, {Louie}, {De Marco},
  {Bond}, {Jones}, \& {Schaub}}]{hfl+15}
{Hillwig}, T.~C., {Frew}, D.~J., {Louie}, M., {et~al.} 2015, \aj, 150, 30,
  \dodoi{10.1088/0004-6256/150/1/30}

\bibitem[{{Hillwig} {et~al.}(2017){Hillwig}, {Frew}, {Reindl}, {Rotter},
  {Webb}, \& {Margheim}}]{hfr+17}
{Hillwig}, T.~C., {Frew}, D.~J., {Reindl}, N., {et~al.} 2017, \aj, 153, 24,
  \dodoi{10.3847/1538-3881/153/1/24}

\bibitem[{{Hillwig} {et~al.}(2016){Hillwig}, {Jones}, {De Marco}, {Bond},
  {Margheim}, \& {Frew}}]{hjd+16}
{Hillwig}, T.~C., {Jones}, D., {De Marco}, O., {et~al.} 2016, \apj, 832, 125,
  \dodoi{10.3847/0004-637X/832/2/125}

\bibitem[{{Hoard} {et~al.}(2004){Hoard}, {Linnell}, {Szkody}, {Fried}, {Sion},
  {Hubeny}, \& {Wolfe}}]{hls+04}
{Hoard}, D.~W., {Linnell}, A.~P., {Szkody}, P., {et~al.} 2004, \apj, 604, 346,
  \dodoi{10.1086/381777}

\bibitem[{{Hogg} {et~al.}(2020){Hogg}, {Casewell}, {Wynn}, {Longstaff},
  {Braker}, {Burleigh}, {Tilbrook}, {Geier}, {Koester}, {Debes}, \&
  {Lodieu}}]{hcw+20}
{Hogg}, M.~A., {Casewell}, S.~L., {Wynn}, G.~A., {et~al.} 2020, \mnras, 498,
  12, \dodoi{10.1093/mnras/staa2233}

\bibitem[{{Holberg} {et~al.}(1995){Holberg}, {Saffer}, {Tweedy}, \&
  {Barstow}}]{hst+95}
{Holberg}, J.~B., {Saffer}, R.~A., {Tweedy}, R.~W., \& {Barstow}, M.~A. 1995,
  \apjl, 452, L133, \dodoi{10.1086/309735}

\bibitem[{{Hong} {et~al.}(2017){Hong}, {Lee}, {Lee}, {Kim}, {Koo}, {Park},
  {Lee}, {Kim}, {Cha}, \& {Lee}}]{hll+17}
{Hong}, K., {Lee}, J.~W., {Lee}, D.-J., {et~al.} 2017, \pasp, 129, 014202,
  \dodoi{10.1088/1538-3873/129/971/014202}

\bibitem[{{Horne} {et~al.}(1986){Horne}, {Wade}, \& {Szkody}}]{hws86}
{Horne}, K., {Wade}, R.~A., \& {Szkody}, P. 1986, \mnras, 219, 791,
  \dodoi{10.1093/mnras/219.4.791}

\bibitem[{{Horne} {et~al.}(1993){Horne}, {Welsh}, \& {Wade}}]{hww93}
{Horne}, K., {Welsh}, W.~F., \& {Wade}, R.~A. 1993, \apj, 410, 357,
  \dodoi{10.1086/172752}

\bibitem[{{Horne} {et~al.}(1991){Horne}, {Wood}, \& {Stiening}}]{hws91}
{Horne}, K., {Wood}, J.~H., \& {Stiening}, R.~F. 1991, \apj, 378, 271,
  \dodoi{10.1086/170426}

\bibitem[{{Howell} {et~al.}(2006){Howell}, {Harrison}, {Campbell}, {Cordova},
  \& {Szkody}}]{hhc+06}
{Howell}, S.~B., {Harrison}, T.~E., {Campbell}, R.~K., {Cordova}, F.~A., \&
  {Szkody}, P. 2006, \aj, 131, 2216, \dodoi{10.1086/500540}

\bibitem[{{Howell} {et~al.}(1993){Howell}, {Schmidt}, {De Young}, {Fried},
  {Schmeer}, \& {Gritz}}]{hsd+93}
{Howell}, S.~B., {Schmidt}, R., {De Young}, J.~A., {et~al.} 1993, \pasp, 105,
  579, \dodoi{10.1086/133198}

\bibitem[{{Hutchings} {et~al.}(1985){Hutchings}, {Cowley}, \&
  {Crampton}}]{hcc85}
{Hutchings}, J.~B., {Cowley}, A.~P., \& {Crampton}, D. 1985, \pasp, 97, 423,
  \dodoi{10.1086/131555}

\bibitem[{{Iaconi} \& {De Marco}(2019)}]{id19}
{Iaconi}, R., \& {De Marco}, O. 2019, \mnras, 490, 2550,
  \dodoi{10.1093/mnras/stz2756}

\bibitem[{{{\.I}bano{\v{g}}lu} {et~al.}(2005){{\.I}bano{\v{g}}lu}, {Evren},
  {Ta{\textcommabelow s}}, \& {{\c{C}}ak{\i}rl{\i}}}]{iet+05}
{{\.I}bano{\v{g}}lu}, C., {Evren}, S., {Ta{\textcommabelow s}}, G., \&
  {{\c{C}}ak{\i}rl{\i}}, {\"O}. 2005, \mnras, 360, 1077,
  \dodoi{10.1111/j.1365-2966.2005.09101.x}

\bibitem[{{Iben} \& {Tutukov}(1984)}]{it1984}
{Iben}, I., J., \& {Tutukov}, A.~V. 1984, \apjs, 54, 335,
  \dodoi{10.1086/190932}

\bibitem[{{Ivanova} {et~al.}(2013{\natexlab{a}}){Ivanova}, {Justham}, {Avendano
  Nandez}, \& {Lombardi}}]{ijn+2013}
{Ivanova}, N., {Justham}, S., {Avendano Nandez}, J.~L., \& {Lombardi}, J.~C.
  2013{\natexlab{a}}, Science, 339, 433, \dodoi{10.1126/science.1225540}

\bibitem[{{Ivanova} {et~al.}(2020){Ivanova}, {Justham}, \& {Ricker}}]{ijr20}
{Ivanova}, N., {Justham}, S., \& {Ricker}, P. 2020, Common Envelope Evolution
  (IOP Publishing), \dodoi{10.1088/2514-3433/abb6f0}

\bibitem[{{Ivanova} {et~al.}(2013{\natexlab{b}}){Ivanova}, {Justham}, {Chen},
  {De Marco}, {Fryer}, {Gaburov}, {Ge}, {Glebbeek}, {Han}, {Li}, {Lu}, {Marsh},
  {Podsiadlowski}, {Potter}, {Soker}, {Taam}, {Tauris}, {van den Heuvel}, \&
  {Webbink}}]{ijc+13}
{Ivanova}, N., {Justham}, S., {Chen}, X., {et~al.} 2013{\natexlab{b}}, \aapr,
  21, 59, \dodoi{10.1007/s00159-013-0059-2}

\bibitem[{{Jacoby} {et~al.}(2007){Jacoby}, {Bailes}, {Ord}, {Knight}, \&
  {Hotan}}]{jbo+07}
{Jacoby}, B.~A., {Bailes}, M., {Ord}, S.~M., {Knight}, H.~S., \& {Hotan}, A.~W.
  2007, \apj, 656, 408, \dodoi{10.1086/509312}

\bibitem[{{Jeffery} \& {Simon}(1997)}]{js97}
{Jeffery}, C.~S., \& {Simon}, T. 1997, \mnras, 286, 487,
  \dodoi{10.1093/mnras/286.2.487}

\bibitem[{{Jones} \& {Lyne}(1988)}]{jl88}
{Jones}, A.~W., \& {Lyne}, A.~G. 1988, \mnras, 232, 473,
  \dodoi{10.1093/mnras/232.3.473}

\bibitem[{{Jones} {et~al.}(2014){Jones}, {Boffin}, {Miszalski}, {Wesson},
  {Corradi}, \& {Tyndall}}]{jbm+14}
{Jones}, D., {Boffin}, H.~M.~J., {Miszalski}, B., {et~al.} 2014, \aap, 562,
  A89, \dodoi{10.1051/0004-6361/201322797}

\bibitem[{{Jones} {et~al.}(2015){Jones}, {Boffin}, {Rodr{\'\i}guez-Gil},
  {Wesson}, {Corradi}, {Miszalski}, \& {Mohamed}}]{jbr+15}
{Jones}, D., {Boffin}, H.~M.~J., {Rodr{\'\i}guez-Gil}, P., {et~al.} 2015, \aap,
  580, A19, \dodoi{10.1051/0004-6361/201425454}

\bibitem[{{Jones} {et~al.}(2019){Jones}, {Boffin}, {Sowicka}, {Miszalski},
  {Rodr{\'\i}guez-Gil}, {Santander-Garc{\'\i}a}, \& {Corradi}}]{jbs+19}
{Jones}, D., {Boffin}, H. M.~J., {Sowicka}, P., {et~al.} 2019, \mnras, 482,
  L75, \dodoi{10.1093/mnrasl/sly142}

\bibitem[{{Kaluzny} {et~al.}(2007){Kaluzny}, {Rucinski}, {Thompson}, {Pych}, \&
  {Krzeminski}}]{krt+07}
{Kaluzny}, J., {Rucinski}, S.~M., {Thompson}, I.~B., {Pych}, W., \&
  {Krzeminski}, W. 2007, \aj, 133, 2457, \dodoi{10.1086/516637}

\bibitem[{{Kaplan} {et~al.}(2014{\natexlab{a}}){Kaplan}, {van Kerkwijk},
  {Koester}, {Stairs}, {Ransom}, {Archibald}, {Hessels}, \& {Boyles}}]{kvk+14}
{Kaplan}, D.~L., {van Kerkwijk}, M.~H., {Koester}, D., {et~al.}
  2014{\natexlab{a}}, \apjl, 783, L23, \dodoi{10.1088/2041-8205/783/1/L23}

\bibitem[{{Kaplan} {et~al.}(2014{\natexlab{b}}){Kaplan}, {Boyles}, {Dunlap},
  {Tendulkar}, {Deller}, {Ransom}, {McLaughlin}, {Lorimer}, \&
  {Stairs}}]{kbd+14}
{Kaplan}, D.~L., {Boyles}, J., {Dunlap}, B.~H., {et~al.} 2014{\natexlab{b}},
  \apj, 789, 119, \dodoi{10.1088/0004-637X/789/2/119}

\bibitem[{{Kaplan} {et~al.}(2014{\natexlab{c}}){Kaplan}, {Marsh}, {Walker},
  {Bildsten}, {Bours}, {Breedt}, {Copperwheat}, {Dhillon}, {Howell},
  {Littlefair}, {Shporer}, \& {Steinfadt}}]{kmw+14}
{Kaplan}, D.~L., {Marsh}, T.~R., {Walker}, A.~N., {et~al.} 2014{\natexlab{c}},
  \apj, 780, 167, \dodoi{10.1088/0004-637X/780/2/167}

\bibitem[{{Karl} {et~al.}(2006){Karl}, {Heber}, {Jeffery}, {Napiwotzki}, \&
  {Geier}}]{khj+06}
{Karl}, C., {Heber}, U., {Jeffery}, S., {Napiwotzki}, R., \& {Geier}, S. 2006,
  Baltic Astronomy, 15, 151.
\newblock \doarXiv{astro-ph/0512388}

\bibitem[{{Karl} {et~al.}(2003){Karl}, {Napiwotzki}, {Nelemans}, {Christlieb},
  {Koester}, {Heber}, \& {Reimers}}]{knn+03}
{Karl}, C.~A., {Napiwotzki}, R., {Nelemans}, G., {et~al.} 2003, \aap, 410, 663,
  \dodoi{10.1051/0004-6361:20031278}

\bibitem[{{Kasian}(2012)}]{kas12}
{Kasian}, L.~E. 2012, PhD thesis, The University of British Columbia

\bibitem[{{Kaspi} {et~al.}(1994){Kaspi}, {Taylor}, \& {Ryba}}]{ktr94}
{Kaspi}, V.~M., {Taylor}, J.~H., \& {Ryba}, M.~F. 1994, \apj, 428, 713,
  \dodoi{10.1086/174280}

\bibitem[{{Kato} {et~al.}(2021){Kato}, {Tampo}, {Kojiguchi}, {Shibata}, {Ito},
  {Isogai}, {Itoh}, {Hambsch}, {Monard}, {Kiyota}, {Vanmunster}, {Sosnovskij},
  {Pavlenko}, {Dubovsky}, {Kudzej}, \& {Medulka}}]{ktk+21}
{Kato}, T., {Tampo}, Y., {Kojiguchi}, N., {et~al.} 2021, arXiv e-prints,
  arXiv:2106.15028.
\newblock \doarXiv{2106.15028}

\bibitem[{{Kawka} {et~al.}(2008){Kawka}, {Vennes}, {Dupuis}, {Chayer}, \&
  {Lanz}}]{kvd+08}
{Kawka}, A., {Vennes}, S., {Dupuis}, J., {Chayer}, P., \& {Lanz}, T. 2008,
  \apj, 675, 1518, \dodoi{10.1086/526411}

\bibitem[{{Kawka} {et~al.}(2002){Kawka}, {Vennes}, {Koch}, \&
  {Williams}}]{kvk+02}
{Kawka}, A., {Vennes}, S., {Koch}, R., \& {Williams}, A. 2002, \aj, 124, 2853,
  \dodoi{10.1086/343836}

\bibitem[{{Kawka} {et~al.}(2015){Kawka}, {Vennes}, {O'Toole}, {N{\'e}meth},
  {Burton}, {Kotze}, \& {Buckley}}]{kvo+15}
{Kawka}, A., {Vennes}, S., {O'Toole}, S., {et~al.} 2015, \mnras, 450, 3514,
  \dodoi{10.1093/mnras/stv821}

\bibitem[{{Kawka} {et~al.}(2012){Kawka}, {Pigulski}, {O'Toole}, {Vennes},
  {N{\'e}meth}, {Williams}, {Iliev}, {Ko{\l}aczkowski}, \&
  {St{\c{e}}{\'s}licki}}]{kpo+12}
{Kawka}, A., {Pigulski}, A., {O'Toole}, S., {et~al.} 2012, in Astronomical
  Society of the Pacific Conference Series, Vol. 452, Fifth Meeting on Hot
  Subdwarf Stars and Related Objects, ed. D.~{Kilkenny}, C.~S. {Jeffery}, \&
  C.~{Koen}, 121.
\newblock \doarXiv{1204.0928}

\bibitem[{{Kilic} {et~al.}(2010{\natexlab{a}}){Kilic}, {Allende Prieto},
  {Brown}, {Ag{\"u}eros}, {Kenyon}, \& {Camilo}}]{kab+10}
{Kilic}, M., {Allende Prieto}, C., {Brown}, W.~R., {et~al.} 2010{\natexlab{a}},
  \apjl, 721, L158, \dodoi{10.1088/2041-8205/721/2/L158}

\bibitem[{{Kilic} {et~al.}(2021){Kilic}, {B{\'e}dard}, \& {Bergeron}}]{kbb21}
{Kilic}, M., {B{\'e}dard}, A., \& {Bergeron}, P. 2021, \mnras,
  \dodoi{10.1093/mnras/stab439}

\bibitem[{{Kilic} {et~al.}(2011{\natexlab{a}}){Kilic}, {Brown}, {Allende
  Prieto}, {Ag{\"u}eros}, {Heinke}, \& {Kenyon}}]{kba+11}
{Kilic}, M., {Brown}, W.~R., {Allende Prieto}, C., {et~al.} 2011{\natexlab{a}},
  \apj, 727, 3, \dodoi{10.1088/0004-637X/727/1/3}

\bibitem[{{Kilic} {et~al.}(2012){Kilic}, {Brown}, {Allende Prieto}, {Kenyon},
  {Heinke}, {Ag{\"u}eros}, \& {Kleinman}}]{kba+12}
---. 2012, \apj, 751, 141, \dodoi{10.1088/0004-637X/751/2/141}

\bibitem[{{Kilic} {et~al.}(2010{\natexlab{b}}){Kilic}, {Brown}, {Allende
  Prieto}, {Kenyon}, \& {Panei}}]{kba+10}
{Kilic}, M., {Brown}, W.~R., {Allende Prieto}, C., {Kenyon}, S.~J., \& {Panei},
  J.~A. 2010{\natexlab{b}}, \apj, 716, 122, \dodoi{10.1088/0004-637X/716/1/122}

\bibitem[{{Kilic} {et~al.}(2007){Kilic}, {Brown}, {Allende Prieto},
  {Pinsonneault}, \& {Kenyon}}]{kba+07}
{Kilic}, M., {Brown}, W.~R., {Allende Prieto}, C., {Pinsonneault}, M.~H., \&
  {Kenyon}, S.~J. 2007, \apj, 664, 1088, \dodoi{10.1086/518735}

\bibitem[{{Kilic} {et~al.}(2009){Kilic}, {Brown}, {Allende Prieto}, {Swift},
  {Kenyon}, {Liebert}, \& {Ag{\"u}eros}}]{kba+09}
{Kilic}, M., {Brown}, W.~R., {Allende Prieto}, C., {et~al.} 2009, \apjl, 695,
  L92, \dodoi{10.1088/0004-637X/695/1/L92}

\bibitem[{{Kilic} {et~al.}(2017){Kilic}, {Brown}, {Gianninas}, {Curd}, {Bell},
  \& {Allende Prieto}}]{kbg+17}
{Kilic}, M., {Brown}, W.~R., {Gianninas}, A., {et~al.} 2017, \mnras, 471, 4218,
  \dodoi{10.1093/mnras/stx1886}

\bibitem[{{Kilic} {et~al.}(2011{\natexlab{b}}){Kilic}, {Brown}, {Hermes},
  {Allende Prieto}, {Kenyon}, {Winget}, \& {Winget}}]{kbh+11}
{Kilic}, M., {Brown}, W.~R., {Hermes}, J.~J., {et~al.} 2011{\natexlab{b}},
  \mnras, 418, L157, \dodoi{10.1111/j.1745-3933.2011.01165.x}

\bibitem[{{Kilic} {et~al.}(2011{\natexlab{c}}){Kilic}, {Brown}, {Kenyon},
  {Allende Prieto}, {Andrews}, {Kleinman}, {Winget}, {Winget}, \&
  {Hermes}}]{Kbk+11}
{Kilic}, M., {Brown}, W.~R., {Kenyon}, S.~J., {et~al.} 2011{\natexlab{c}},
  \mnras, 413, L101, \dodoi{10.1111/j.1745-3933.2011.01044.x}

\bibitem[{{Kilic} {et~al.}(2014){Kilic}, {Hermes}, {Gianninas}, {Brown},
  {Heinke}, {Ag{\"u}eros}, {Chote}, {Sullivan}, {Bell}, \& {Harrold}}]{khg+14}
{Kilic}, M., {Hermes}, J.~J., {Gianninas}, A., {et~al.} 2014, \mnras, 438, L26,
  \dodoi{10.1093/mnrasl/slt151}

\bibitem[{{Kilkenny}(2011)}]{kil11}
{Kilkenny}, D. 2011, \mnras, 412, 487, \dodoi{10.1111/j.1365-2966.2010.17919.x}

\bibitem[{{Kilkenny} {et~al.}(1988){Kilkenny}, {Spencer Jones}, \&
  {Marang}}]{ksm88}
{Kilkenny}, D., {Spencer Jones}, J.~H., \& {Marang}, F. 1988, The Observatory,
  108, 88

\bibitem[{{Kippenhahn} {et~al.}(2012){Kippenhahn}, {Weigert}, \&
  {Weiss}}]{kww12}
{Kippenhahn}, R., {Weigert}, A., \& {Weiss}, A. 2012, {Stellar Structure and
  Evolution} (Springer Berlin), \dodoi{10.1007/978-3-642-30304-3}

\bibitem[{{Kirichenko} {et~al.}(2020){Kirichenko}, {Karpova}, {Zyuzin},
  {Zharikov}, {L{\'o}pez}, {Shibanov}, {Freire}, {Fonseca}, \&
  {Cabrera-Lavers}}]{kkz+20}
{Kirichenko}, A.~Y., {Karpova}, A.~V., {Zyuzin}, D.~A., {et~al.} 2020, \mnras,
  492, 3032, \dodoi{10.1093/mnras/staa066}

\bibitem[{{Klencki} {et~al.}(2021){Klencki}, {Nelemans}, {Istrate}, \&
  {Chruslinska}}]{kni+21}
{Klencki}, J., {Nelemans}, G., {Istrate}, A.~G., \& {Chruslinska}, M. 2021,
  \aap, 645, A54, \dodoi{10.1051/0004-6361/202038707}

\bibitem[{{Klepp} \& {Rauch}(2011)}]{kr11}
{Klepp}, S., \& {Rauch}, T. 2011, \aap, 531, L7,
  \dodoi{10.1051/0004-6361/201116887}

\bibitem[{{Kobayashi} {et~al.}(2020){Kobayashi}, {Karakas}, \&
  {Lugaro}}]{kkl20}
{Kobayashi}, C., {Karakas}, A.~I., \& {Lugaro}, M. 2020, \apj, 900, 179,
  \dodoi{10.3847/1538-4357/abae65}

\bibitem[{{Koen} {et~al.}(2010){Koen}, {Kilkenny}, {Pretorius}, \&
  {Frew}}]{kkp+10}
{Koen}, C., {Kilkenny}, D., {Pretorius}, M.~L., \& {Frew}, D.~J. 2010, \mnras,
  401, 1850, \dodoi{10.1111/j.1365-2966.2009.15761.x}

\bibitem[{{Kolb} \& {Ritter}(1990)}]{kr1990}
{Kolb}, U., \& {Ritter}, H. 1990, \aap, 236, 385

\bibitem[{{Kosakowski} {et~al.}(2021){Kosakowski}, {Kilic}, \& {Brown}}]{kkb20}
{Kosakowski}, A., {Kilic}, M., \& {Brown}, W. 2021, \mnras, 500, 5098,
  \dodoi{10.1093/mnras/staa3571}

\bibitem[{{Kozai}(1962)}]{koz62}
{Kozai}, Y. 1962, \aj, 67, 591, \dodoi{10.1086/108790}

\bibitem[{{Kramer} {et~al.}(2020){Kramer}, {Schneider}, {Ohlmann}, {Geier},
  {Schaffenroth}, {Pakmor}, \& {R{\"o}pke}}]{kso+20}
{Kramer}, M., {Schneider}, F.~R.~N., {Ohlmann}, S.~T., {et~al.} 2020, \aap,
  642, A97, \dodoi{10.1051/0004-6361/202038702}

\bibitem[{{Kromer} {et~al.}(2010){Kromer}, {Sim}, {Fink}, {R{\"o}pke},
  {Seitenzahl}, \& {Hillebrandt}}]{KSF2010}
{Kromer}, M., {Sim}, S.~A., {Fink}, M., {et~al.} 2010, \apj, 719, 1067,
  \dodoi{10.1088/0004-637X/719/2/1067}

\bibitem[{{Kruckow} {et~al.}(2018){Kruckow}, {Tauris}, {Langer}, {Kramer}, \&
  {Izzard}}]{ktl+18}
{Kruckow}, M.~U., {Tauris}, T.~M., {Langer}, N., {Kramer}, M., \& {Izzard},
  R.~G. 2018, \mnras, 481, 1908, \dodoi{10.1093/mnras/sty2190}

\bibitem[{{Kruckow} {et~al.}(2016){Kruckow}, {Tauris}, {Langer}, {Sz{\'e}csi},
  {Marchant}, \& {Podsiadlowski}}]{ktl+16}
{Kruckow}, M.~U., {Tauris}, T.~M., {Langer}, N., {et~al.} 2016, \aap, 596, A58,
  \dodoi{10.1051/0004-6361/201629420}

\bibitem[{{Kruse} \& {Agol}(2014)}]{ka14}
{Kruse}, E., \& {Agol}, E. 2014, Science, 344, 275,
  \dodoi{10.1126/science.1251999}

\bibitem[{{Krzeminski}(1962)}]{krz62}
{Krzeminski}, W. 1962, \pasp, 74, 66, \dodoi{10.1086/127758}

\bibitem[{{Kudritzki} \& {Simon}(1978)}]{ks78}
{Kudritzki}, R.~P., \& {Simon}, K.~P. 1978, \aap, 70, 653

\bibitem[{{Kuerster} \& {Barwig}(1988)}]{kb88}
{Kuerster}, M., \& {Barwig}, H. 1988, \aap, 199, 201

\bibitem[{{Kupfer} {et~al.}(2016){Kupfer}, {Steeghs}, {Groot}, {Marsh},
  {Nelemans}, \& {Roelofs}}]{ksg+16}
{Kupfer}, T., {Steeghs}, D., {Groot}, P.~J., {et~al.} 2016, \mnras, 457, 1828,
  \dodoi{10.1093/mnras/stw126}

\bibitem[{{Kupfer} {et~al.}(2014){Kupfer}, {Geier}, {McLeod}, {Groot},
  {Verbeek}, {Schaffenroth}, {Heber}, {Heuser}, {Ziegerer}, {{\"O}stensen},
  {Nemeth}, {Dhillon}, {Butterley}, {Littlefair}, {Wilson}, {Telting},
  {Shporer}, \& {Fulton}}]{kgm+14}
{Kupfer}, T., {Geier}, S., {McLeod}, A., {et~al.} 2014, in Astronomical Society
  of the Pacific Conference Series, Vol. 481, 6th Meeting on Hot Subdwarf Stars
  and Related Objects, ed. V.~{van Grootel}, E.~{Green}, G.~{Fontaine}, \&
  S.~{Charpinet}, 293.
\newblock \doarXiv{1308.2447}

\bibitem[{{Kupfer} {et~al.}(2015{\natexlab{a}}){Kupfer}, {Geier}, {Heber},
  {{\O}stensen}, {Barlow}, {Maxted}, {Heuser}, {Schaffenroth}, \&
  {G{\"a}nsicke}}]{kgh+15}
{Kupfer}, T., {Geier}, S., {Heber}, U., {et~al.} 2015{\natexlab{a}}, \aap, 576,
  A44, \dodoi{10.1051/0004-6361/201425213}

\bibitem[{{Kupfer} {et~al.}(2015{\natexlab{b}}){Kupfer}, {Groot}, {Bloemen},
  {Levitan}, {Steeghs}, {Marsh}, {Rutten}, {Nelemans}, {Prince}, {F{\"u}rst},
  \& {Geier}}]{kgb+15}
{Kupfer}, T., {Groot}, P.~J., {Bloemen}, S., {et~al.} 2015{\natexlab{b}},
  \mnras, 453, 483, \dodoi{10.1093/mnras/stv1609}

\bibitem[{{Kupfer} {et~al.}(2018){Kupfer}, {Korol}, {Shah}, {Nelemans},
  {Marsh}, {Ramsay}, {Groot}, {Steeghs}, \& {Rossi}}]{kks+18}
{Kupfer}, T., {Korol}, V., {Shah}, S., {et~al.} 2018, \mnras, 480, 302,
  \dodoi{10.1093/mnras/sty1545}

\bibitem[{{Kupfer} {et~al.}(2020{\natexlab{a}}){Kupfer}, {Bauer}, {Burdge},
  {Roestel}, {Bellm}, {Fuller}, {Hermes}, {Marsh}, {Bildsten}, {Kulkarni},
  {Phinney}, {Prince}, {Szkody}, {Yao}, {Irrgang}, {Heber}, {Schneider},
  {Dhillon}, {Murawski}, {Drake}, {Duev}, {Feeney}, {Graham}, {Laher},
  {Littlefair}, {Mahabal}, {Masci}, {Porter}, {Reiley}, {Rodriguez},
  {Rusholme}, {Shupe}, \& {Soumagnac}}]{kbb+20}
{Kupfer}, T., {Bauer}, E.~B., {Burdge}, K.~B., {et~al.} 2020{\natexlab{a}},
  \apjl, 898, L25, \dodoi{10.3847/2041-8213/aba3c2}

\bibitem[{{Kupfer} {et~al.}(2020{\natexlab{b}}){Kupfer}, {Bauer}, {Marsh}, {van
  Roestel}, {Bellm}, {Burdge}, {Coughlin}, {Fuller}, {Hermes}, {Bildsten},
  {Kulkarni}, {Prince}, {Szkody}, {Dhillon}, {Murawski}, {Burruss}, {Dekany},
  {Delacroix}, {Drake}, {Duev}, {Feeney}, {Graham}, {Kaplan}, {Laher},
  {Littlefair}, {Masci}, {Riddle}, {Rusholme}, {Serabyn}, {Smith}, {Shupe}, \&
  {Soumagnac}}]{kbm+20}
{Kupfer}, T., {Bauer}, E.~B., {Marsh}, T.~R., {et~al.} 2020{\natexlab{b}},
  \apj, 891, 45, \dodoi{10.3847/1538-4357/ab72ff}

\bibitem[{{Kára} {et~al.}(2021){Kára}, {Zharikov}, {Wolf}, {Kučáková},
  {Cagaš}, {Medina Rodriguez}, \& {Mašek}}]{kzw+21}
{Kára}, J., {Zharikov}, S., {Wolf}, M., {et~al.} 2021, arXiv e-prints,
  arXiv:2107.02664.
\newblock \doarXiv{2107.02664}

\bibitem[{{Landau} \& {Lifshitz}(1975)}]{ll1975}
{Landau}, L.~D., \& {Lifshitz}, E.~M. 1975, {The classical theory of fields},
  Course of theoretical physics (Butterworth Heinemann)

\bibitem[{{Landsman} {et~al.}(1997){Landsman}, {Aparicio}, {Bergeron}, {Di
  Stefano}, \& {Stecher}}]{lab+97}
{Landsman}, W., {Aparicio}, J., {Bergeron}, P., {Di Stefano}, R., \& {Stecher},
  T.~P. 1997, \apjl, 481, L93, \dodoi{10.1086/310654}

\bibitem[{{Landsman} {et~al.}(1993){Landsman}, {Simon}, \& {Bergeron}}]{lsb+93}
{Landsman}, W., {Simon}, T., \& {Bergeron}, P. 1993, \pasp, 105, 841,
  \dodoi{10.1086/133242}

\bibitem[{{Lanning}(1982)}]{lan82}
{Lanning}, H.~H. 1982, \apj, 253, 752, \dodoi{10.1086/159676}

\bibitem[{{Lanning} \& {Pesch}(1981)}]{lp81}
{Lanning}, H.~H., \& {Pesch}, P. 1981, \apj, 244, 280, \dodoi{10.1086/158705}

\bibitem[{{Law} {et~al.}(2012){Law}, {Kraus}, {Street}, {Fulton},
  {Hillenbrand}, {Shporer}, {Lister}, {Baranec}, {Bloom}, {Bui}, {Burse},
  {Cenko}, {Das}, {Davis}, {Dekany}, {Filippenko}, {Kasliwal}, {Kulkarni},
  {Nugent}, {Ofek}, {Poznanski}, {Quimby}, {Ramaprakash}, {Riddle},
  {Silverman}, {Sivanandam}, \& {Tendulkar}}]{lks+12}
{Law}, N.~M., {Kraus}, A.~L., {Street}, R., {et~al.} 2012, \apj, 757, 133,
  \dodoi{10.1088/0004-637X/757/2/133}

\bibitem[{{Lee} {et~al.}(2009){Lee}, {Kim}, {Kim}, {Koch}, {Lee}, {Kim}, \&
  {Park}}]{lkk+09}
{Lee}, J.~W., {Kim}, S.-L., {Kim}, C.-H., {et~al.} 2009, \aj, 137, 3181,
  \dodoi{10.1088/0004-6256/137/2/3181}

\bibitem[{{Levitan} {et~al.}(2014){Levitan}, {Kupfer}, {Groot}, {Margon},
  {Prince}, {Kulkarni}, {Hallinan}, {Harding}, {Kyne}, {Laher}, {Ofek},
  {Rutten}, {Sesar}, \& {Surace}}]{lkg+14}
{Levitan}, D., {Kupfer}, T., {Groot}, P.~J., {et~al.} 2014, \apj, 785, 114,
  \dodoi{10.1088/0004-637X/785/2/114}

\bibitem[{{Lindegren} {et~al.}(2021){Lindegren}, {Bastian}, {Biermann},
  {Bombrun}, {de Torres}, {Gerlach}, {Geyer}, {Hern{\'a}ndez}, {Hilger},
  {Hobbs}, {Klioner}, {Lammers}, {McMillan}, {Ramos-Lerate},
  {Steidelm{\"u}ller}, {Stephenson}, \& {van Leeuwen}}]{lbb+21}
{Lindegren}, L., {Bastian}, U., {Biermann}, M., {et~al.} 2021, \aap, 649, A4,
  \dodoi{10.1051/0004-6361/202039653}

\bibitem[{{Linnell} {et~al.}(2009){Linnell}, {Godon}, {Hubeny}, {Sion},
  {Szkody}, \& {Barrett}}]{lgh+09}
{Linnell}, A.~P., {Godon}, P., {Hubeny}, I., {et~al.} 2009, \apj, 703, 1839,
  \dodoi{10.1088/0004-637X/703/2/1839}

\bibitem[{{Lisker} {et~al.}(2005){Lisker}, {Heber}, {Napiwotzki}, {Christlieb},
  {Han}, {Homeier}, \& {Reimers}}]{lhn+05}
{Lisker}, T., {Heber}, U., {Napiwotzki}, R., {et~al.} 2005, \aap, 430, 223,
  \dodoi{10.1051/0004-6361:20040232}

\bibitem[{{Littlefair} {et~al.}(2014{\natexlab{a}}){Littlefair}, {Dhillon},
  {G{\"a}nsicke}, {Bours}, {Copperwheat}, \& {Marsh}}]{ldg+14}
{Littlefair}, S.~P., {Dhillon}, V.~S., {G{\"a}nsicke}, B.~T., {et~al.}
  2014{\natexlab{a}}, \mnras, 443, 718, \dodoi{10.1093/mnras/stu1158}

\bibitem[{{Littlefair} {et~al.}(2006{\natexlab{a}}){Littlefair}, {Dhillon},
  {Marsh}, \& {G{\"a}nsicke}}]{ldmg06}
{Littlefair}, S.~P., {Dhillon}, V.~S., {Marsh}, T.~R., \& {G{\"a}nsicke}, B.~T.
  2006{\natexlab{a}}, \mnras, 371, 1435,
  \dodoi{10.1111/j.1365-2966.2006.10771.x}

\bibitem[{{Littlefair} {et~al.}(2007){Littlefair}, {Dhillon}, {Marsh},
  {G{\"a}nsicke}, {Baraffe}, \& {Watson}}]{ldm+07}
{Littlefair}, S.~P., {Dhillon}, V.~S., {Marsh}, T.~R., {et~al.} 2007, \mnras,
  381, 827, \dodoi{10.1111/j.1365-2966.2007.12285.x}

\bibitem[{{Littlefair} {et~al.}(2008){Littlefair}, {Dhillon}, {Marsh},
  {G{\"a}nsicke}, {Southworth}, {Baraffe}, {Watson}, \& {Copperwheat}}]{ldm+08}
---. 2008, \mnras, 388, 1582, \dodoi{10.1111/j.1365-2966.2008.13539.x}

\bibitem[{{Littlefair} {et~al.}(2006{\natexlab{b}}){Littlefair}, {Dhillon},
  {Marsh}, {G{\"a}nsicke}, {Southworth}, \& {Watson}}]{ldm+06}
---. 2006{\natexlab{b}}, Science, 314, 1578, \dodoi{10.1126/science.1133333}

\bibitem[{{Littlefair} {et~al.}(2014{\natexlab{b}}){Littlefair}, {Casewell},
  {Parsons}, {Dhillon}, {Marsh}, {G{\"a}nsicke}, {Bloemen}, {Catalan},
  {Irawati}, {Hardy}, {Mcallister}, {Bours}, {Richichi}, {Burleigh},
  {Burningham}, {Breedt}, \& {Kerry}}]{lcp+14}
{Littlefair}, S.~P., {Casewell}, S.~L., {Parsons}, S.~G., {et~al.}
  2014{\natexlab{b}}, \mnras, 445, 2106, \dodoi{10.1093/mnras/stu1895}

\bibitem[{{Lynch} {et~al.}(2012){Lynch}, {Freire}, {Ransom}, \&
  {Jacoby}}]{lfr+12}
{Lynch}, R.~S., {Freire}, P. C.~C., {Ransom}, S.~M., \& {Jacoby}, B.~A. 2012,
  \apj, 745, 109, \dodoi{10.1088/0004-637X/745/2/109}

\bibitem[{{MacLeod} {et~al.}(2017){MacLeod}, {Macias}, {Ramirez-Ruiz},
  {Grindlay}, {Batta}, \& {Montes}}]{mmr+17}
{MacLeod}, M., {Macias}, P., {Ramirez-Ruiz}, E., {et~al.} 2017, \apj, 835, 282,
  \dodoi{10.3847/1538-4357/835/2/282}

\bibitem[{{Manchester} {et~al.}(2005){Manchester}, {Hobbs}, {Teoh}, \&
  {Hobbs}}]{mht+05}
{Manchester}, R.~N., {Hobbs}, G.~B., {Teoh}, A., \& {Hobbs}, M. 2005, \aj, 129,
  1993, \dodoi{10.1086/428488}

\bibitem[{{Mapelli} {et~al.}(2019){Mapelli}, {Giacobbo}, {Santoliquido}, \&
  {Artale}}]{mgs+19}
{Mapelli}, M., {Giacobbo}, N., {Santoliquido}, F., \& {Artale}, M.~C. 2019,
  \mnras, 487, 2, \dodoi{10.1093/mnras/stz1150}

\bibitem[{{Marino} \& {Walker}(1974)}]{mw74}
{Marino}, B.~F., \& {Walker}, W.~S.~G. 1974, Information Bulletin on Variable
  Stars, 864, 1

\bibitem[{{Marsh}(1995)}]{mar95}
{Marsh}, T.~R. 1995, \mnras, 275, L1, \dodoi{10.1093/mnras/275.1.L1}

\bibitem[{{Marsh}(2000{\natexlab{a}})}]{m2000}
---. 2000{\natexlab{a}}, \nar, 44, 119, \dodoi{10.1016/S1387-6473(00)00024-5}

\bibitem[{{Marsh}(2000{\natexlab{b}})}]{mar00}
---. 2000{\natexlab{b}}, \nar, 44, 119, \dodoi{10.1016/S1387-6473(00)00024-5}

\bibitem[{{Marsh} {et~al.}(1995){Marsh}, {Dhillon}, \& {Duck}}]{mdd95}
{Marsh}, T.~R., {Dhillon}, V.~S., \& {Duck}, S.~R. 1995, \mnras, 275, 828,
  \dodoi{10.1093/mnras/275.3.828}

\bibitem[{{Marsh} {et~al.}(2011){Marsh}, {G{\"a}nsicke}, {Steeghs},
  {Southworth}, {Koester}, {Harris}, \& {Merry}}]{mgs+11}
{Marsh}, T.~R., {G{\"a}nsicke}, B.~T., {Steeghs}, D., {et~al.} 2011, \apj, 736,
  95, \dodoi{10.1088/0004-637X/736/2/95}

\bibitem[{{Marsh} {et~al.}(2016){Marsh}, {G{\"a}nsicke}, {H{\"u}mmerich},
  {Hambsch}, {Bernhard}, {Lloyd}, {Breedt}, {Stanway}, {Steeghs}, {Parsons},
  {Toloza}, {Schreiber}, {Jonker}, {van Roestel}, {Kupfer}, {Pala}, {Dhillon},
  {Hardy}, {Littlefair}, {Aungwerojwit}, {Arjyotha}, {Koester}, {Bochinski},
  {Haswell}, {Frank}, \& {Wheatley}}]{mgh+16}
{Marsh}, T.~R., {G{\"a}nsicke}, B.~T., {H{\"u}mmerich}, S., {et~al.} 2016,
  \nat, 537, 374, \dodoi{10.1038/nature18620}

\bibitem[{{Mart{\'\i}nez-Pais} {et~al.}(2000){Mart{\'\i}nez-Pais},
  {Mart{\'\i}n-Hern{\'a}ndez}, {Casares}, \& {Rodr{\'\i}guez-Gil}}]{mmc+00}
{Mart{\'\i}nez-Pais}, I.~G., {Mart{\'\i}n-Hern{\'a}ndez}, N.~L., {Casares}, J.,
  \& {Rodr{\'\i}guez-Gil}, P. 2000, \apj, 538, 315, \dodoi{10.1086/309092}

\bibitem[{{Mason} {et~al.}(2001){Mason}, {Skidmore}, {Howell}, \&
  {Mennickent}}]{msh+01}
{Mason}, E., {Skidmore}, W., {Howell}, S.~B., \& {Mennickent}, R.~E. 2001,
  \apj, 563, 351, \dodoi{10.1086/323431}

\bibitem[{{Mason} {et~al.}(2019){Mason}, {Wells}, {Motsoaledi}, {Szkody}, \&
  {Gonzalez}}]{mwm+19}
{Mason}, P.~A., {Wells}, N.~K., {Motsoaledi}, M., {Szkody}, P., \& {Gonzalez},
  E. 2019, \mnras, 488, 2881, \dodoi{10.1093/mnras/stz1863}

\bibitem[{{Maxted} {et~al.}(2002{\natexlab{a}}){Maxted}, {Burleigh}, {Marsh},
  \& {Bannister}}]{mbm+02}
{Maxted}, P.~F.~L., {Burleigh}, M.~R., {Marsh}, T.~R., \& {Bannister}, N.~P.
  2002{\natexlab{a}}, \mnras, 334, 833,
  \dodoi{10.1046/j.1365-8711.2002.05545.x}

\bibitem[{{Maxted} {et~al.}(2001){Maxted}, {Heber}, {Marsh}, \&
  {North}}]{mhm+01}
{Maxted}, P.~F.~L., {Heber}, U., {Marsh}, T.~R., \& {North}, R.~C. 2001,
  \mnras, 326, 1391, \dodoi{10.1111/j.1365-2966.2001.04714.x}

\bibitem[{{Maxted} \& {Marsh}(1999)}]{mm99}
{Maxted}, P.~F.~L., \& {Marsh}, T.~R. 1999, \mnras, 307, 122,
  \dodoi{10.1046/j.1365-8711.1999.02635.x}

\bibitem[{{Maxted} {et~al.}(2002{\natexlab{b}}){Maxted}, {Marsh}, {Heber},
  {Morales-Rueda}, {North}, \& {Lawson}}]{mmh+02}
{Maxted}, P.~F.~L., {Marsh}, T.~R., {Heber}, U., {et~al.} 2002{\natexlab{b}},
  \mnras, 333, 231, \dodoi{10.1046/j.1365-8711.2002.05406.x}

\bibitem[{{Maxted} {et~al.}(2004){Maxted}, {Marsh}, {Morales-Rueda}, {Barstow},
  {Dobbie}, {Schreiber}, {Dhillon}, \& {Brinkworth}}]{mmm+04}
{Maxted}, P.~F.~L., {Marsh}, T.~R., {Morales-Rueda}, L., {et~al.} 2004, \mnras,
  355, 1143, \dodoi{10.1111/j.1365-2966.2004.08393.x}

\bibitem[{{Maxted} {et~al.}(1998){Maxted}, {Marsh}, {Moran}, {Dhillon}, \&
  {Hilditch}}]{mmm+98}
{Maxted}, P.~F.~L., {Marsh}, T.~R., {Moran}, C., {Dhillon}, V.~S., \&
  {Hilditch}, R.~W. 1998, \mnras, 300, 1225,
  \dodoi{10.1046/j.1365-8711.1998.02036.x}

\bibitem[{{Maxted} {et~al.}(2002{\natexlab{c}}){Maxted}, {Marsh}, \&
  {Moran}}]{mmm02}
{Maxted}, P.~F.~L., {Marsh}, T.~R., \& {Moran}, C.~K.~J. 2002{\natexlab{c}},
  \mnras, 332, 745, \dodoi{10.1046/j.1365-8711.2002.05368.x}

\bibitem[{{Maxted} {et~al.}(2000{\natexlab{a}}){Maxted}, {Marsh}, {Moran}, \&
  {Han}}]{mmm+00}
{Maxted}, P.~F.~L., {Marsh}, T.~R., {Moran}, C.~K.~J., \& {Han}, Z.
  2000{\natexlab{a}}, \mnras, 314, 334,
  \dodoi{10.1046/j.1365-8711.2000.03343.x}

\bibitem[{{Maxted} {et~al.}(2000{\natexlab{b}}){Maxted}, {Marsh}, \&
  {North}}]{mmn00}
{Maxted}, P.~F.~L., {Marsh}, T.~R., \& {North}, R.~C. 2000{\natexlab{b}},
  \mnras, 317, L41, \dodoi{10.1046/j.1365-8711.2000.03856.x}

\bibitem[{{Maxted} {et~al.}(2000{\natexlab{c}}){Maxted}, {Moran}, {Marsh}, \&
  {Gatti}}]{mmmg00}
{Maxted}, P.~F.~L., {Moran}, C.~K.~J., {Marsh}, T.~R., \& {Gatti}, A.~A.
  2000{\natexlab{c}}, \mnras, 311, 877,
  \dodoi{10.1046/j.1365-8711.2000.03102.x}

\bibitem[{{Maxted} {et~al.}(2006){Maxted}, {Napiwotzki}, {Dobbie}, \&
  {Burleigh}}]{mnd+06}
{Maxted}, P.~F.~L., {Napiwotzki}, R., {Dobbie}, P.~D., \& {Burleigh}, M.~R.
  2006, \nat, 442, 543, \dodoi{10.1038/nature04987}

\bibitem[{{Maxted} {et~al.}(2007){Maxted}, {O'Donoghue}, {Morales-Rueda},
  {Napiwotzki}, \& {Smalley}}]{mom+07}
{Maxted}, P.~F.~L., {O'Donoghue}, D., {Morales-Rueda}, L., {Napiwotzki}, R., \&
  {Smalley}, B. 2007, \mnras, 376, 919,
  \dodoi{10.1111/j.1365-2966.2007.11564.x}

\bibitem[{{Maxted} {et~al.}(2014{\natexlab{a}}){Maxted}, {Serenelli}, {Marsh},
  {Catal{\'a}n}, {Mahtani}, \& {Dhillon}}]{msm+14}
{Maxted}, P.~F.~L., {Serenelli}, A.~M., {Marsh}, T.~R., {et~al.}
  2014{\natexlab{a}}, \mnras, 444, 208, \dodoi{10.1093/mnras/stu1465}

\bibitem[{{Maxted} {et~al.}(2011){Maxted}, {Anderson}, {Burleigh}, {Collier
  Cameron}, {Heber}, {G{\"a}nsicke}, {Geier}, {Kupfer}, {Marsh}, {Nelemans},
  {O'Toole}, {{\O}stensen}, {Smalley}, \& {West}}]{mab+11}
{Maxted}, P.~F.~L., {Anderson}, D.~R., {Burleigh}, M.~R., {et~al.} 2011,
  \mnras, 418, 1156, \dodoi{10.1111/j.1365-2966.2011.19567.x}

\bibitem[{{Maxted} {et~al.}(2013){Maxted}, {Serenelli}, {Miglio}, {Marsh},
  {Heber}, {Dhillon}, {Littlefair}, {Copperwheat}, {Smalley}, {Breedt}, \&
  {Schaffenroth}}]{msm+13}
{Maxted}, P. F.~L., {Serenelli}, A.~M., {Miglio}, A., {et~al.} 2013, \nat, 498,
  463, \dodoi{10.1038/nature12192}

\bibitem[{{Maxted} {et~al.}(2014{\natexlab{b}}){Maxted}, {Bloemen}, {Heber},
  {Geier}, {Wheatley}, {Marsh}, {Breedt}, {Sebastian}, {Faillace}, {Owen},
  {Pulley}, {Smith}, {Kolb}, {Haswell}, {Southworth}, {Anderson}, {Smalley},
  {Collier Cameron}, {Hebb}, {Simpson}, {West}, {Bochinski}, {Busuttil}, \&
  {Hadigal}}]{mbh+14}
{Maxted}, P.~F.~L., {Bloemen}, S., {Heber}, U., {et~al.} 2014{\natexlab{b}},
  \mnras, 437, 1681, \dodoi{10.1093/mnras/stt2007}

\bibitem[{{McAllister} {et~al.}(2019){McAllister}, {Littlefair}, {Parsons},
  {Dhillon}, {Marsh}, {G{\"a}nsicke}, {Breedt}, {Copperwheat}, {Green},
  {Knigge}, {Sahman}, {Dyer}, {Kerry}, {Ashley}, {Irawati}, \&
  {Rattanasoon}}]{mlp+19}
{McAllister}, M., {Littlefair}, S.~P., {Parsons}, S.~G., {et~al.} 2019, \mnras,
  486, 5535, \dodoi{10.1093/mnras/stz976}

\bibitem[{{McAllister} {et~al.}(2015){McAllister}, {Littlefair}, {Baraffe},
  {Dhillon}, {Marsh}, {Bento}, {Bochinski}, {Bours}, {Breedt}, {Copperwheat},
  {Hardy}, {Kerry}, {Parsons}, {Rostron}, {Sahman}, {Savoury}, \&
  {Tunnicliffe}}]{mlb+15}
{McAllister}, M.~J., {Littlefair}, S.~P., {Baraffe}, I., {et~al.} 2015, \mnras,
  451, 114, \dodoi{10.1093/mnras/stv956}

\bibitem[{{McAllister} {et~al.}(2017{\natexlab{a}}){McAllister}, {Littlefair},
  {Dhillon}, {Marsh}, {G{\"a}nsicke}, {Bochinski}, {Bours}, {Breedt}, {Hardy},
  {Hermes}, {Kengkriangkrai}, {Kerry}, {Parsons}, \& {Rattanasoon}}]{mld+17a}
{McAllister}, M.~J., {Littlefair}, S.~P., {Dhillon}, V.~S., {et~al.}
  2017{\natexlab{a}}, \mnras, 467, 1024, \dodoi{10.1093/mnras/stx253}

\bibitem[{{McAllister} {et~al.}(2017{\natexlab{b}}){McAllister}, {Littlefair},
  {Dhillon}, {Marsh}, {Ashley}, {Bours}, {Breedt}, {Hardy}, {Hermes},
  {Kengkriangkrai}, {Kerry}, {Rattanasoon}, \& {Sahman}}]{mld+17b}
---. 2017{\natexlab{b}}, \mnras, 464, 1353, \dodoi{10.1093/mnras/stw2417}

\bibitem[{{McKee} {et~al.}(2020){McKee}, {Freire}, {Berezina}, {Champion},
  {Cognard}, {Graikou}, {Guillemot}, {Keith}, {Kramer}, {Lyne}, {Stappers},
  {Tauris}, \& {Theureau}}]{mfb+20}
{McKee}, J.~W., {Freire}, P.~C.~C., {Berezina}, M., {et~al.} 2020, \mnras, 499,
  4082, \dodoi{10.1093/mnras/staa2994}

\bibitem[{{Mendez} \& {Niemela}(1981)}]{mn81}
{Mendez}, R.~H., \& {Niemela}, V.~S. 1981, \apj, 250, 240,
  \dodoi{10.1086/159368}

\bibitem[{{Mennekens} \& {Vanbeveren}(2014)}]{mv14}
{Mennekens}, N., \& {Vanbeveren}, D. 2014, \aap, 564, A134,
  \dodoi{10.1051/0004-6361/201322198}

\bibitem[{{Mennekens} {et~al.}(2010){Mennekens}, {Vanbeveren}, {De Greve}, \&
  {De Donder}}]{mvd+10}
{Mennekens}, N., {Vanbeveren}, D., {De Greve}, J.~P., \& {De Donder}, E. 2010,
  \aap, 515, A89, \dodoi{10.1051/0004-6361/201014115}

\bibitem[{{Mennickent} \& {Arenas}(1998)}]{ma98}
{Mennickent}, R.~E., \& {Arenas}, J. 1998, \pasj, 50, 333,
  \dodoi{10.1093/pasj/50.3.333}

\bibitem[{{Mennickent} \& {Diaz}(1996)}]{md96}
{Mennickent}, R.~E., \& {Diaz}, M. 1996, \aap, 309, 147

\bibitem[{{Mennickent} \& {Sterken}(1998)}]{ms98}
{Mennickent}, R.~E., \& {Sterken}, C. 1998, \pasp, 110, 1032,
  \dodoi{10.1086/316226}

\bibitem[{{Mereghetti} {et~al.}(2010){Mereghetti}, {Tiengo}, {Esposito}, {La
  Palombara}, {Israel}, \& {Stella}}]{mte+10}
{Mereghetti}, S., {Tiengo}, A., {Esposito}, P., {et~al.} 2010, in American
  Institute of Physics Conference Series, Vol. 1248, X-ray Astronomy 2009;
  Present Status, Multi-Wavelength Approach and Future Perspectives, ed.
  A.~{Comastri}, L.~{Angelini}, \& M.~{Cappi}, 85--88,
  \dodoi{10.1063/1.3475362}

\bibitem[{{Michaely} \& {Perets}(2014)}]{mp14}
{Michaely}, E., \& {Perets}, H.~B. 2014, \apj, 794, 122,
  \dodoi{10.1088/0004-637X/794/2/122}

\bibitem[{{Miszalski} {et~al.}(2016){Miszalski}, {Woudt}, {Littlefair},
  {Warner}, {Boffin}, {Corradi}, {Jones}, {Motsoaledi}, {Rodr{\'\i}guez-Gil},
  {Sabin}, \& {Santander-Garc{\'\i}a}}]{mwl+16}
{Miszalski}, B., {Woudt}, P.~A., {Littlefair}, S.~P., {et~al.} 2016, \mnras,
  456, 633, \dodoi{10.1093/mnras/stv2689}

\bibitem[{{Moe} \& {Di Stefano}(2017)}]{md17}
{Moe}, M., \& {Di Stefano}, R. 2017, \apjs, 230, 15,
  \dodoi{10.3847/1538-4365/aa6fb6}

\bibitem[{{Morales-Rueda} {et~al.}(2005){Morales-Rueda}, {Marsh}, {Maxted},
  {Nelemans}, {Karl}, {Napiwotzki}, \& {Moran}}]{mmm+05}
{Morales-Rueda}, L., {Marsh}, T.~R., {Maxted}, P.~F.~L., {et~al.} 2005, \mnras,
  359, 648, \dodoi{10.1111/j.1365-2966.2005.08943.x}

\bibitem[{{Morales-Rueda} {et~al.}(2003){Morales-Rueda}, {Maxted}, {Marsh},
  {North}, \& {Heber}}]{mmm+03}
{Morales-Rueda}, L., {Maxted}, P.~F.~L., {Marsh}, T.~R., {North}, R.~C., \&
  {Heber}, U. 2003, \mnras, 338, 752, \dodoi{10.1046/j.1365-8711.2003.06088.x}

\bibitem[{{Morales-Rueda} {et~al.}(2002){Morales-Rueda}, {Still}, {Roche},
  {Wood}, \& {Lockley}}]{msr+02}
{Morales-Rueda}, L., {Still}, M.~D., {Roche}, P., {Wood}, J.~H., \& {Lockley},
  J.~J. 2002, \mnras, 329, 597, \dodoi{10.1046/j.1365-8711.2002.05013.x}

\bibitem[{{Moran} {et~al.}(1997){Moran}, {Marsh}, \& {Bragaglia}}]{mmb97}
{Moran}, C., {Marsh}, T.~R., \& {Bragaglia}, A. 1997, \mnras, 288, 538,
  \dodoi{10.1093/mnras/288.2.538}

\bibitem[{{Moran} {et~al.}(1999){Moran}, {Maxted}, {Marsh}, {Saffer}, \&
  {Livio}}]{mmm+99}
{Moran}, C., {Maxted}, P., {Marsh}, T.~R., {Saffer}, R.~A., \& {Livio}, M.
  1999, \mnras, 304, 535, \dodoi{10.1046/j.1365-8711.1999.02314.x}

\bibitem[{{Muirhead} {et~al.}(2013){Muirhead}, {Vanderburg}, {Shporer},
  {Becker}, {Swift}, {Lloyd}, {Fuller}, {Zhao}, {Hinkley}, {Pineda}, {Bottom},
  {Howard}, {von Braun}, {Boyajian}, {Law}, {Baranec}, {Riddle}, {Ramaprakash},
  {Tendulkar}, {Bui}, {Burse}, {Chordia}, {Das}, {Dekany}, {Punnadi}, \&
  {Johnson}}]{mvs+13}
{Muirhead}, P.~S., {Vanderburg}, A., {Shporer}, A., {et~al.} 2013, \apj, 767,
  111, \dodoi{10.1088/0004-637X/767/2/111}

\bibitem[{{Mumford}(1969)}]{mum69}
{Mumford}, G.~S. 1969, Information Bulletin on Variable Stars, 337, 1

\bibitem[{{Naoz}(2016)}]{nao16}
{Naoz}, S. 2016, \araa, 54, 441, \dodoi{10.1146/annurev-astro-081915-023315}

\bibitem[{{Napiwotzki} {et~al.}(2001){Napiwotzki}, {Edelmann}, {Heber}, {Karl},
  {Drechsel}, {Pauli}, \& {Christlieb}}]{neh+01}
{Napiwotzki}, R., {Edelmann}, H., {Heber}, U., {et~al.} 2001, \aap, 378, L17,
  \dodoi{10.1051/0004-6361:20011223}

\bibitem[{{Napiwotzki} {et~al.}(2004){Napiwotzki}, {Karl}, {Lisker}, {Heber},
  {Christlieb}, {Reimers}, {Nelemans}, \& {Homeier}}]{nkl+04}
{Napiwotzki}, R., {Karl}, C.~A., {Lisker}, T., {et~al.} 2004, \apss, 291, 321,
  \dodoi{10.1023/B:ASTR.0000044362.07416.6c}

\bibitem[{{Napiwotzki} {et~al.}(2002){Napiwotzki}, {Koester}, {Nelemans},
  {Yungelson}, {Christlieb}, {Renzini}, {Reimers}, {Drechsel}, \&
  {Leibundgut}}]{nkn+02}
{Napiwotzki}, R., {Koester}, D., {Nelemans}, G., {et~al.} 2002, \aap, 386, 957,
  \dodoi{10.1051/0004-6361:20020361}

\bibitem[{{Napiwotzki} {et~al.}(2007){Napiwotzki}, {Karl}, {Nelemans},
  {Yungelson}, {Christlieb}, {Drechsel}, {Heber}, {Homeier}, {Koester},
  {Leibundgut}, {Marsh}, {Moehler}, {Renzini}, \& {Reimers}}]{nkn+07}
{Napiwotzki}, R., {Karl}, C.~A., {Nelemans}, G., {et~al.} 2007, in Astronomical
  Society of the Pacific Conference Series, Vol. 372, 15th European Workshop on
  White Dwarfs, ed. R.~{Napiwotzki} \& M.~R. {Burleigh}, 387

\bibitem[{{Napiwotzki} {et~al.}(2020){Napiwotzki}, {Karl}, {Lisker},
  {Catal{\'a}n}, {Drechsel}, {Heber}, {Homeier}, {Koester}, {Leibundgut},
  {Marsh}, {Moehler}, {Nelemans}, {Reimers}, {Renzini}, {Str{\"o}er}, \&
  {Yungelson}}]{nkl+20}
{Napiwotzki}, R., {Karl}, C.~A., {Lisker}, T., {et~al.} 2020, \aap, 638, A131,
  \dodoi{10.1051/0004-6361/201629648}

\bibitem[{{Nebot G{\'o}mez-Mor{\'a}n} {et~al.}(2011){Nebot
  G{\'o}mez-Mor{\'a}n}, {G{\"a}nsicke}, {Schreiber}, {Rebassa-Mansergas},
  {Schwope}, {Southworth}, {Aungwerojwit}, {Bothe}, {Davis}, {Kolb},
  {M{\"u}ller}, {Papadaki}, {Pyrzas}, {Rabitz}, {Rodr{\'\i}guez-Gil},
  {Schmidtobreick}, {Schwarz}, {Tappert}, {Toloza}, {Vogel}, \&
  {Zorotovic}}]{ngs+11}
{Nebot G{\'o}mez-Mor{\'a}n}, A., {G{\"a}nsicke}, B.~T., {Schreiber}, M.~R.,
  {et~al.} 2011, \aap, 536, A43, \dodoi{10.1051/0004-6361/201117514}

\bibitem[{{Nelemans} \& {Tout}(2005)}]{nt05}
{Nelemans}, G., \& {Tout}, C.~A. 2005, \mnras, 356, 753,
  \dodoi{10.1111/j.1365-2966.2004.08496.x}

\bibitem[{{Nelemans} {et~al.}(2000){Nelemans}, {Verbunt}, {Yungelson}, \&
  {Portegies Zwart}}]{nvyp00}
{Nelemans}, G., {Verbunt}, F., {Yungelson}, L.~R., \& {Portegies Zwart}, S.~F.
  2000, \aap, 360, 1011.
\newblock \doarXiv{astro-ph/0006216}

\bibitem[{{Nelemans} {et~al.}(2005){Nelemans}, {Napiwotzki}, {Karl}, {Marsh},
  {Voss}, {Roelofs}, {Izzard}, {Montgomery}, {Reerink}, {Christlieb}, \&
  {Reimers}}]{nnk+05}
{Nelemans}, G., {Napiwotzki}, R., {Karl}, C., {et~al.} 2005, \aap, 440, 1087,
  \dodoi{10.1051/0004-6361:20053174}

\bibitem[{{Neunteufel}(2020)}]{n2020}
{Neunteufel}, P. 2020, \aap, 641, A52, \dodoi{10.1051/0004-6361/202037792}

\bibitem[{{Neunteufel} {et~al.}(2016){Neunteufel}, {Yoon}, \&
  {Langer}}]{nyl2016}
{Neunteufel}, P., {Yoon}, S.~C., \& {Langer}, N. 2016, \aap, 589, A43,
  \dodoi{10.1051/0004-6361/201527845}

\bibitem[{{Neunteufel} {et~al.}(2017){Neunteufel}, {Yoon}, \&
  {Langer}}]{NYL2017}
{Neunteufel}, P., {Yoon}, S.-C., \& {Langer}, N. 2017, \aap, 602, A55,
  \dodoi{10.1051/0004-6361/201630121}

\bibitem[{{Neustroev} {et~al.}(2011){Neustroev}, {Suleimanov}, {Borisov},
  {Belyakov}, \& {Shearer}}]{nsb+11}
{Neustroev}, V.~V., {Suleimanov}, V.~F., {Borisov}, N.~V., {Belyakov}, K.~V.,
  \& {Shearer}, A. 2011, \mnras, 410, 963,
  \dodoi{10.1111/j.1365-2966.2010.17495.x}

\bibitem[{{Neustroev} \& {Zharikov}(2008)}]{nz08}
{Neustroev}, V.~V., \& {Zharikov}, S. 2008, \mnras, 386, 1366,
  \dodoi{10.1111/j.1365-2966.2008.12930.x}

\bibitem[{{Nice} {et~al.}(2003){Nice}, {Splaver}, \& {Stairs}}]{nss03}
{Nice}, D.~J., {Splaver}, E.~M., \& {Stairs}, I.~H. 2003, in Astronomical
  Society of the Pacific Conference Series, Vol. 302, Radio Pulsars, ed.
  M.~{Bailes}, D.~J. {Nice}, \& S.~E. {Thorsett}, 75.
\newblock \doarXiv{astro-ph/0210637}

\bibitem[{{Nice} {et~al.}(2008){Nice}, {Stairs}, \& {Kasian}}]{nsk08}
{Nice}, D.~J., {Stairs}, I.~H., \& {Kasian}, L.~E. 2008, in American Institute
  of Physics Conference Series, Vol. 983, 40 Years of Pulsars: Millisecond
  Pulsars, Magnetars and More, ed. C.~{Bassa}, Z.~{Wang}, A.~{Cumming}, \&
  V.~M. {Kaspi}, 453--458, \dodoi{10.1063/1.2900273}

\bibitem[{{Nomoto}(1982{\natexlab{a}})}]{N1982a}
{Nomoto}, K. 1982{\natexlab{a}}, \apj, 253, 798, \dodoi{10.1086/159682}

\bibitem[{{Nomoto}(1982{\natexlab{b}})}]{N1982b}
---. 1982{\natexlab{b}}, \apj, 257, 780, \dodoi{10.1086/160031}

\bibitem[{{O'Brien} {et~al.}(2001){O'Brien}, {Bond}, \& {Sion}}]{obs01}
{O'Brien}, M.~S., {Bond}, H.~E., \& {Sion}, E.~M. 2001, \apj, 563, 971,
  \dodoi{10.1086/324040}

\bibitem[{{Ochsenbein} {et~al.}(2000){Ochsenbein}, {Bauer}, \&
  {Marcout}}]{obm00}
{Ochsenbein}, F., {Bauer}, P., \& {Marcout}, J. 2000, \aaps, 143, 23,
  \dodoi{10.1051/aas:2000169}

\bibitem[{{O'Donoghue} {et~al.}(2003){O'Donoghue}, {Koen}, {Kilkenny},
  {Stobie}, {Koester}, {Bessell}, {Hambly}, \& {MacGillivray}}]{okk+03}
{O'Donoghue}, D., {Koen}, C., {Kilkenny}, D., {et~al.} 2003, \mnras, 345, 506,
  \dodoi{10.1046/j.1365-8711.2003.06973.x}

\bibitem[{{Orosz} {et~al.}(2001){Orosz}, {Thorstensen}, \& {Kent
  Honeycutt}}]{otk01}
{Orosz}, J.~A., {Thorstensen}, J.~R., \& {Kent Honeycutt}, R. 2001, \mnras,
  326, 1134, \dodoi{10.1046/j.1365-8711.2001.04682.x}

\bibitem[{{Orosz} \& {Wade}(1999)}]{ow99}
{Orosz}, J.~A., \& {Wade}, R.~A. 1999, \mnras, 310, 773,
  \dodoi{10.1046/j.1365-8711.1999.02989.x}

\bibitem[{{Orosz} {et~al.}(1999){Orosz}, {Wade}, {Harlow}, {Thorstensen},
  {Taylor}, \& {Eracleous}}]{owh+99}
{Orosz}, J.~A., {Wade}, R.~A., {Harlow}, J. J.~B., {et~al.} 1999, \aj, 117,
  1598, \dodoi{10.1086/300790}

\bibitem[{{{\O}stensen} {et~al.}(2014){{\O}stensen}, {Telting}, {Reed},
  {Baran}, {Nemeth}, \& {Kiaeerad}}]{otr+14}
{{\O}stensen}, R.~H., {Telting}, J.~H., {Reed}, M.~D., {et~al.} 2014, \aap,
  569, A15, \dodoi{10.1051/0004-6361/201423611}

\bibitem[{{{\O}stensen} {et~al.}(2010){{\O}stensen}, {Green}, {Bloemen},
  {Marsh}, {Laird}, {Morris}, {Moriyama}, {Oreiro}, {Reed}, {Kawaler}, {Aerts},
  {Vu{\v{c}}kovi{\'c}}, {Degroote}, {Telting}, {Kjeldsen}, {Gilliland},
  {Christensen-Dalsgaard}, {Borucki}, \& {Koch}}]{ogb+10}
{{\O}stensen}, R.~H., {Green}, E.~M., {Bloemen}, S., {et~al.} 2010, \mnras,
  408, L51, \dodoi{10.1111/j.1745-3933.2010.00926.x}

\bibitem[{{{\O}stensen} {et~al.}(2013){{\O}stensen}, {Geier}, {Schaffenroth},
  {Telting}, {Bloemen}, {N{\'e}meth}, {Beck}, {Lombaert}, {P{\'a}pics},
  {Tillich}, {Ziegerer}, {Fox Machado}, {Littlefair}, {Dhillon}, {Aerts},
  {Heber}, {Maxted}, {G{\"a}nsicke}, \& {Marsh}}]{ogs+13}
{{\O}stensen}, R.~H., {Geier}, S., {Schaffenroth}, V., {et~al.} 2013, \aap,
  559, A35, \dodoi{10.1051/0004-6361/201322022}

\bibitem[{{O'Toole} {et~al.}(2006){O'Toole}, {Napiwotzki}, {Heber}, {Drechsel},
  {Frandsen}, {Grundahl}, \& {Bruntt}}]{onh+06}
{O'Toole}, S.~J., {Napiwotzki}, R., {Heber}, U., {et~al.} 2006, Baltic
  Astronomy, 15, 61.
\newblock \doarXiv{astro-ph/0605441}

\bibitem[{{Pablo} {et~al.}(2011){Pablo}, {Kawaler}, \& {Green}}]{pkg11}
{Pablo}, H., {Kawaler}, S.~D., \& {Green}, E.~M. 2011, \apjl, 740, L47,
  \dodoi{10.1088/2041-8205/740/2/L47}

\bibitem[{{Paczynski}(1976)}]{pac76}
{Paczynski}, B. 1976, in IAU Symposium, Vol.~73, Structure and Evolution of
  Close Binary Systems, ed. P.~{Eggleton}, S.~{Mitton}, \& J.~{Whelan}, 75

\bibitem[{{Pakmor} {et~al.}(2010){Pakmor}, {Kromer}, {R{\"o}pke}, {Sim},
  {Ruiter}, \& {Hillebrandt}}]{pkr2010}
{Pakmor}, R., {Kromer}, M., {R{\"o}pke}, F.~K., {et~al.} 2010, \nat, 463, 61,
  \dodoi{10.1038/nature08642}

\bibitem[{{Pala} {et~al.}(2018){Pala}, {Schmidtobreick}, {Tappert},
  {G{\"a}nsicke}, \& {Mehner}}]{pst+18}
{Pala}, A.~F., {Schmidtobreick}, L., {Tappert}, C., {G{\"a}nsicke}, B.~T., \&
  {Mehner}, A. 2018, \mnras, 481, 2523, \dodoi{10.1093/mnras/sty2434}

\bibitem[{{Pallanca} {et~al.}(2013){Pallanca}, {Lanzoni}, {Dalessandro},
  {Ferraro}, {Possenti}, {Salaris}, \& {Burgay}}]{pld+13}
{Pallanca}, C., {Lanzoni}, B., {Dalessandro}, E., {et~al.} 2013, \apj, 773,
  127, \dodoi{10.1088/0004-637X/773/2/127}

\bibitem[{{Pandel} {et~al.}(2002){Pandel}, {Cordova}, {Shirey}, {Ramsay},
  {Cropper}, {Mason}, {Much}, \& {Kilkenny}}]{pcs+02}
{Pandel}, D., {Cordova}, F.~A., {Shirey}, R.~E., {et~al.} 2002, \mnras, 332,
  116, \dodoi{10.1046/j.1365-8711.2002.05279.x}

\bibitem[{{Parsons} {et~al.}(2021){Parsons}, {G{\"a}nsicke}, {Schreiber},
  {Marsh}, {Ashley}, {Breedt}, {Littlefair}, \& {Meusinger}}]{pgs+21}
{Parsons}, S.~G., {G{\"a}nsicke}, B.~T., {Schreiber}, M.~R., {et~al.} 2021,
  \mnras, \dodoi{10.1093/mnras/stab284}

\bibitem[{{Parsons} {et~al.}(2010{\natexlab{a}}){Parsons}, {Marsh},
  {Copperwheat}, {Dhillon}, {Littlefair}, {G{\"a}nsicke}, \&
  {Hickman}}]{pmc+10a}
{Parsons}, S.~G., {Marsh}, T.~R., {Copperwheat}, C.~M., {et~al.}
  2010{\natexlab{a}}, \mnras, 402, 2591,
  \dodoi{10.1111/j.1365-2966.2009.16072.x}

\bibitem[{{Parsons} {et~al.}(2010{\natexlab{b}}){Parsons}, {Marsh},
  {Copperwheat}, {Dhillon}, {Littlefair}, {Hickman}, {Maxted}, {G{\"a}nsicke},
  {Unda-Sanzana}, {Colque}, {Barraza}, {S{\'a}nchez}, \& {Monard}}]{pmc+10b}
---. 2010{\natexlab{b}}, \mnras, 407, 2362,
  \dodoi{10.1111/j.1365-2966.2010.17063.x}

\bibitem[{{Parsons} {et~al.}(2012{\natexlab{a}}){Parsons}, {Marsh},
  {G{\"a}nsicke}, {Rebassa-Mansergas}, {Dhillon}, {Littlefair}, {Copperwheat},
  {Hickman}, {Burleigh}, {Kerry}, {Koester}, {Nebot G{\'o}mez-Mor{\'a}n},
  {Pyrzas}, {Savoury}, {Schreiber}, {Schmidtobreick}, {Schwope}, {Steele}, \&
  {Tappert}}]{pmg+12}
{Parsons}, S.~G., {Marsh}, T.~R., {G{\"a}nsicke}, B.~T., {et~al.}
  2012{\natexlab{a}}, \mnras, 420, 3281,
  \dodoi{10.1111/j.1365-2966.2011.20251.x}

\bibitem[{{Parsons} {et~al.}(2012{\natexlab{b}}){Parsons}, {Marsh},
  {G{\"a}nsicke}, {Dhillon}, {Copperwheat}, {Littlefair}, {Pyrzas}, {Drake},
  {Koester}, {Schreiber}, \& {Rebassa-Mansergas}}]{pmgd+12}
---. 2012{\natexlab{b}}, \mnras, 419, 304,
  \dodoi{10.1111/j.1365-2966.2011.19691.x}

\bibitem[{{Parsons} {et~al.}(2013){Parsons}, {G{\"a}nsicke}, {Marsh}, {Drake},
  {Dhillon}, {Littlefair}, {Pyrzas}, {Rebassa-Mansergas}, \&
  {Schreiber}}]{pgm+13}
{Parsons}, S.~G., {G{\"a}nsicke}, B.~T., {Marsh}, T.~R., {et~al.} 2013, \mnras,
  429, 256, \dodoi{10.1093/mnras/sts332}

\bibitem[{{Parsons} {et~al.}(2015){Parsons}, {Schreiber}, {G{\"a}nsicke},
  {Rebassa-Mansergas}, {Brahm}, {Zorotovic}, {Toloza}, {Pala}, {Tappert},
  {Bayo}, \& {Jord{\'a}n}}]{psg+15}
{Parsons}, S.~G., {Schreiber}, M.~R., {G{\"a}nsicke}, B.~T., {et~al.} 2015,
  \mnras, 452, 1754, \dodoi{10.1093/mnras/stv1395}

\bibitem[{{Parsons} {et~al.}(2017){Parsons}, {Hermes}, {Marsh}, {G{\"a}nsicke},
  {Tremblay}, {Littlefair}, {Sahman}, {Ashley}, {Green}, {Rattanasoon},
  {Dhillon}, {Burleigh}, {Casewell}, {Buckley}, {Braker}, {Irawati}, {Dennihy},
  {Rodr{\'\i}guez-Gil}, {Winget}, {Winget}, {Bell}, \& {Kilic}}]{phm+17}
{Parsons}, S.~G., {Hermes}, J.~J., {Marsh}, T.~R., {et~al.} 2017, \mnras, 471,
  976, \dodoi{10.1093/mnras/stx1610}

\bibitem[{{Passy} {et~al.}(2012){Passy}, {De Marco}, {Fryer}, {Herwig},
  {Diehl}, {Oishi}, {Mac Low}, {Bryan}, \& {Rockefeller}}]{pdf+12}
{Passy}, J.-C., {De Marco}, O., {Fryer}, C.~L., {et~al.} 2012, \apj, 744, 52,
  \dodoi{10.1088/0004-637X/744/1/52}

\bibitem[{{Patterson} {et~al.}(2002){Patterson}, {Fried}, {Rea}, {Kemp},
  {Espaillat}, {Skillman}, {Harvey}, {O'Donoghue}, {McCormick}, {Velthuis},
  {Walker}, {Retter}, {Lipkin}, {Butterworth}, {McGee}, \& {Cook}}]{pfr+02}
{Patterson}, J., {Fried}, R.~E., {Rea}, R., {et~al.} 2002, \pasp, 114, 65,
  \dodoi{10.1086/339450}

\bibitem[{{Patterson} {et~al.}(2005){Patterson}, {Kemp}, {Harvey}, {Fried},
  {Rea}, {Monard}, {Cook}, {Skillman}, {Vanmunster}, {Bolt}, {Armstrong},
  {McCormick}, {Krajci}, {Jensen}, {Gunn}, {Butterworth}, {Foote}, {Bos},
  {Masi}, \& {Warhurst}}]{pkh+05}
{Patterson}, J., {Kemp}, J., {Harvey}, D.~A., {et~al.} 2005, \pasp, 117, 1204,
  \dodoi{10.1086/447771}

\bibitem[{{Pavlenko} {et~al.}(2021){Pavlenko}, {Kato}, {Antonyuk}, {Pit},
  {Keir}, {Udovichenko}, {Dubovsk'y}, {Sosnovskij}, {Antonyuk}, {Shimansky},
  {Gabdeev}, {Rakhmatullaeva}, {Kokhirova}, {Belan}, {Simon}, {Baklanov},
  {Kojiguchi}, \& {Godunova}}]{pka+21}
{Pavlenko}, E., {Kato}, T., {Antonyuk}, K., {et~al.} 2021, arXiv e-prints,
  arXiv:2103.14369.
\newblock \doarXiv{2103.14369}

\bibitem[{{Penning}(1985)}]{pen85}
{Penning}, W.~R. 1985, \apj, 289, 300, \dodoi{10.1086/162889}

\bibitem[{{Peters} \& {Thorstensen}(2006)}]{pt06}
{Peters}, C.~S., \& {Thorstensen}, J.~R. 2006, \pasp, 118, 687,
  \dodoi{10.1086/504641}

\bibitem[{{Peters}(1964)}]{pet64}
{Peters}, P.~C. 1964, Physical Review, 136, 1224,
  \dodoi{10.1103/PhysRev.136.B1224}

\bibitem[{{Pietrzy{\'n}ski} {et~al.}(2012){Pietrzy{\'n}ski}, {Thompson},
  {Gieren}, {Graczyk}, {St{\k{e}}pie{\'n}}, {Bono}, {Moroni}, {Pilecki},
  {Udalski}, {Soszy{\'n}ski}, {Preston}, {Nardetto}, {McWilliam}, {Roederer},
  {G{\'o}rski}, {Konorski}, \& {Storm}}]{ptg+12}
{Pietrzy{\'n}ski}, G., {Thompson}, I.~B., {Gieren}, W., {et~al.} 2012, \nat,
  484, 75, \dodoi{10.1038/nature10966}

\bibitem[{{Pol} {et~al.}(2019){Pol}, {McLaughlin}, \& {Lorimer}}]{pll+2019}
{Pol}, N., {McLaughlin}, M., \& {Lorimer}, D.~R. 2019, \apj, 870, 71,
  \dodoi{10.3847/1538-4357/aaf006}

\bibitem[{{Pollacco} \& {Bell}(1994)}]{pb94}
{Pollacco}, D.~L., \& {Bell}, S.~A. 1994, \mnras, 267, 452,
  \dodoi{10.1093/mnras/267.2.452}

\bibitem[{{Polubek} {et~al.}(2007){Polubek}, {Pigulski}, {Baran}, \&
  {Udalski}}]{ppb+07}
{Polubek}, G., {Pigulski}, A., {Baran}, A., \& {Udalski}, A. 2007, in
  Astronomical Society of the Pacific Conference Series, Vol. 372, 15th
  European Workshop on White Dwarfs, ed. R.~{Napiwotzki} \& M.~R. {Burleigh},
  487

\bibitem[{{Prodan} \& {Murray}(2015)}]{pm15}
{Prodan}, S., \& {Murray}, N. 2015, \apj, 798, 117,
  \dodoi{10.1088/0004-637X/798/2/117}

\bibitem[{{Provencal} {et~al.}(1997){Provencal}, {Winget}, {Nather},
  {Robinson}, {Clemens}, {Bradley}, {Claver}, {Kleinman}, {Grauer}, {Hine},
  {Ferrario}, {Warner}, {Vauclair}, {Chevreton}, {Kepler}, {Wood}, \&
  {Henry}}]{pwn+97}
{Provencal}, J.~L., {Winget}, D.~E., {Nather}, R.~E., {et~al.} 1997, \apj, 480,
  383, \dodoi{10.1086/303971}

\bibitem[{{Pyrzas} {et~al.}(2009){Pyrzas}, {G{\"a}nsicke}, {Marsh},
  {Aungwerojwit}, {Rebassa-Mansergas}, {Rodr{\'\i}guez-Gil}, {Southworth},
  {Schreiber}, {Nebot Gomez-Moran}, \& {Koester}}]{pgm+09}
{Pyrzas}, S., {G{\"a}nsicke}, B.~T., {Marsh}, T.~R., {et~al.} 2009, \mnras,
  394, 978, \dodoi{10.1111/j.1365-2966.2008.14378.x}

\bibitem[{{Pyrzas} {et~al.}(2012){Pyrzas}, {G{\"a}nsicke}, {Brady}, {Parsons},
  {Marsh}, {Koester}, {Breedt}, {Copperwheat}, {Nebot G{\'o}mez-Mor{\'a}n},
  {Rebassa-Mansergas}, {Schreiber}, \& {Zorotovic}}]{pgb+12}
{Pyrzas}, S., {G{\"a}nsicke}, B.~T., {Brady}, S., {et~al.} 2012, \mnras, 419,
  817, \dodoi{10.1111/j.1365-2966.2011.19746.x}

\bibitem[{{Qian} {et~al.}(2007){Qian}, {Dai}, {He}, {Yuan}, {Xiang}, \&
  {Zejda}}]{qdh+07}
{Qian}, S.~B., {Dai}, Z.~B., {He}, J.~J., {et~al.} 2007, \aap, 466, 589,
  \dodoi{10.1051/0004-6361:20065970}

\bibitem[{{Qian} {et~al.}(2009){Qian}, {Zhu}, {Zola}, {Liao}, {Liu}, {Li},
  {Winiarski}, {Kuligowska}, \& {Kreiner}}]{qzz+09}
{Qian}, S.~B., {Zhu}, L.~Y., {Zola}, S., {et~al.} 2009, \apjl, 695, L163,
  \dodoi{10.1088/0004-637X/695/2/L163}

\bibitem[{{Raghavan} {et~al.}(2010){Raghavan}, {McAlister}, {Henry}, {Latham},
  {Marcy}, {Mason}, {Gies}, {White}, \& {ten Brummelaar}}]{rmh+10}
{Raghavan}, D., {McAlister}, H.~A., {Henry}, T.~J., {et~al.} 2010, \apjs, 190,
  1, \dodoi{10.1088/0067-0049/190/1/1}

\bibitem[{{Ramsay} {et~al.}(2002){Ramsay}, {Hakala}, \& {Cropper}}]{rhc02}
{Ramsay}, G., {Hakala}, P., \& {Cropper}, M. 2002, \mnras, 332, L7,
  \dodoi{10.1046/j.1365-8711.2002.05471.x}

\bibitem[{{Ransom} {et~al.}(2014){Ransom}, {Stairs}, {Archibald}, {Hessels},
  {Kaplan}, {van Kerkwijk}, {Boyles}, {Deller}, {Chatterjee},
  {Schechtman-Rook}, {Berndsen}, {Lynch}, {Lorimer}, {Karako-Argaman}, {Kaspi},
  {Kondratiev}, {McLaughlin}, {van Leeuwen}, {Rosen}, {Roberts}, \&
  {Stovall}}]{rsa+14}
{Ransom}, S.~M., {Stairs}, I.~H., {Archibald}, A.~M., {et~al.} 2014, \nat, 505,
  520, \dodoi{10.1038/nature12917}

\bibitem[{{Rappaport} {et~al.}(2015){Rappaport}, {Nelson}, {Levine},
  {Sanchis-Ojeda}, {Gandolfi}, {Nowak}, {Palle}, \& {Prsa}}]{rnl+15}
{Rappaport}, S., {Nelson}, L., {Levine}, A., {et~al.} 2015, \apj, 803, 82,
  \dodoi{10.1088/0004-637X/803/2/82}

\bibitem[{{Rappaport} {et~al.}(2009){Rappaport}, {Podsiadlowski}, \&
  {Horev}}]{rph09}
{Rappaport}, S., {Podsiadlowski}, P., \& {Horev}, I. 2009, \apj, 698, 666,
  \dodoi{10.1088/0004-637X/698/1/666}

\bibitem[{{Ratti} {et~al.}(2013{\natexlab{a}}){Ratti}, {van Grunsven},
  {Jonker}, {Britt}, {Hynes}, {Steeghs}, {Greiss}, {Torres}, {Maccarone},
  {Groot}, {Knigge}, {Gossen}, {Mikles}, {Villar}, \& {Collazzi}}]{rvj+13}
{Ratti}, E.~M., {van Grunsven}, T.~F.~J., {Jonker}, P.~G., {et~al.}
  2013{\natexlab{a}}, \mnras, 428, 3543, \dodoi{10.1093/mnras/sts292}

\bibitem[{{Ratti} {et~al.}(2013{\natexlab{b}}){Ratti}, {van Grunsven},
  {Torres}, {Jonker}, {Miller-Jones}, {Hessels}, {Van Winckel}, {van der
  Sluys}, \& {Nelemans}}]{rvt+13}
{Ratti}, E.~M., {van Grunsven}, T.~F.~J., {Torres}, M.~A.~P., {et~al.}
  2013{\natexlab{b}}, \mnras, 431, L10, \dodoi{10.1093/mnrasl/sls052}

\bibitem[{{Reardon} {et~al.}(2016){Reardon}, {Hobbs}, {Coles}, {Levin},
  {Keith}, {Bailes}, {Bhat}, {Burke-Spolaor}, {Dai}, {Kerr}, {Lasky},
  {Manchester}, {Os{\l}owski}, {Ravi}, {Shannon}, {van Straten}, {Toomey},
  {Wang}, {Wen}, {You}, \& {Zhu}}]{rhc+16}
{Reardon}, D.~J., {Hobbs}, G., {Coles}, W., {et~al.} 2016, \mnras, 455, 1751,
  \dodoi{10.1093/mnras/stv2395}

\bibitem[{{Rebassa-Mansergas} {et~al.}(2014){Rebassa-Mansergas}, {Parsons},
  {Copperwheat}, {Justham}, {G{\"a}nsicke}, {Schreiber}, {Marsh}, \&
  {Dhillon}}]{rpc+14}
{Rebassa-Mansergas}, A., {Parsons}, S.~G., {Copperwheat}, C.~M., {et~al.} 2014,
  \apj, 790, 28, \dodoi{10.1088/0004-637X/790/1/28}

\bibitem[{{Rebassa-Mansergas} {et~al.}(2017){Rebassa-Mansergas}, {Parsons},
  {Garc{\'\i}a-Berro}, {G{\"a}nsicke}, {Schreiber}, {Rybicka}, \&
  {Koester}}]{rpg+17}
{Rebassa-Mansergas}, A., {Parsons}, S.~G., {Garc{\'\i}a-Berro}, E., {et~al.}
  2017, \mnras, 466, 1575, \dodoi{10.1093/mnras/stw3155}

\bibitem[{{Rebassa-Mansergas} {et~al.}(2008){Rebassa-Mansergas},
  {G{\"a}nsicke}, {Schreiber}, {Southworth}, {Schwope}, {Nebot Gomez-Moran},
  {Aungwerojwit}, {Rodr{\'\i}guez-Gil}, {Karamanavis}, {Krumpe}, {Tremou},
  {Schwarz}, {Staude}, \& {Vogel}}]{rgs+08}
{Rebassa-Mansergas}, A., {G{\"a}nsicke}, B.~T., {Schreiber}, M.~R., {et~al.}
  2008, \mnras, 390, 1635, \dodoi{10.1111/j.1365-2966.2008.13850.x}

\bibitem[{{Rebassa-Mansergas} {et~al.}(2012){Rebassa-Mansergas}, {Zorotovic},
  {Schreiber}, {G{\"a}nsicke}, {Southworth}, {Nebot G{\'o}mez-Mor{\'a}n},
  {Tappert}, {Koester}, {Pyrzas}, {Papadaki}, {Schmidtobreick}, {Schwope}, \&
  {Toloza}}]{rzs+12}
{Rebassa-Mansergas}, A., {Zorotovic}, M., {Schreiber}, M.~R., {et~al.} 2012,
  \mnras, 423, 320, \dodoi{10.1111/j.1365-2966.2012.20880.x}

\bibitem[{{Rebassa-Mansergas} {et~al.}(2021){Rebassa-Mansergas}, {Maldonado},
  {Raddi}, {Knowles}, {Torres}, {Hoskin}, {Cunningham}, {Hollands}, {Ren},
  {Gaensicke}, {Tremblay}, {Castro-Rodriguez}, {Camisassa}, \&
  {Koester}}]{rmr+21}
{Rebassa-Mansergas}, A., {Maldonado}, J., {Raddi}, R., {et~al.} 2021, arXiv
  e-prints, arXiv:2105.13379.
\newblock \doarXiv{2105.13379}

\bibitem[{{Reed} {et~al.}(2010){Reed}, {Terndrup}, {{\O}stensen}, {Geier},
  {Gilker}, {Beck}, {Degroote}, {Vanautgaerden}, \& {Waelkens}}]{rto+10}
{Reed}, M.~D., {Terndrup}, D.~M., {{\O}stensen}, R., {et~al.} 2010, \apss, 329,
  83, \dodoi{10.1007/s10509-010-0323-0}

\bibitem[{{Reimers} {et~al.}(1988){Reimers}, {Griffin}, \& {Brown}}]{rgb88}
{Reimers}, D., {Griffin}, R.~F., \& {Brown}, A. 1988, \aap, 193, 180

\bibitem[{{Ribeiro} \& {Baptista}(2011)}]{rb11}
{Ribeiro}, T., \& {Baptista}, R. 2011, \aap, 526, A150,
  \dodoi{10.1051/0004-6361/201015724}

\bibitem[{{Ricker} \& {Taam}(2012)}]{rt12}
{Ricker}, P.~M., \& {Taam}, R.~E. 2012, \apj, 746, 74,
  \dodoi{10.1088/0004-637X/746/1/74}

\bibitem[{{Ritter}(1988)}]{r1988}
{Ritter}, H. 1988, \aap, 202, 93

\bibitem[{{Ritter} \& {Kolb}(2003)}]{rk03}
{Ritter}, H., \& {Kolb}, U. 2003, \aap, 404, 301,
  \dodoi{10.1051/0004-6361:20030330}

\bibitem[{{Robinson}(1976)}]{r1976}
{Robinson}, E.~L. 1976, \araa, 14, 119,
  \dodoi{10.1146/annurev.aa.14.090176.001003}

\bibitem[{{Rodr{\'\i}guez-Gil} \& {Mart{\'\i}nez-Pais}(2002)}]{rm02}
{Rodr{\'\i}guez-Gil}, P., \& {Mart{\'\i}nez-Pais}, I.~G. 2002, \mnras, 337,
  209, \dodoi{10.1046/j.1365-8711.2002.05904.x}

\bibitem[{{Rodr{\'\i}guez-Gil} {et~al.}(2001){Rodr{\'\i}guez-Gil},
  {Mart{\'\i}nez-Pais}, {Casares}, {Villada}, \& {van Zyl}}]{rmc+01}
{Rodr{\'\i}guez-Gil}, P., {Mart{\'\i}nez-Pais}, I.~G., {Casares}, J.,
  {Villada}, M., \& {van Zyl}, L. 2001, \mnras, 328, 903,
  \dodoi{10.1046/j.1365-8711.2001.04965.x}

\bibitem[{{Rodr{\'\i}guez-Gil} {et~al.}(2009){Rodr{\'\i}guez-Gil}, {Torres},
  {G{\"a}nsicke}, {Mu{\~n}oz-Darias}, {Steeghs}, {Schwarz}, {Rau}, \&
  {Hagen}}]{rtg+09}
{Rodr{\'\i}guez-Gil}, P., {Torres}, M.~A.~P., {G{\"a}nsicke}, B.~T., {et~al.}
  2009, \aap, 496, 805, \dodoi{10.1051/0004-6361/200811312}

\bibitem[{{Rodr{\'\i}guez-Gil} {et~al.}(2015){Rodr{\'\i}guez-Gil}, {Shahbaz},
  {Marsh}, {G{\"a}nsicke}, {Steeghs}, {Long}, {Mart{\'\i}nez-Pais}, {Armas
  Padilla}, {Schwarz}, {Schreiber}, {Torres}, {Koester}, {Dhillon},
  {Castellano}, \& {Rodr{\'\i}guez}}]{rsm+15}
{Rodr{\'\i}guez-Gil}, P., {Shahbaz}, T., {Marsh}, T.~R., {et~al.} 2015, \mnras,
  452, 146, \dodoi{10.1093/mnras/stv1244}

\bibitem[{{Rodr{\'\i}guez-Gil} {et~al.}(2020){Rodr{\'\i}guez-Gil}, {Shahbaz},
  {Torres}, {G{\"a}nsicke}, {Izquierdo}, {Toloza}, {{\'A}lvarez-Hern{\'a}ndez},
  {Steeghs}, {van Spaandonk}, {Koester}, \& {Rodr{\'\i}guez}}]{rst+20}
{Rodr{\'\i}guez-Gil}, P., {Shahbaz}, T., {Torres}, M.~A.~P., {et~al.} 2020,
  \mnras, 494, 425, \dodoi{10.1093/mnras/staa612}

\bibitem[{{Roelofs} {et~al.}(2007{\natexlab{a}}){Roelofs}, {Groot}, {Benedict},
  {McArthur}, {Steeghs}, {Morales-Rueda}, {Marsh}, \& {Nelemans}}]{rgb+07}
{Roelofs}, G.~H.~A., {Groot}, P.~J., {Benedict}, G.~F., {et~al.}
  2007{\natexlab{a}}, \apj, 666, 1174, \dodoi{10.1086/520491}

\bibitem[{{Roelofs} {et~al.}(2005){Roelofs}, {Groot}, {Marsh}, {Steeghs},
  {Barros}, \& {Nelemans}}]{rgm+05}
{Roelofs}, G.~H.~A., {Groot}, P.~J., {Marsh}, T.~R., {et~al.} 2005, \mnras,
  361, 487, \dodoi{10.1111/j.1365-2966.2005.09186.x}

\bibitem[{{Roelofs} {et~al.}(2007{\natexlab{b}}){Roelofs}, {Groot}, {Nelemans},
  {Marsh}, \& {Steeghs}}]{rgn+07}
{Roelofs}, G.~H.~A., {Groot}, P.~J., {Nelemans}, G., {Marsh}, T.~R., \&
  {Steeghs}, D. 2007{\natexlab{b}}, \mnras, 379, 176,
  \dodoi{10.1111/j.1365-2966.2007.11931.x}

\bibitem[{{Roelofs} {et~al.}(2010){Roelofs}, {Rau}, {Marsh}, {Steeghs},
  {Groot}, \& {Nelemans}}]{rrm+10}
{Roelofs}, G. H.~A., {Rau}, A., {Marsh}, T.~R., {et~al.} 2010, \apjl, 711,
  L138, \dodoi{10.1088/2041-8205/711/2/L138}

\bibitem[{{Rolfe} {et~al.}(2000){Rolfe}, {Haswell}, \& {Patterson}}]{rhp00}
{Rolfe}, D.~J., {Haswell}, C.~A., \& {Patterson}, J. 2000, \mnras, 317, 759,
  \dodoi{10.1046/j.1365-8711.2000.03571.x}

\bibitem[{{Rosen} {et~al.}(1987){Rosen}, {Mason}, \& {Cordova}}]{rmc87}
{Rosen}, S.~R., {Mason}, K.~O., \& {Cordova}, F.~A. 1987, \mnras, 224, 987,
  \dodoi{10.1093/mnras/224.4.987}

\bibitem[{{Ruiter}(2020)}]{r2020}
{Ruiter}, A.~J. 2020, IAU Symposium, 357, 1, \dodoi{10.1017/S1743921320000587}

\bibitem[{{Ruiter} {et~al.}(2009){Ruiter}, {Belczynski}, \& {Fryer}}]{rbf09}
{Ruiter}, A.~J., {Belczynski}, K., \& {Fryer}, C. 2009, \apj, 699, 2026,
  \dodoi{10.1088/0004-637X/699/2/2026}

\bibitem[{{Ruiz} {et~al.}(2001){Ruiz}, {Rojo}, {Garay}, \& {Maza}}]{rrg+01}
{Ruiz}, M.~T., {Rojo}, P.~M., {Garay}, G., \& {Maza}, J. 2001, \apj, 552, 679,
  \dodoi{10.1086/320578}

\bibitem[{{Saffer} {et~al.}(1994){Saffer}, {Bergeron}, {Koester}, \&
  {Liebert}}]{sbk+94}
{Saffer}, R.~A., {Bergeron}, P., {Koester}, D., \& {Liebert}, J. 1994, \apj,
  432, 351, \dodoi{10.1086/174573}

\bibitem[{{Saffer} {et~al.}(1988){Saffer}, {Liebert}, \& {Olszewski}}]{slo88}
{Saffer}, R.~A., {Liebert}, J., \& {Olszewski}, E.~W. 1988, \apj, 334, 947,
  \dodoi{10.1086/166888}

\bibitem[{{Saffer} {et~al.}(1993){Saffer}, {Wade}, {Liebert}, {Green}, {Sion},
  {Bechtold}, {Foss}, \& {Kidder}}]{swl+93}
{Saffer}, R.~A., {Wade}, R.~A., {Liebert}, J., {et~al.} 1993, \aj, 105, 1945,
  \dodoi{10.1086/116569}

\bibitem[{{Sahman} {et~al.}(2013){Sahman}, {Dhillon}, {Marsh}, {Moll},
  {Thoroughgood}, {Watson}, \& {Littlefair}}]{sdm+13}
{Sahman}, D.~I., {Dhillon}, V.~S., {Marsh}, T.~R., {et~al.} 2013, \mnras, 433,
  1588, \dodoi{10.1093/mnras/stt830}

\bibitem[{{Salazar} {et~al.}(2017){Salazar}, {LeBleu}, {Schaefer}, {Landolt},
  \& {Dvorak}}]{sls+17}
{Salazar}, I.~V., {LeBleu}, A., {Schaefer}, B.~E., {Landolt}, A.~U., \&
  {Dvorak}, S. 2017, \mnras, 469, 4116, \dodoi{10.1093/mnras/stx1161}

\bibitem[{{Sand} {et~al.}(2020){Sand}, {Ohlmann}, {Schneider}, {Pakmor}, \&
  {R{\"o}pke}}]{sos+20}
{Sand}, C., {Ohlmann}, S.~T., {Schneider}, F. R.~N., {Pakmor}, R., \&
  {R{\"o}pke}, F.~K. 2020, \aap, 644, A60, \dodoi{10.1051/0004-6361/202038992}

\bibitem[{{Santander-Garc{\'\i}a} {et~al.}(2015){Santander-Garc{\'\i}a},
  {Rodr{\'\i}guez-Gil}, {Corradi}, {Jones}, {Miszalski}, {Boffin},
  {Rubio-D{\'\i}ez}, \& {Kotze}}]{src+15}
{Santander-Garc{\'\i}a}, M., {Rodr{\'\i}guez-Gil}, P., {Corradi}, R.~L.~M.,
  {et~al.} 2015, \nat, 519, 63, \dodoi{10.1038/nature14124}

\bibitem[{{Savoury} {et~al.}(2011){Savoury}, {Littlefair}, {Dhillon}, {Marsh},
  {G{\"a}nsicke}, {Copperwheat}, {Kerry}, {Hickman}, \& {Parsons}}]{sld+11}
{Savoury}, C.~D.~J., {Littlefair}, S.~P., {Dhillon}, V.~S., {et~al.} 2011,
  \mnras, 415, 2025, \dodoi{10.1111/j.1365-2966.2011.18707.x}

\bibitem[{{Scaringi} {et~al.}(2013){Scaringi}, {Groot}, \& {Still}}]{sgs13}
{Scaringi}, S., {Groot}, P.~J., \& {Still}, M. 2013, \mnras, 435, L68,
  \dodoi{10.1093/mnrasl/slt099}

\bibitem[{{Schaefer}(2020)}]{sch20}
{Schaefer}, B.~E. 2020, \mnras, 492, 3343, \dodoi{10.1093/mnras/stz3424}

\bibitem[{{Schaffenroth} {et~al.}(2015){Schaffenroth}, {Barlow}, {Drechsel}, \&
  {Dunlap}}]{sbd+15}
{Schaffenroth}, V., {Barlow}, B.~N., {Drechsel}, H., \& {Dunlap}, B.~H. 2015,
  \aap, 576, A123, \dodoi{10.1051/0004-6361/201525701}

\bibitem[{{Schaffenroth} {et~al.}(2014{\natexlab{a}}){Schaffenroth}, {Geier},
  {Barbu-Barna}, {Heber}, {Kupfer}, \& {Cordes}}]{sgb+14}
{Schaffenroth}, V., {Geier}, S., {Barbu-Barna}, I., {et~al.}
  2014{\natexlab{a}}, in Astronomical Society of the Pacific Conference Series,
  Vol. 481, 6th Meeting on Hot Subdwarf Stars and Related Objects, ed. V.~{van
  Grootel}, E.~{Green}, G.~{Fontaine}, \& S.~{Charpinet}, 253.
\newblock \doarXiv{1405.4714}

\bibitem[{{Schaffenroth} {et~al.}(2013){Schaffenroth}, {Geier}, {Drechsel},
  {Heber}, {Wils}, {{\O}stensen}, {Maxted}, \& {di Scala}}]{sgd+13}
{Schaffenroth}, V., {Geier}, S., {Drechsel}, H., {et~al.} 2013, \aap, 553, A18,
  \dodoi{10.1051/0004-6361/201220929}

\bibitem[{{Schaffenroth} {et~al.}(2014{\natexlab{b}}){Schaffenroth}, {Geier},
  {Heber}, {Kupfer}, {Ziegerer}, {Heuser}, {Classen}, \& {Cordes}}]{sgh+14}
{Schaffenroth}, V., {Geier}, S., {Heber}, U., {et~al.} 2014{\natexlab{b}},
  \aap, 564, A98, \dodoi{10.1051/0004-6361/201423377}

\bibitem[{{Schaffenroth} {et~al.}(2021){Schaffenroth}, {Casewell}, {Schneider},
  {Kilkenny}, {Geier}, {Heber}, {Irrgang}, {Przybilla}, {Marsh}, {Littlefair},
  \& {Dhillon}}]{scs+20}
{Schaffenroth}, V., {Casewell}, S.~L., {Schneider}, D., {et~al.} 2021, \mnras,
  501, 3847, \dodoi{10.1093/mnras/staa3661}

\bibitem[{{Schindewolf} {et~al.}(2015){Schindewolf}, {Levitan}, {Heber},
  {Drechsel}, {Schaffenroth}, {Kupfer}, \& {Prince}}]{slh+15}
{Schindewolf}, M., {Levitan}, D., {Heber}, U., {et~al.} 2015, \aap, 580, A117,
  \dodoi{10.1051/0004-6361/201425581}

\bibitem[{{Schmidt} {et~al.}(2007){Schmidt}, {Szkody}, {Henden}, {Anderson},
  {Lamb}, {Margon}, \& {Schneider}}]{ssh+07}
{Schmidt}, G.~D., {Szkody}, P., {Henden}, A., {et~al.} 2007, \apj, 654, 521,
  \dodoi{10.1086/509613}

\bibitem[{{Schmidt} {et~al.}(2005{\natexlab{a}}){Schmidt}, {Szkody},
  {Silvestri}, {Cushing}, {Liebert}, \& {Smith}}]{sss+05}
{Schmidt}, G.~D., {Szkody}, P., {Silvestri}, N.~M., {et~al.}
  2005{\natexlab{a}}, \apjl, 630, L173, \dodoi{10.1086/491702}

\bibitem[{{Schmidt} {et~al.}(2005{\natexlab{b}}){Schmidt}, {Szkody},
  {Vanlandingham}, {Anderson}, {Barentine}, {Brewington}, {Hall}, {Harvanek},
  {Kleinman}, {Krzesinski}, {Long}, {Margon}, {Neilsen}, {Newman}, {Nitta},
  {Schneider}, \& {Snedden}}]{ssv+05}
{Schmidt}, G.~D., {Szkody}, P., {Vanlandingham}, K.~M., {et~al.}
  2005{\natexlab{b}}, \apj, 630, 1037, \dodoi{10.1086/431969}

\bibitem[{{Schoembs} \& {Vogt}(1981)}]{sv81}
{Schoembs}, R., \& {Vogt}, N. 1981, \aap, 97, 185

\bibitem[{{Schreiber} {et~al.}(2008){Schreiber}, {G{\"a}nsicke}, {Southworth},
  {Schwope}, \& {Koester}}]{sgs+08}
{Schreiber}, M.~R., {G{\"a}nsicke}, B.~T., {Southworth}, J., {Schwope}, A.~D.,
  \& {Koester}, D. 2008, \aap, 484, 441, \dodoi{10.1051/0004-6361:20078765}

\bibitem[{{Schwope} {et~al.}(2011){Schwope}, {Horne}, {Steeghs}, \&
  {Still}}]{shs+11}
{Schwope}, A.~D., {Horne}, K., {Steeghs}, D., \& {Still}, M. 2011, \aap, 531,
  A34, \dodoi{10.1051/0004-6361/201016373}

\bibitem[{{Schwope} \& {Mengel}(1997)}]{sm97}
{Schwope}, A.~D., \& {Mengel}, S. 1997, Astronomische Nachrichten, 318, 25,
  \dodoi{10.1002/asna.2113180104}

\bibitem[{{Schwope} {et~al.}(1993){Schwope}, {Thomas}, {Beuermann}, \&
  {Reinsch}}]{stb+93}
{Schwope}, A.~D., {Thomas}, H.~C., {Beuermann}, K., \& {Reinsch}, K. 1993,
  \aap, 267, 103

\bibitem[{{Shafter}(1983)}]{sha83}
{Shafter}, A.~W. 1983, PhD thesis, California Univ., Los Angeles.

\bibitem[{{Shafter} \& {Hessman}(1988)}]{sh88}
{Shafter}, A.~W., \& {Hessman}, F.~V. 1988, \aj, 95, 178,
  \dodoi{10.1086/114626}

\bibitem[{{Shafter} \& {Holland}(2003)}]{sh03}
{Shafter}, A.~W., \& {Holland}, J.~N. 2003, \pasp, 115, 1105,
  \dodoi{10.1086/376934}

\bibitem[{{Shahbaz} \& {Wood}(1996)}]{sw96}
{Shahbaz}, T., \& {Wood}, J.~H. 1996, \mnras, 282, 362,
  \dodoi{10.1093/mnras/282.2.362}

\bibitem[{{Shimanskii}(2002)}]{shi02}
{Shimanskii}, V.~V. 2002, Astronomy Reports, 46, 127, \dodoi{10.1134/1.1451926}

\bibitem[{{Shimanskii} {et~al.}(2004){Shimanskii}, {Borisov}, {Sakhibullin}, \&
  {Surkov}}]{sbs+04}
{Shimanskii}, V.~V., {Borisov}, N.~V., {Sakhibullin}, N.~A., \& {Surkov}, A.~E.
  2004, Astronomy Reports, 48, 563, \dodoi{10.1134/1.1777274}

\bibitem[{{Shimanskii} {et~al.}(2012){Shimanskii}, {Yakin}, {Borisov}, \&
  {Bikmaev}}]{syb+12}
{Shimanskii}, V.~V., {Yakin}, D.~G., {Borisov}, N.~V., \& {Bikmaev}, I.~F.
  2012, Astronomy Reports, 56, 867, \dodoi{10.1134/S1063772912110066}

\bibitem[{{Shimansky} {et~al.}(2013){Shimansky}, {Borisov}, {Bikmaev},
  {Sakhibullin}, {Shimanskaya}, {Spiridonova}, \& {Irtuganov}}]{sbb+13}
{Shimansky}, V.~V., {Borisov}, N.~V., {Bikmaev}, I.~F., {et~al.} 2013,
  Astronomy Reports, 57, 212, \dodoi{10.1134/S1063772913030050}

\bibitem[{{Shimansky} {et~al.}(2012){Shimansky}, {Borisov}, {Nurtdinova},
  {Mitrofanova}, {Vlasyuk}, \& {Spiridonova}}]{sbn+12}
{Shimansky}, V.~V., {Borisov}, N.~V., {Nurtdinova}, D.~N., {et~al.} 2012,
  Astronomy Reports, 56, 441, \dodoi{10.1134/S1063772912050058}

\bibitem[{{Shimansky} {et~al.}(2015){Shimansky}, {Borisov}, {Nurtdinova},
  {Solovyeva}, {Sakhibullin}, \& {Spiridonova}}]{sbn+15}
---. 2015, Astronomy Reports, 59, 199, \dodoi{10.1134/S1063772915030063}

\bibitem[{{Shimansky} {et~al.}(2002){Shimansky}, {Borisov}, {Sakhibullin},
  {Suleimanov}, \& {Stupalov}}]{sbs+02}
{Shimansky}, V.~V., {Borisov}, N.~V., {Sakhibullin}, N.~A., {Suleimanov},
  V.~F., \& {Stupalov}, M.~S. 2002, Astronomy Reports, 46, 656,
  \dodoi{10.1134/1.1502226}

\bibitem[{{Shimansky} {et~al.}(2003){Shimansky}, {Borisov}, \&
  {Shimanskaya}}]{sbs03}
{Shimansky}, V.~V., {Borisov}, N.~V., \& {Shimanskaya}, N.~N. 2003, Astronomy
  Reports, 47, 763, \dodoi{10.1134/1.1611217}

\bibitem[{{Shimansky} {et~al.}(2008){Shimansky}, {Pozdnyakova}, {Borisov},
  {Bikmaev}, {Galeev}, {Sakhibullin}, \& {Spiridonova}}]{spb+08}
{Shimansky}, V.~V., {Pozdnyakova}, S.~A., {Borisov}, N.~V., {et~al.} 2008,
  Astronomy Letters, 34, 423, \dodoi{10.1134/S1063773708060078}

\bibitem[{{Shimansky} {et~al.}(2009){Shimansky}, {Pozdnyakova}, {Borisov},
  {Bikmaev}, {Vlasyuk}, {Spiridonova}, {Galeev}, \& {Mel'Nikov}}]{spb+09}
---. 2009, Astrophysical Bulletin, 64, 349, \dodoi{10.1134/S1990341309040051}

\bibitem[{{Silvotti} {et~al.}(2012){Silvotti}, {{\O}stensen}, {Bloemen},
  {Telting}, {Heber}, {Oreiro}, {Reed}, {Farris}, {O'Toole}, {Lanteri},
  {Degroote}, {Hu}, {Baran}, {Hermes}, {Althaus}, {Marsh}, {Charpinet}, {Li},
  {Morris}, \& {Sanderfer}}]{sob+12}
{Silvotti}, R., {{\O}stensen}, R.~H., {Bloemen}, S., {et~al.} 2012, \mnras,
  424, 1752, \dodoi{10.1111/j.1365-2966.2012.21232.x}

\bibitem[{{Silvotti} {et~al.}(2020){Silvotti}, {Schaffenroth}, {Heber},
  {{\O}stensen}, {Telting}, {Vos}, {Kilkenny}, {Mancini}, {Ciceri}, {Irrgang},
  \& {Drechsel}}]{ssh+20}
{Silvotti}, R., {Schaffenroth}, V., {Heber}, U., {et~al.} 2020, \mnras,
  \dodoi{10.1093/mnras/staa3332}

\bibitem[{{Sim} {et~al.}(2010){Sim}, {R{\"o}pke}, {Hillebrandt}, {Kromer},
  {Pakmor}, {Fink}, {Ruiter}, \& {Seitenzahl}}]{SRH2010}
{Sim}, S.~A., {R{\"o}pke}, F.~K., {Hillebrandt}, W., {et~al.} 2010, \apjl, 714,
  L52, \dodoi{10.1088/2041-8205/714/1/L52}

\bibitem[{{Simon} {et~al.}(1985){Simon}, {Fekel}, \& {Gibson}}]{sfg85}
{Simon}, T., {Fekel}, F.~C., J., \& {Gibson}, D.~M. 1985, \apj, 295, 153,
  \dodoi{10.1086/163360}

\bibitem[{{Sing} {et~al.}(2007){Sing}, {Green}, {Howell}, {Holberg},
  {Lopez-Morales}, {Shaw}, \& {Schmidt}}]{sgh+07}
{Sing}, D.~K., {Green}, E.~M., {Howell}, S.~B., {et~al.} 2007, \aap, 474, 951,
  \dodoi{10.1051/0004-6361:20078026}

\bibitem[{{Sing} {et~al.}(2004){Sing}, {Holberg}, {Burleigh}, {Good},
  {Barstow}, {Oswalt}, {Howell}, {Brinkworth}, {Rudkin}, {Johnston}, \&
  {Rafferty}}]{shb+04}
{Sing}, D.~K., {Holberg}, J.~B., {Burleigh}, M.~R., {et~al.} 2004, \aj, 127,
  2936, \dodoi{10.1086/386354}

\bibitem[{{Sion} {et~al.}(2001){Sion}, {Szkody}, {Gaensicke}, {Cheng}, {La
  Dous}, \& {Hassall}}]{ssg+01}
{Sion}, E.~M., {Szkody}, P., {Gaensicke}, B., {et~al.} 2001, \apj, 555, 834,
  \dodoi{10.1086/321529}

\bibitem[{{Skillman} {et~al.}(1999){Skillman}, {Patterson}, {Kemp}, {Harvey},
  {Fried}, {Retter}, {Lipkin}, \& {Vanmunster}}]{spk+99}
{Skillman}, D.~R., {Patterson}, J., {Kemp}, J., {et~al.} 1999, \pasp, 111,
  1281, \dodoi{10.1086/316437}

\bibitem[{{Skillman} {et~al.}(2002){Skillman}, {Krajci}, {Beshore},
  {Patterson}, {Kemp}, {Starkey}, {Oksanen}, {Vanmunster}, {Martin}, \&
  {Rea}}]{skb+02}
{Skillman}, D.~R., {Krajci}, T., {Beshore}, E., {et~al.} 2002, \pasp, 114, 630,
  \dodoi{10.1086/341692}

\bibitem[{{Smak} {et~al.}(2001){Smak}, {Belczynski}, \& {Zola}}]{sbz01}
{Smak}, J.~I., {Belczynski}, K., \& {Zola}, S. 2001, \actaa, 51, 117

\bibitem[{{Smith} {et~al.}(2006){Smith}, {Haswell}, \& {Hynes}}]{shh06}
{Smith}, A.~J., {Haswell}, C.~A., \& {Hynes}, R.~I. 2006, \mnras, 369, 1537,
  \dodoi{10.1111/j.1365-2966.2006.10409.x}

\bibitem[{{Smith} {et~al.}(1998){Smith}, {Dhillon}, \& {Marsh}}]{sdm98}
{Smith}, D.~A., {Dhillon}, V.~S., \& {Marsh}, T.~R. 1998, \mnras, 296, 465,
  \dodoi{10.1046/j.1365-8711.1998.00743.x}

\bibitem[{{Spogli} \& {Claudi}(1994)}]{sc94}
{Spogli}, C., \& {Claudi}, R.~U. 1994, \aap, 281, 808

\bibitem[{{Spruit} \& {Ritter}(1983)}]{sr1983}
{Spruit}, H.~C., \& {Ritter}, H. 1983, \aap, 124, 267

\bibitem[{{Staude} {et~al.}(2001){Staude}, {Schwope}, \& {Schwarz}}]{sss01}
{Staude}, A., {Schwope}, A.~D., \& {Schwarz}, R. 2001, \aap, 374, 588,
  \dodoi{10.1051/0004-6361:20010695}

\bibitem[{{Stauffer}(1987)}]{sta87}
{Stauffer}, J.~R. 1987, \aj, 94, 996, \dodoi{10.1086/114533}

\bibitem[{{Steeghs} {et~al.}(2007){Steeghs}, {Howell}, {Knigge},
  {G{\"a}nsicke}, {Sion}, \& {Welsh}}]{shk+07}
{Steeghs}, D., {Howell}, S.~B., {Knigge}, C., {et~al.} 2007, \apj, 667, 442,
  \dodoi{10.1086/520702}

\bibitem[{{Steeghs} {et~al.}(2003){Steeghs}, {Perryman}, {Reynolds}, {de
  Bruijne}, {Marsh}, {Dhillon}, \& {Peacock}}]{spr+03}
{Steeghs}, D., {Perryman}, M.~A.~C., {Reynolds}, A., {et~al.} 2003, \mnras,
  339, 810, \dodoi{10.1046/j.1365-8711.2003.06226.x}

\bibitem[{{Steele} {et~al.}(2011){Steele}, {Burleigh}, {Dobbie}, {Jameson},
  {Barstow}, \& {Satterthwaite}}]{sbd+11}
{Steele}, P.~R., {Burleigh}, M.~R., {Dobbie}, P.~D., {et~al.} 2011, \mnras,
  416, 2768, \dodoi{10.1111/j.1365-2966.2011.19225.x}

\bibitem[{{Steele} {et~al.}(2013){Steele}, {Saglia}, {Burleigh}, {Marsh},
  {G{\"a}nsicke}, {Lawrie}, {Cappetta}, {Girven}, \& {Napiwotzki}}]{ssb+13}
{Steele}, P.~R., {Saglia}, R.~P., {Burleigh}, M.~R., {et~al.} 2013, \mnras,
  429, 3492, \dodoi{10.1093/mnras/sts620}

\bibitem[{{Stefanov}(2021)}]{ste21}
{Stefanov}, S.~Y. 2021, arXiv e-prints, arXiv:2106.03568.
\newblock \doarXiv{2106.03568}

\bibitem[{{Stella} {et~al.}(1987){Stella}, {Priedhorsky}, \& {White}}]{spw87}
{Stella}, L., {Priedhorsky}, W., \& {White}, N.~E. 1987, \apjl, 312, L17,
  \dodoi{10.1086/184811}

\bibitem[{{Stevenson} {et~al.}(2017){Stevenson}, {Vigna-G{\'o}mez}, {Mandel},
  {Barrett}, {Neijssel}, {Perkins}, \& {de Mink}}]{svm+17}
{Stevenson}, S., {Vigna-G{\'o}mez}, A., {Mandel}, I., {et~al.} 2017, Nature
  Communications, 8, 14906, \dodoi{10.1038/ncomms14906}

\bibitem[{{Stroeer} {et~al.}(2007){Stroeer}, {Heber}, {Lisker}, {Napiwotzki},
  {Dreizler}, {Christlieb}, \& {Reimers}}]{shl+07}
{Stroeer}, A., {Heber}, U., {Lisker}, T., {et~al.} 2007, \aap, 462, 269,
  \dodoi{10.1051/0004-6361:20065564}

\bibitem[{{Strohmayer}(2005)}]{str05}
{Strohmayer}, T.~E. 2005, \apj, 627, 920, \dodoi{10.1086/430439}

\bibitem[{{Subebekova} {et~al.}(2020){Subebekova}, {Zharikov}, {Tovmassian},
  {Neustroev}, {Wolf}, {Hernandez}, {Ku{\v{c}}{\'a}kov{\'a}}, \&
  {Khokhlov}}]{szt+20}
{Subebekova}, G., {Zharikov}, S., {Tovmassian}, G., {et~al.} 2020, \mnras, 497,
  1475, \dodoi{10.1093/mnras/staa2091}

\bibitem[{{Szkody} \& {Ingram}(1994)}]{si94}
{Szkody}, P., \& {Ingram}, D. 1994, \apj, 420, 830, \dodoi{10.1086/173607}

\bibitem[{{Szkody} {et~al.}(2002){Szkody}, {Anderson}, {Ag{\"u}eros},
  {Covarrubias}, {Bentz}, {Hawley}, {Margon}, {Voges}, {Henden}, {Knapp},
  {Vanden Berk}, {Rest}, {Miknaitis}, {Magnier}, {Brinkmann}, {Csabai},
  {Harvanek}, {Hindsley}, {Hennessy}, {Ivezic}, {Kleinman}, {Lamb}, {Long},
  {Newman}, {Neilsen}, {Nichol}, {Nitta}, {Schneider}, {Snedden}, \&
  {York}}]{saa+2002}
{Szkody}, P., {Anderson}, S.~F., {Ag{\"u}eros}, M., {et~al.} 2002, \aj, 123,
  430, \dodoi{10.1086/324734}

\bibitem[{{Tappert} {et~al.}(2007){Tappert}, {G{\"a}nsicke}, {Schmidtobreick},
  {Aungwerojwit}, {Mennickent}, \& {Koester}}]{tgs+07}
{Tappert}, C., {G{\"a}nsicke}, B.~T., {Schmidtobreick}, L., {et~al.} 2007,
  \aap, 474, 205, \dodoi{10.1051/0004-6361:20077316}

\bibitem[{{Tappert} {et~al.}(2009){Tappert}, {G{\"a}nsicke}, {Zorotovic},
  {Toledo}, {Southworth}, {Papadaki}, \& {Mennickent}}]{tgz+09}
{Tappert}, C., {G{\"a}nsicke}, B.~T., {Zorotovic}, M., {et~al.} 2009, \aap,
  504, 491, \dodoi{10.1051/0004-6361/200912049}

\bibitem[{{Tappert} {et~al.}(2013){Tappert}, {Vogt}, {Schmidtobreick},
  {Ederoclite}, \& {Vanderbeke}}]{tvs+13}
{Tappert}, C., {Vogt}, N., {Schmidtobreick}, L., {Ederoclite}, A., \&
  {Vanderbeke}, J. 2013, \mnras, 431, 92, \dodoi{10.1093/mnras/stt139}

\bibitem[{{Tappert} {et~al.}(1997){Tappert}, {Wargau}, {Hanuschik}, \&
  {Vogt}}]{twh+97}
{Tappert}, C., {Wargau}, W.~F., {Hanuschik}, R.~W., \& {Vogt}, N. 1997, \aap,
  327, 231

\bibitem[{{Telting} {et~al.}(2012){Telting}, {{\O}stensen}, {Baran}, {Bloemen},
  {Reed}, {Oreiro}, {Farris}, {Ottosen}, {Aerts}, {Kawaler}, {Heber}, {Prins},
  {Green}, {Kalomeni}, {O'Toole}, {Mullally}, {Sanderfer}, {Smith}, \&
  {Kjeldsen}}]{tob+12}
{Telting}, J.~H., {{\O}stensen}, R.~H., {Baran}, A.~S., {et~al.} 2012, \aap,
  544, A1, \dodoi{10.1051/0004-6361/201219458}

\bibitem[{{Telting} {et~al.}(2014){Telting}, {Baran}, {Nemeth}, {{\O}stensen},
  {Kupfer}, {Macfarlane}, {Heber}, {Aerts}, \& {Geier}}]{tbn+14}
{Telting}, J.~H., {Baran}, A.~S., {Nemeth}, P., {et~al.} 2014, \aap, 570, A129,
  \dodoi{10.1051/0004-6361/201424169}

\bibitem[{{Thompson}(2011)}]{tho11}
{Thompson}, T.~A. 2011, \apj, 741, 82, \dodoi{10.1088/0004-637X/741/2/82}

\bibitem[{{Thoroughgood} {et~al.}(2001){Thoroughgood}, {Dhillon}, {Littlefair},
  {Marsh}, \& {Smith}}]{tdl+01}
{Thoroughgood}, T.~D., {Dhillon}, V.~S., {Littlefair}, S.~P., {Marsh}, T.~R.,
  \& {Smith}, D.~A. 2001, \mnras, 327, 1323,
  \dodoi{10.1046/j.1365-8711.2001.04828.x}

\bibitem[{{Thoroughgood} {et~al.}(2004){Thoroughgood}, {Dhillon}, {Watson},
  {Buckley}, {Steeghs}, \& {Stevenson}}]{tdw+04}
{Thoroughgood}, T.~D., {Dhillon}, V.~S., {Watson}, C.~A., {et~al.} 2004,
  \mnras, 353, 1135, \dodoi{10.1111/j.1365-2966.2004.08135.x}

\bibitem[{{Thoroughgood} {et~al.}(2005){Thoroughgood}, {Dhillon}, {Steeghs},
  {Watson}, {Buckley}, {Littlefair}, {Smith}, {Still}, {van der Heyden}, \&
  {Warner}}]{tds+05}
{Thoroughgood}, T.~D., {Dhillon}, V.~S., {Steeghs}, D., {et~al.} 2005, \mnras,
  357, 881, \dodoi{10.1111/j.1365-2966.2004.08613.x}

\bibitem[{{Thorsett} {et~al.}(1993){Thorsett}, {Arzoumanian}, {McKinnon}, \&
  {Taylor}}]{tam+93}
{Thorsett}, S.~E., {Arzoumanian}, Z., {McKinnon}, M.~M., \& {Taylor}, J.~H.
  1993, \apjl, 405, L29, \dodoi{10.1086/186758}

\bibitem[{{Thorstensen} {et~al.}(2004){Thorstensen}, {Fenton}, \&
  {Taylor}}]{tft04}
{Thorstensen}, J.~R., {Fenton}, W.~H., \& {Taylor}, C.~J. 2004, \pasp, 116,
  300, \dodoi{10.1086/382792}

\bibitem[{{Tokovinin}(1997)}]{tok97}
{Tokovinin}, A.~A. 1997, \aaps, 121, 71, \dodoi{10.1051/aas:1997114}

\bibitem[{{Toonen} {et~al.}(2012){Toonen}, {Nelemans}, \& {Portegies
  Zwart}}]{tnp12}
{Toonen}, S., {Nelemans}, G., \& {Portegies Zwart}, S. 2012, \aap, 546, A70,
  \dodoi{10.1051/0004-6361/201218966}

\bibitem[{{Tovmassian} {et~al.}(2014){Tovmassian}, {Stephania Hernandez},
  {Gonz{\'a}lez-Buitrago}, {Zharikov}, \& {Garc{\'\i}a-D{\'\i}az}}]{tsg+14}
{Tovmassian}, G., {Stephania Hernandez}, M., {Gonz{\'a}lez-Buitrago}, D.,
  {Zharikov}, S., \& {Garc{\'\i}a-D{\'\i}az}, M.~T. 2014, \aj, 147, 68,
  \dodoi{10.1088/0004-6256/147/3/68}

\bibitem[{{Tovmassian} {et~al.}(2010){Tovmassian}, {Yungelson}, {Rauch},
  {Suleimanov}, {Napiwotzki}, {Stasi{\'n}ska}, {Tomsick}, {Wilms}, {Morisset},
  {Pe{\~n}a}, \& {Richer}}]{tyr+10}
{Tovmassian}, G., {Yungelson}, L., {Rauch}, T., {et~al.} 2010, \apj, 714, 178,
  \dodoi{10.1088/0004-637X/714/1/178}

\bibitem[{{Tovmassian} {et~al.}(2004){Tovmassian}, {Napiwotzki}, {Richer},
  {Stasi{\'n}ska}, {Fullerton}, \& {Rauch}}]{tnr+04}
{Tovmassian}, G.~H., {Napiwotzki}, R., {Richer}, M.~G., {et~al.} 2004, \apj,
  616, 485, \dodoi{10.1086/424795}

\bibitem[{{Tovmassian} {et~al.}(1999){Tovmassian}, {Szkody}, {Greiner},
  {Vrielmann}, {Kroll}, {Howell}, {Saxton}, {Ciardi}, {Mason}, \&
  {Hastings}}]{tsg+99}
{Tovmassian}, G.~H., {Szkody}, P., {Greiner}, J., {et~al.} 1999, in
  Astronomical Society of the Pacific Conference Series, Vol. 157, Annapolis
  Workshop on Magnetic Cataclysmic Variables, ed. C.~{Hellier} \& K.~{Mukai},
  133.
\newblock \doarXiv{astro-ph/9809306}

\bibitem[{{Tutukov} \& {Yungelson}(1979)}]{ty1979}
{Tutukov}, A.~V., \& {Yungelson}, L.~R. 1979, \actaa, 29, 665

\bibitem[{{Tutukov} \& {Yungelson}(1994)}]{ty1994}
---. 1994, \mnras, 268, 871, \dodoi{10.1093/mnras/268.4.871}

\bibitem[{{Uthas} {et~al.}(2010){Uthas}, {Knigge}, \& {Steeghs}}]{uks10}
{Uthas}, H., {Knigge}, C., \& {Steeghs}, D. 2010, \mnras, 409, 237,
  \dodoi{10.1111/j.1365-2966.2010.17046.x}

\bibitem[{{Vaccaro} \& {Wilson}(2003)}]{vw03}
{Vaccaro}, T.~R., \& {Wilson}, R.~E. 2003, \mnras, 342, 564,
  \dodoi{10.1046/j.1365-8711.2003.06565.x}

\bibitem[{{Vaccaro} {et~al.}(2015){Vaccaro}, {Wilson}, {Van Hamme}, \&
  {Terrell}}]{vwv+15}
{Vaccaro}, T.~R., {Wilson}, R.~E., {Van Hamme}, W., \& {Terrell}, D. 2015,
  \apj, 810, 157, \dodoi{10.1088/0004-637X/810/2/157}

\bibitem[{{van den Besselaar} {et~al.}(2007){van den Besselaar}, {Greimel},
  {Morales-Rueda}, {Nelemans}, {Thorstensen}, {Marsh}, {Dhillon}, {Robb},
  {Balam}, {Guenther}, {Kemp}, {Augusteijn}, \& {Groot}}]{vgm+07}
{van den Besselaar}, E.~J.~M., {Greimel}, R., {Morales-Rueda}, L., {et~al.}
  2007, \aap, 466, 1031, \dodoi{10.1051/0004-6361:20066246}

\bibitem[{{Van Grootel} {et~al.}(2008){Van Grootel}, {Charpinet}, {Fontaine},
  \& {Brassard}}]{vcf+08}
{Van Grootel}, V., {Charpinet}, S., {Fontaine}, G., \& {Brassard}, P. 2008,
  \aap, 483, 875, \dodoi{10.1051/0004-6361:200809554}

\bibitem[{{Van Grootel} {et~al.}(2014){Van Grootel}, {Charpinet}, {Fontaine},
  {Brassard}, \& {Green}}]{vcf+14}
{Van Grootel}, V., {Charpinet}, S., {Fontaine}, G., {Brassard}, P., \& {Green},
  E.~M. 2014, in Precision Asteroseismology, ed. J.~A. {Guzik}, W.~J.
  {Chaplin}, G.~{Handler}, \& A.~{Pigulski}, Vol. 301, 305--308,
  \dodoi{10.1017/S174392131301449X}

\bibitem[{{van Kerkwijk} {et~al.}(2000){van Kerkwijk}, {Bell}, {Kaspi}, \&
  {Kulkarni}}]{vbk+00}
{van Kerkwijk}, M.~H., {Bell}, J.~F., {Kaspi}, V.~M., \& {Kulkarni}, S.~R.
  2000, \apjl, 530, L37, \dodoi{10.1086/312478}

\bibitem[{{van Kerkwijk} {et~al.}(1996){van Kerkwijk}, {Bergeron}, \&
  {Kulkarni}}]{vbk96}
{van Kerkwijk}, M.~H., {Bergeron}, P., \& {Kulkarni}, S.~R. 1996, \apjl, 467,
  L89, \dodoi{10.1086/310209}

\bibitem[{{van Kerkwijk} {et~al.}(2010){van Kerkwijk}, {Rappaport}, {Breton},
  {Justham}, {Podsiadlowski}, \& {Han}}]{vrb+10}
{van Kerkwijk}, M.~H., {Rappaport}, S.~A., {Breton}, R.~P., {et~al.} 2010,
  \apj, 715, 51, \dodoi{10.1088/0004-637X/715/1/51}

\bibitem[{{van Roestel} {et~al.}(2018){van Roestel}, {Kupfer}, {Ruiz-Carmona},
  {Groot}, {Prince}, {Burdge}, {Laher}, {Shupe}, \& {Bellm}}]{vkr+18}
{van Roestel}, J., {Kupfer}, T., {Ruiz-Carmona}, R., {et~al.} 2018, \mnras,
  475, 2560, \dodoi{10.1093/mnras/stx3291}

\bibitem[{{van Roestel} {et~al.}(2021){van Roestel}, {Kupfer}, {Bell},
  {Burdge}, {Mr{\'o}z}, {Prince}, {Bellm}, {Drake}, {Dekany}, {Mahabal},
  {Porter}, {Riddle}, {Shin}, \& {Shupe}}]{vkb+21}
{van Roestel}, J., {Kupfer}, T., {Bell}, K.~J., {et~al.} 2021, arXiv e-prints,
  arXiv:2105.08687.
\newblock \doarXiv{2105.08687}

\bibitem[{{van Spaandonk} {et~al.}(2010){van Spaandonk}, {Steeghs}, {Marsh}, \&
  {Parsons}}]{vsm+10}
{van Spaandonk}, L., {Steeghs}, D., {Marsh}, T.~R., \& {Parsons}, S.~G. 2010,
  \apjl, 715, L109, \dodoi{10.1088/2041-8205/715/2/L109}

\bibitem[{{van Straten} {et~al.}(2001){van Straten}, {Bailes}, {Britton},
  {Kulkarni}, {Anderson}, {Manchester}, \& {Sarkissian}}]{vbb+01}
{van Straten}, W., {Bailes}, M., {Britton}, M., {et~al.} 2001, \nat, 412, 158.
\newblock \doarXiv{astro-ph/0108254}

\bibitem[{{Vande Putte} {et~al.}(2003){Vande Putte}, {Smith}, {Hawkins}, \&
  {Martin}}]{vsh+03}
{Vande Putte}, D., {Smith}, R.~C., {Hawkins}, N.~A., \& {Martin}, J.~S. 2003,
  \mnras, 342, 151, \dodoi{10.1046/j.1365-8711.2003.06524.x}

\bibitem[{{Vanderburg} {et~al.}(2020){Vanderburg}, {Rappaport}, {Xu},
  {Crossfield}, {Becker}, {Gary}, {Murgas}, {Blouin}, {Kaye}, {Palle}, {Melis},
  {Morris}, {Kreidberg}, {Gorjian}, {Morley}, {Mann}, {Parviainen}, {Pearce},
  {Newton}, {Carrillo}, {Zuckerman}, {Nelson}, {Zeimann}, {Brown},
  {Tronsgaard}, {Klein}, {Ricker}, {Vanderspek}, {Latham}, {Seager}, {Winn},
  {Jenkins}, {Adams}, {Benneke}, {Berardo}, {Buchhave}, {Caldwell},
  {Christiansen}, {Collins}, {Col{\'o}n}, {Daylan}, {Doty}, {Doyle},
  {Dragomir}, {Dressing}, {Dufour}, {Fukui}, {Glidden}, {Guerrero}, {Guo},
  {Heng}, {Henriksen}, {Huang}, {Kaltenegger}, {Kane}, {Lewis}, {Lissauer},
  {Morales}, {Narita}, {Pepper}, {Rose}, {Smith}, {Stassun}, \& {Yu}}]{vrx+20}
{Vanderburg}, A., {Rappaport}, S.~A., {Xu}, S., {et~al.} 2020, \nat, 585, 363,
  \dodoi{10.1038/s41586-020-2713-y}

\bibitem[{{Vennes} {et~al.}(2012){Vennes}, {Kawka}, {O'Toole}, {N{\'e}meth}, \&
  {Burton}}]{vko+12}
{Vennes}, S., {Kawka}, A., {O'Toole}, S.~J., {N{\'e}meth}, P., \& {Burton}, D.
  2012, \apjl, 759, L25, \dodoi{10.1088/2041-8205/759/1/L25}

\bibitem[{{Vennes} {et~al.}(1999){Vennes}, {Thorstensen}, \&
  {Polomski}}]{vtp99}
{Vennes}, S., {Thorstensen}, J.~R., \& {Polomski}, E.~F. 1999, \apj, 523, 386,
  \dodoi{10.1086/307722}

\bibitem[{{Vennes} {et~al.}(2011){Vennes}, {Thorstensen}, {Kawka},
  {N{\'e}meth}, {Skinner}, {Pigulski}, {Ste\&{\textcommabelow s}acute},
  {licki}, {Ko{\l}aczkowski}, \& {{\'S}r{\'o}dka}}]{vtk+11}
{Vennes}, S., {Thorstensen}, J.~R., {Kawka}, A., {et~al.} 2011, \apjl, 737,
  L16, \dodoi{10.1088/2041-8205/737/1/L16}

\bibitem[{{Vogt}(1981)}]{vog81}
{Vogt}, N. 1981, {SU-UMA Sterne und andere Zwergnovae. Eine Untersuchung ihrer
  Eruptionsmechanismen, ihrer Struktur und entwicklungsgeschichtlichen Stellung
  unter den kataklysmischen Doppelsternen} (Bochum: Ruhr-Universitaet)

\bibitem[{{Voss} \& {Tauris}(2003)}]{vt03}
{Voss}, R., \& {Tauris}, T.~M. 2003, \mnras, 342, 1169,
  \dodoi{10.1046/j.1365-8711.2003.06616.x}

\bibitem[{{Vu{\v{c}}kovi{\'c}} {et~al.}(2007){Vu{\v{c}}kovi{\'c}}, {Aerts},
  {{\"O}stensen}, {Nelemans}, {Hu}, {Jeffery}, {Dhillon}, \& {Marsh}}]{vao+07}
{Vu{\v{c}}kovi{\'c}}, M., {Aerts}, C., {{\"O}stensen}, R., {et~al.} 2007, \aap,
  471, 605, \dodoi{10.1051/0004-6361:20077179}

\bibitem[{{Wade} \& {Horne}(1988)}]{wh88}
{Wade}, R.~A., \& {Horne}, K. 1988, \apj, 324, 411, \dodoi{10.1086/165905}

\bibitem[{{Wakamatsu} {et~al.}(2021){Wakamatsu}, {Thorstensen}, {Kojiguchi},
  {Isogai}, {Kimura}, {Ohnishi}, {Kato}, {Itoh}, {Sugiura}, {Sumiya},
  {Matsumoto}, {Ito}, {Nikai}, {Akitaya}, {Ishioka}, {Oide}, {Kanai}, {Uzawa},
  {Oasa}, {Tordai}, {Vanmunster}, {Shugarov}, {Yamanaka}, {Sasada}, {Takagi},
  {Nishinaka}, {Yamazaki}, {Otsubo}, {Nakaoka}, {Murata}, {Ohsawa}, {Morita},
  {Ichiki}, {Dufoer}, {Mizutani}, {Horiuchi}, {Tozuka}, {Takayama}, {Ohshima},
  {Saito}, {Dubovsky}, {Stone}, {Miller}, \& {Nogami}}]{wtk+21}
{Wakamatsu}, Y., {Thorstensen}, J.~R., {Kojiguchi}, N., {et~al.} 2021, \pasj,
  \dodoi{10.1093/pasj/psab003}

\bibitem[{{Wang} {et~al.}(2013){Wang}, {Justham}, \& {Han}}]{wjh2013}
{Wang}, B., {Justham}, S., \& {Han}, Z. 2013, \aap, 559, A94,
  \dodoi{10.1051/0004-6361/201322298}

\bibitem[{{Wang} {et~al.}(2018){Wang}, {Luo}, {Zhang}, {Zhang}, {Deng}, \&
  {Luo}}]{wlz+18}
{Wang}, K., {Luo}, C., {Zhang}, X., {et~al.} 2018, \aj, 156, 187,
  \dodoi{10.3847/1538-3881/aade52}

\bibitem[{{Wang} \& {Chakrabarty}(2004)}]{wc04}
{Wang}, Z., \& {Chakrabarty}, D. 2004, \apjl, 616, L139, \dodoi{10.1086/426787}

\bibitem[{{Warner} \& {Thackeray}(1975)}]{wt75}
{Warner}, B., \& {Thackeray}, A.~D. 1975, \mnras, 172, 433,
  \dodoi{10.1093/mnras/172.2.433}

\bibitem[{{Watson} {et~al.}(2003){Watson}, {Dhillon}, {Rutten}, \&
  {Schwope}}]{wdr+03}
{Watson}, C.~A., {Dhillon}, V.~S., {Rutten}, R.~G.~M., \& {Schwope}, A.~D.
  2003, \mnras, 341, 129, \dodoi{10.1046/j.1365-8711.2003.06381.x}

\bibitem[{{Watson} {et~al.}(2007){Watson}, {Steeghs}, {Shahbaz}, \&
  {Dhillon}}]{wss+07}
{Watson}, C.~A., {Steeghs}, D., {Shahbaz}, T., \& {Dhillon}, V.~S. 2007,
  \mnras, 382, 1105, \dodoi{10.1111/j.1365-2966.2007.12173.x}

\bibitem[{{Webbink}(1984)}]{w1984}
{Webbink}, R.~F. 1984, \apj, 277, 355, \dodoi{10.1086/161701}

\bibitem[{{Welsh} {et~al.}(2007){Welsh}, {Froning}, {Marsh}, {Reimer},
  {Robinson}, \& {Wood}}]{wfm+07}
{Welsh}, W.~F., {Froning}, C.~S., {Marsh}, T.~R., {et~al.} 2007, in
  Astronomical Society of the Pacific Conference Series, Vol. 362, The Seventh
  Pacific Rim Conference on Stellar Astrophysics, ed. Y.~W. {Kang}, H.~W.
  {Lee}, K.~C. {Leung}, \& K.~S. {Cheng}, 241

\bibitem[{{Wenger} {et~al.}(2000){Wenger}, {Ochsenbein}, {Egret}, {Dubois},
  {Bonnarel}, {Borde}, {Genova}, {Jasniewicz}, {Lalo{\"e}}, {Lesteven}, \&
  {Monier}}]{woe+00}
{Wenger}, M., {Ochsenbein}, F., {Egret}, D., {et~al.} 2000, \aaps, 143, 9,
  \dodoi{10.1051/aas:2000332}

\bibitem[{{Wevers} {et~al.}(2016){Wevers}, {Torres}, {Jonker}, {Wetuski},
  {Nelemans}, {Steeghs}, {Maccarone}, {Heinke}, {Hynes}, {Udalski},
  {Kostrzewa-Rutkowska}, {Groot}, {Gazer}, {Szyma{\'n}ski}, {Britt},
  {Wyrzykowski}, \& {Poleski}}]{wtj+16}
{Wevers}, T., {Torres}, M.~A.~P., {Jonker}, P.~G., {et~al.} 2016, \mnras, 462,
  L106, \dodoi{10.1093/mnrasl/slw141}

\bibitem[{{Whelan} \& {Iben}(1973)}]{wi1973}
{Whelan}, J., \& {Iben}, Icko, J. 1973, \apj, 186, 1007, \dodoi{10.1086/152565}

\bibitem[{{Wolff} {et~al.}(1999){Wolff}, {Wood}, {Imamura}, {Middleditch}, \&
  {Steiman-Cameron}}]{wwi+99}
{Wolff}, M.~T., {Wood}, K.~S., {Imamura}, J.~N., {Middleditch}, J., \&
  {Steiman-Cameron}, T.~Y. 1999, \apj, 526, 435, \dodoi{10.1086/307982}

\bibitem[{{Wood} {et~al.}(1989){Wood}, {Horne}, {Berriman}, \& {Wade}}]{whb+89}
{Wood}, J.~H., {Horne}, K., {Berriman}, G., \& {Wade}, R.~A. 1989, \apj, 341,
  974, \dodoi{10.1086/167557}

\bibitem[{{Wood} \& {Saffer}(1999)}]{ws99}
{Wood}, J.~H., \& {Saffer}, R. 1999, \mnras, 305, 820,
  \dodoi{10.1046/j.1365-8711.1999.02501.x}

\bibitem[{{Wood} {et~al.}(2002){Wood}, {Casey}, {Garnavich}, \&
  {Haag}}]{wcg+02}
{Wood}, M.~A., {Casey}, M.~J., {Garnavich}, P.~M., \& {Haag}, B. 2002, \mnras,
  334, 87, \dodoi{10.1046/j.1365-8711.2002.05484.x}

\bibitem[{{Woosley} \& {Kasen}(2011)}]{WK2011}
{Woosley}, S.~E., \& {Kasen}, D. 2011, \apj, 734, 38,
  \dodoi{10.1088/0004-637X/734/1/38}

\bibitem[{{Woosley} \& {Weaver}(1994)}]{WW1994}
{Woosley}, S.~E., \& {Weaver}, T.~A. 1994, \apj, 423, 371,
  \dodoi{10.1086/173813}

\bibitem[{{Wu} {et~al.}(2002){Wu}, {Li}, {Ding}, {Zhang}, \& {Li}}]{wld+02}
{Wu}, X., {Li}, Z., {Ding}, Y., {Zhang}, Z., \& {Li}, Z. 2002, \apj, 569, 418,
  \dodoi{10.1086/339278}

\bibitem[{{Yoon} {et~al.}(2010){Yoon}, {Woosley}, \& {Langer}}]{ywl10}
{Yoon}, S., {Woosley}, S.~E., \& {Langer}, N. 2010, \apj, 725, 940,
  \dodoi{10.1088/0004-637X/725/1/940}

\bibitem[{{Zahn} \& {Bouchet}(1989)}]{zb1989}
{Zahn}, J.~P., \& {Bouchet}, L. 1989, \aap, 223, 112

\bibitem[{{Zhang} \& {Jeffery}(2012)}]{zj2012}
{Zhang}, X., \& {Jeffery}, C.~S. 2012, \mnras, 426, L81,
  \dodoi{10.1111/j.1745-3933.2012.01330.x}

\bibitem[{{Zhang} {et~al.}(2016){Zhang}, {Fu}, {Li}, {Ren}, \& {Luo}}]{zfl+16}
{Zhang}, X.~B., {Fu}, J.~N., {Li}, Y., {Ren}, A.~B., \& {Luo}, C.~Q. 2016,
  \apjl, 821, L32, \dodoi{10.3847/2041-8205/821/2/L32}

\bibitem[{{Zhang} {et~al.}(2017){Zhang}, {Fu}, {Liu}, {Luo}, \& {Ren}}]{zfl+17}
{Zhang}, X.~B., {Fu}, J.~N., {Liu}, N., {Luo}, C.~Q., \& {Ren}, A.~B. 2017,
  \apj, 850, 125, \dodoi{10.3847/1538-4357/aa9577}

\bibitem[{{Zhu} {et~al.}(2011){Zhu}, {Qian}, {Liu}, {Liao}, {He}, {Li}, {Zhao},
  {Dai}, {Zhang}, \& {Li}}]{zql+11}
{Zhu}, L., {Qian}, S., {Liu}, L., {et~al.} 2011, in Astronomical Society of the
  Pacific Conference Series, Vol. 451, 9th Pacific Rim Conference on Stellar
  Astrophysics, ed. S.~{Qain}, K.~{Leung}, L.~{Zhu}, \& S.~{Kwok}, 155

\bibitem[{{Zhu} {et~al.}(2015){Zhu}, {Stairs}, {Demorest}, {Nice}, {Ellis},
  {Ransom}, {Arzoumanian}, {Crowter}, {Dolch}, {Ferdman}, {Fonseca},
  {Gonzalez}, {Jones}, {Jones}, {Lam}, {Levin}, {McLaughlin}, {Pennucci},
  {Stovall}, \& {Swiggum}}]{zsd+15}
{Zhu}, W.~W., {Stairs}, I.~H., {Demorest}, P.~B., {et~al.} 2015, \apj, 809, 41,
  \dodoi{10.1088/0004-637X/809/1/41}

\bibitem[{{Zorotovic} {et~al.}(2011){Zorotovic}, {Schreiber}, \&
  {G{\"a}nsicke}}]{zsg11}
{Zorotovic}, M., {Schreiber}, M.~R., \& {G{\"a}nsicke}, B.~T. 2011, \aap, 536,
  A42, \dodoi{10.1051/0004-6361/201116626}

\bibitem[{{Zorotovic} {et~al.}(2016){Zorotovic}, {Schreiber}, {Parsons},
  {G{\"a}nsicke}, {Hardy}, {Agurto-Gangas}, {Nebot G{\'o}mez-Mor{\'a}n},
  {Rebassa-Mansergas}, \& {Schwope}}]{zsp+16}
{Zorotovic}, M., {Schreiber}, M.~R., {Parsons}, S.~G., {et~al.} 2016, \mnras,
  457, 3867, \dodoi{10.1093/mnras/stw246}

\end{thebibliography}
\bibliographystyle{aasjournal}



\appendix


\section{A Shorthand Catalogue Version}
\label{sec:shorthandVersion}
\begin{deluxetable}{cl}
    \tabletypesize{\scriptsize}
    \tablecaption{
        Content of the columns in the catalogue. The column numbers in brackets are the columns with the low and upper limits, respectively. We refer to the common-envelope primary as the donor of the possible CE and the secondary its in-spiralling companion. A list of the type and flags is given in table~\ref{tab:dataObsShort} and explained in \S~\ref{sec:overview} and \S~\ref{sec:usage}.\label{tab:columns}
    }
    \tablewidth{\columnwidth}
    \tablehead{
        \colhead{column} & \colhead{description}
    }
    \startdata
        1 & names\\
        2 & period in days\\
        3 (7,8) & primary mass in $\Msun$\\
        4 (9,10) & secondary mass in $\Msun$\\
        5 & primary type\\
        6 & secondary type\\
        11 (12,13) & mass ratio = primary over secondary mass\\
        14 (15,16) & semi-major axis in $\Rsun$\\
\enddata
    \tablenotetext{S}{\url{http://simbad.u-strasbg.fr/simbad/}}
    \tablenotetext{G}{\url{https://www.cosmos.esa.int/web/gaia/earlydr3}}
\end{deluxetable}
Table~\ref{tab:columns} gives a summary description of the columns which are available in the catalogue. The first 6 columns plus column 50 and 59 are given in table~\ref{tab:dataObsShort}. If there is no value in the columns 3 or 4 a lower or upper limit is taken from the columns 7 to 10 accordingly.

\startlongtable
\begin{deluxetable*}{cDDDcchcc}
    \tabletypesize{\scriptsize}
    \tablecaption{
        Shorthand version of the catalogue showing the name(s), the period, the masses (sometimes only limits) and types of the two stars, a flag to indicate possible issues, and the reference(s).\label{tab:dataObsShort}
    }
    \tablewidth{\textwidth}
    \tablehead{
        \colhead{name(s)} & \multicolumn2c{period} & \multicolumn2c{mass$_1$} & \multicolumn2c{mass$_2$} & \colhead{type$_1$} & \colhead{type$_2$} & \nocolhead{...} & \colhead{flag(s)} & \colhead{reference(s)}\\
        \colhead{} & \multicolumn2c{$[\mathrm{d}]$} & \multicolumn2c{$[\Msun]$} & \multicolumn2c{$[\Msun]$} & \colhead{} & \colhead{} & \nocolhead{} & \colhead{} & \colhead{}
    }
    \decimals
    \startdata
        RXJ0806.3+1527/HMCnc & $0.0037214020$ & $0.55$ & $0.27$ & WD & WD &  & MT & 330,407,343\\
ZTFJ1539+5027 & $0.004800828014$ & $0.610$ & $0.210$ & WD & WD &  & - & 15\\
ZTFJ2243+5242 & $0.00611035664$ & $0.349$ & $0.384$ & WD & WD &  & - & 16\\
V407Vul/RXJ1914.4+2456 & $0.00659022$ & $0.8$ & $0.177$ & WD & - &  & MT/S & 330,200\\
ESCet & $0.007178376$ & $0.8$ & $0.161$ & WD & - &  & MT/S & 99,200\\
4U1820-303 & $0.00793$ & $0.055$ & $1.58$ & WD & NS &  & MT & 396,138\\
J0651+2844 & $0.00885655721$ & $0.247$ & $0.49$ & WD & WD &  & - & 32,154,22,15\\
ZTFJ0538+1953 & $0.010030131$ & $0.45$ & $0.32$ & WD & WD &  & - & 38\\
\enddata
\end{deluxetable*}%


\section{Additional Figures}
\label{sec:addFigures}
\begin{figure}
    \centering
    \includegraphics[]{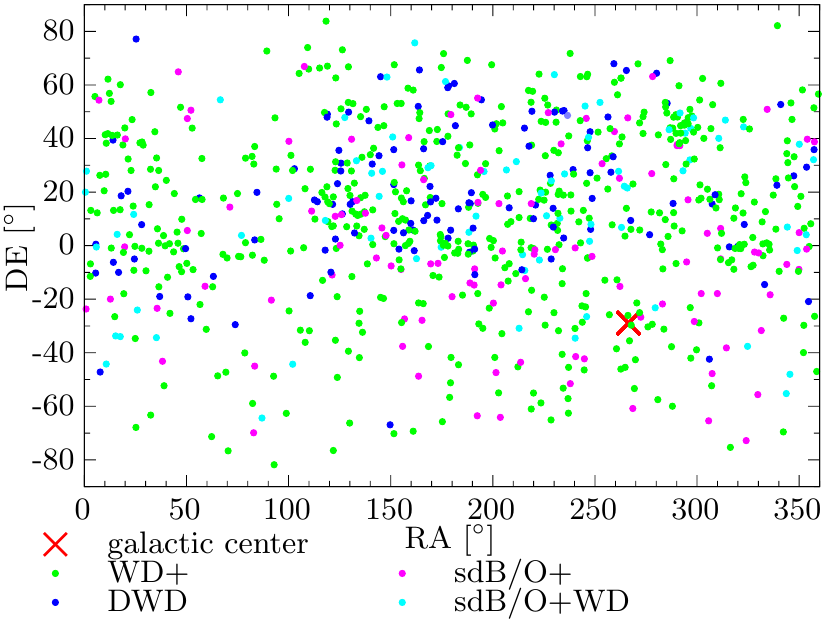}
    \caption{Positions of the catalogue members, in J2000 coordinates. The type of the system is colour coded. The red cross marks the position of the galactic centre.}
    \label{fig:location}
\end{figure}
The collection from many different observational campaigns result in a good sky coverage, see Fig.~\ref{fig:location}. 

\begin{figure*}
    \centering
    \includegraphics[]{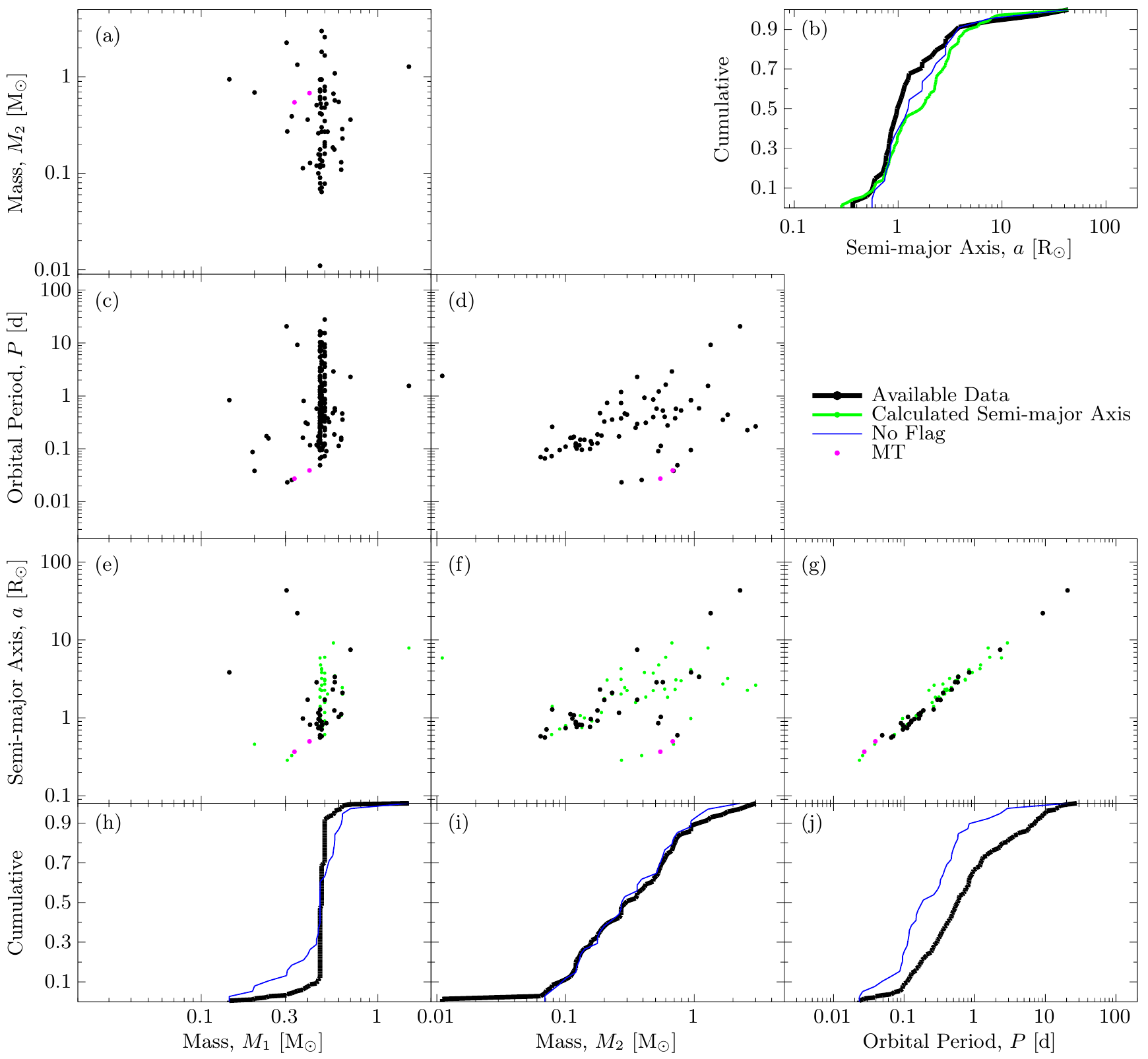}
    \caption{Similar to Fig.~\ref{fig:dataObs} but showing only systems with an sdB/O star, cf. second column in table~\ref{tab:dataObsOveriew}. Additionally, the mass transferring systems are marked in purple.}
    \label{fig:dataObssdBp}
\end{figure*}
\begin{figure*}
    \centering
    \includegraphics[]{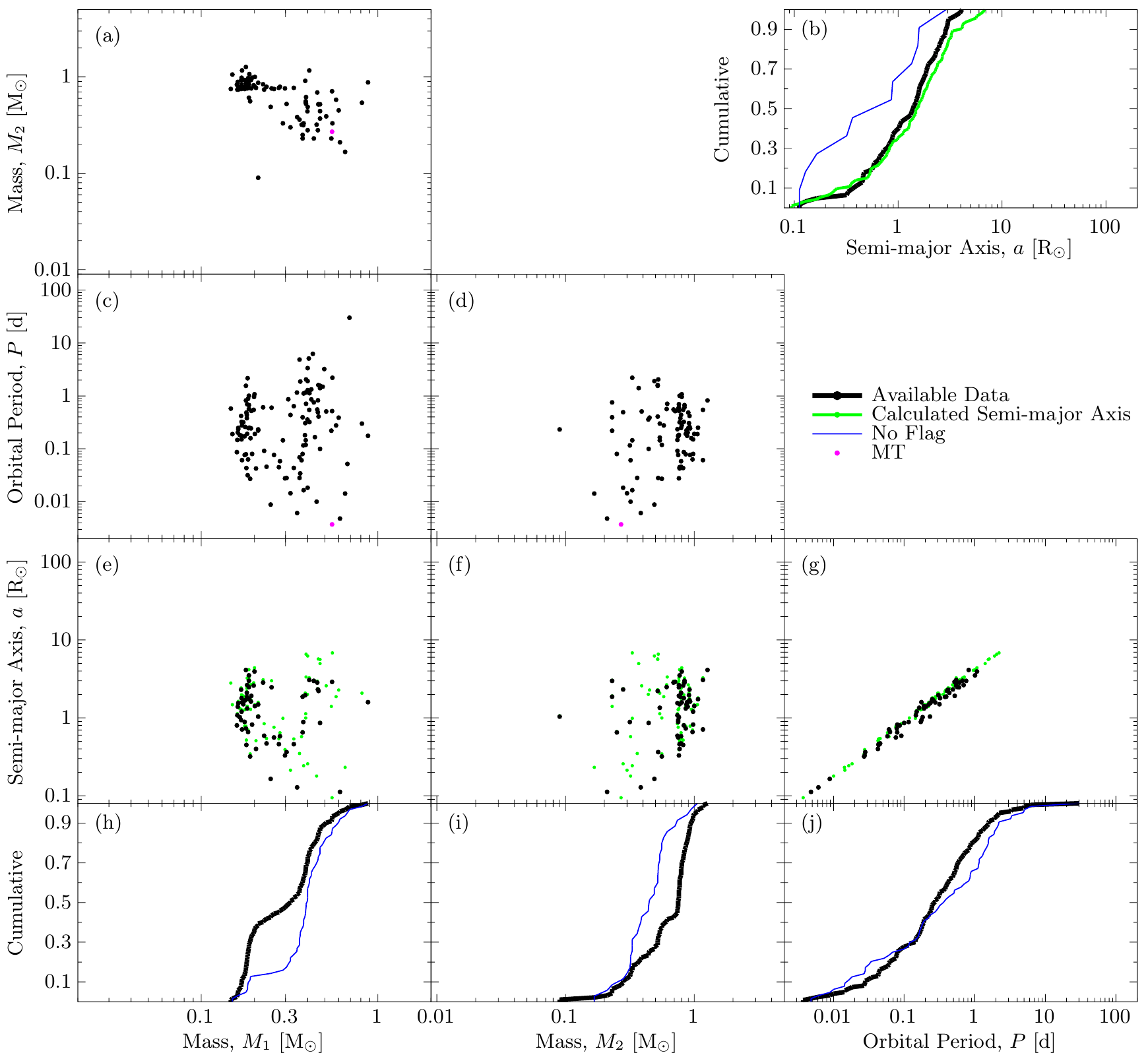}
    \caption{Similar to Fig.~\ref{fig:dataObs} but showing only double white dwarf systems, cf. third column in table~\ref{tab:dataObsOveriew}. Additionally, the mass transferring systems are marked in purple.}
    \label{fig:dataObsDWD}
\end{figure*}
\begin{figure*}
    \centering
    \includegraphics[]{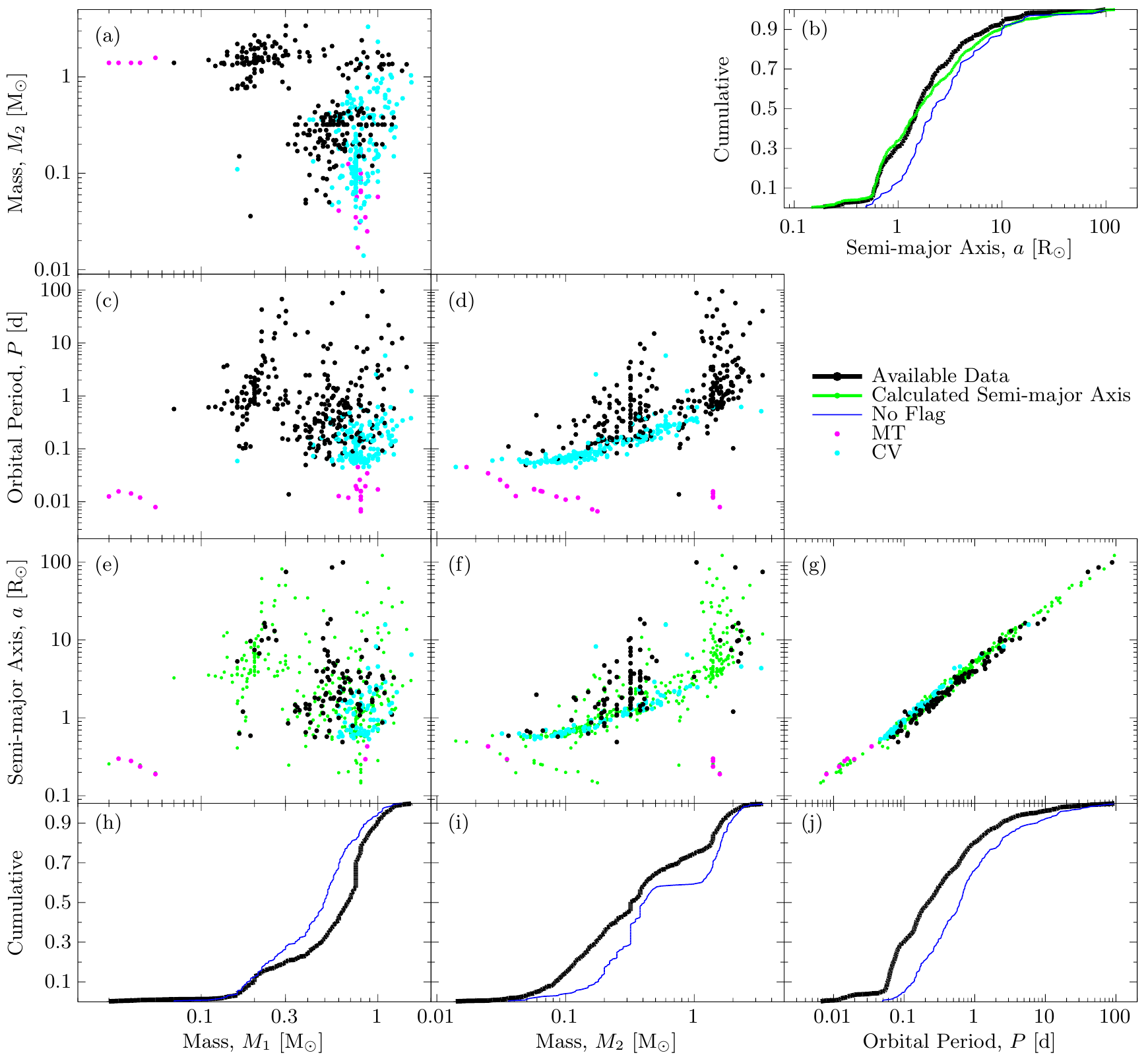}
    \caption{Similar to Fig.~\ref{fig:dataObs} but showing only WD systems with non WD companions, cf. fourth column in table~\ref{tab:dataObsOveriew}. Additionally, the mass transferring systems are marked in purple and the systems showing observational features of CVs are marked in teal.}
    \label{fig:dataObsWDp}
\end{figure*}
Figures~\ref{fig:dataObssdBp} to \ref{fig:dataObsWDp} show again the basic parameter but for the three main classes of systems separately.


\section{Very Close Triples}
\label{sec:CloseTriples}
Lidov-Kozai Resonance requires an inclination between the inner and outer orbital plane. To date, this is perhaps the most thoroughly-studied effect a tertiary can have.

For closer tertiaries, which are almost but not quite close enough to exchange matter with the inner binary, tertiary tidal effects can drive the inner binary into tighter orbits \citep{gce+18,gtg+20,fdb+13}. Even closer tertiaries can directly undergo RLOF, resulting in an even greater impact on the inner binary \citep{dis19,dis20}. For those tertiaries that lie yet even closer to their companion inner binaries, they themselves may be responsible for the CE, although whether or not the subsequent system can survive as a triple is questionable according to recent studies \citep{ci20,gp21}. However, as these processes were only recently brought to the attention of the astrophysical community, frantic efforts to make sense of them are still underway, and one should exercise caution when estimating the magnitude of their influence.

\section{Supernova Progenitor Scenarios Applicable to the Systems in This Catalogue}
\label{sec:SN}
Some of the tighter systems contained in this catalogue are likely to interact again within a Hubble time, which will, depending on the current structure of the system, lead to the formation of either stellar remnants of different sorts or transient events. This section is intended to provide a brief overview of the most important transients and remnants to be expected, though we encourage the user of this catalogue to analyse individual systems with regard to their most likely outcome.

Any conceivable SN mechanism applicable to the systems contained in this catalogue, due to none of the companions lying in a mass range subject to any SN mechanism accessible to single stars, must involve mass transfer or violent mergers of the companion stars. This results in either a thermonuclear supernova or (in ONeMg WDs) electron capture and collapse into a NS (accretion induced collapse). 
Progenitor scenarios for thermonuclear SNe are usually grouped, depending on the terminal state of the system, into single degenerate \citep[see, e.g. the classical paper by][and newer sources below]{wi1973} and double degenerate channels \citep[see e.g. classical papers by][and newer sources below]{it1984,w1984}, the former denoting involvement of one WD and one non-degenerate star, the latter suggesting involvement of two WDs. In the literature, total system masses both exceeding and falling short of the Chandrasekhar mass have been proposed as possible SN type Ia (and related) progenitors \citep[see e.g.][]{ty1994,N1982a,N1982b}. We refrain from making any predictions on the eventual outcome of our systems, but instead give a short summary of possible progenitor channels.

\begin{description}
    \item[Single degenerate hydrogen donor] In this scenario, the donor star is assumed to be either a hydrogen-rich main sequence star or an evolved star with a hydrogen rich envelope. This star then donates material to its companion WD. Unstable ignition of the accumulated material is associated with cataclysmic variable evolution. The hydrogen may also be processed through stable burning into helium, which may then ignite, triggering a secondary detonation in the WD's CO core (this is known as the double detonation scenario) or, when the WD approaches the Chandrasekhar mass, a detonation is triggered in the WD's centre as the degenerate electron gas becomes unstable against further gravitational collapse (i.e. the classic mechanism according to Chandrasekhar) \citep[see e.g.][the latter being a recent review]{hn2000,r2020}.
    \item[Single degenerate helium donor] Alternatively, the non-degenerate component is a hydrogen depleted star. The physical radii of these stars are about one order of magnitude smaller than those of hydrogen rich stars of the same mass. In this case, the system is required to be much closer than in the hydrogen donor case, which is conducive to the argument that most SN progenitors of this type are necessarily post CE systems. 
    At high mass transfer rates, the helium will either be processed stably into carbon and oxygen or undergo a sequence of unstable helium ignitions, resulting in successively more and more massive He novae, which may, if sufficient CO can be built up, result in a Chandrasekhar-like detonation. At low rates, the helium may be accumulated quiescently, building up until it ignites explosively, which may then trigger a secondary explosion of the CO core \citep[][]{N1982a,N1982b,r2020,WW1994,WK2011,nyl2016,NYL2017}.
    \item[Double degenerate] Systems containing two WDs have been proposed as potential progenitors of thermonuclear SNe \citep[][]{it1984,w1984,pkr2010,FHR2007,FRH2010,SRH2010,KSF2010}. Here, mass transfer is invariably initiated through the effects of GWR, with simulations predicting either merging of the two WDs, followed by a detonation (the 'classical' double degenerate mechanism) or dynamical ignition of the WD by the infalling matter stream. In the latter case, ignition may have to be catalysed by the presence of an unburnt layer of helium, either previously accreted during a post-CE mass transfer phase (e.g. with the less massive donor in the state of an sdB star), or a remnant from other stages of stellar evolution prior to becoming a WD.
\end{description}




\end{document}